\documentclass[11pt,a4paper]{article}

\usepackage{amsmath}
\usepackage{amssymb}
\usepackage{geometry}
\usepackage{booktabs}
\usepackage{tabularx}
\usepackage{hyperref}
\usepackage{lineno}
\usepackage[numbers,sort&compress]{natbib}
\usepackage{siunitx}
\usepackage{wasysym}
\usepackage{graphicx}
\usepackage{xcolor}
\usepackage{enumitem}
\usepackage{fancyhdr}
\usepackage{array}
\usepackage{longtable}
\usepackage{caption}
\usepackage{subcaption}
\usepackage{tikz}
\usetikzlibrary{arrows.meta, decorations.pathmorphing, decorations.markings,
	shapes.geometric, positioning, calc, backgrounds}

\hypersetup{
	colorlinks=true,
	linkcolor=blue!60!black,
	citecolor=magenta,
	urlcolor=blue!60!black
}

\title{Quantum noninvasive three-component beam-spin polarimetry \\in the Hadron Storage Ring of the Electron-Ion Collider}

\author{Frank Rathmann\\[0.3em]
Brookhaven National Laboratory} 
\date{\today}

\newcommand{\ex}{\vec{e}_x}
\newcommand{\ey}{\vec{e}_y}
\newcommand{\ez}{\vec{e}_z}
\newcommand{\ephi}{\vec{e}_\phi}
\newcommand{\rcoil}{r_\mathrm{coil}}
\newcommand{\dd}{\mathrm{d}}
\newcommand{\Rmat}{\mathbf{R}}

\begin{document}
	
\maketitle

\begin{abstract}
	We propose a noninvasive SQUID-based polarimeter for the polarized proton beam in the Electron-Ion Collider (EIC) Hadron Storage Ring (HSR), exploiting the collective magnetic dipole moment of the bunches rather than scattering. The six-snake HSR lattice has synchronous-particle spin tune $\nu_s = 1/2$, placing the in-plane spin-precession signal at half the revolution frequency ($\sim$39 kHz), in the DC SQUID band. Three pickup channels (cosine-$\theta$ and sine-$\theta$ saddle loops for the transverse components, a coaxial axial gradiometer for the longitudinal one) reconstruct the full polarization vector $(P_x, P_y, P_z)$ in two complementary modes. Static mode, the default for continuous noninvasive monitoring, reads all three components: $P_y$ at the revolution frequency and the residual in-plane components at $\nu_s f_\mathrm{rev}$, bunch by bunch over an hours-long fill, including $P_z$, inaccessible to single-spin scattering polarimetry by parity conservation. Dynamic mode gives a precise polarization-magnitude measurement: a longitudinal kicker tips a small fraction of the polarization into the horizontal (ring) plane to produce a free-induction-decay (FID) signal, and many phase-locked tip-$\pi$-echo-restore cycles are summed coherently via a matched filter across all bunches, with $O(\alpha^2/\pi^2) \sim 10^{-4}$ loss per cycle, negligible over a full $\delta P/P = 1\%$ measurement. For tipping angle $\alpha = 30$ mrad, polarization $P = 0.7$, and effective rms spin-tune spread $\sigma_{\nu_s}^\mathrm{eff} = 10^{-3}$ (coherence time $\sim$2 ms), the integration time to reach $\delta P/P = 1\%$ is about 18 s at injection and 5 min at flattop. The architecture extends to deuteron and $^3$He beams via species-specific spin-magnetic factors, with applications to storage-ring EDM searches.
	\end{abstract}

\tableofcontents

\section{Introduction}

Conventional proton beam polarimetry at high-energy colliders relies on nuclear scattering reactions: elastic proton-carbon (pC) scattering for fast relative measurements, or Coulomb-Nuclear Interference (CNI) scattering from a polarized atomic-hydrogen jet (HJET) for absolute calibration~\cite{Rathmann2026PRAB}. Both methods intercept the stored beam with a target (a thin carbon strip for pC, an atomic-hydrogen jet traversed by the full stored beam for HJET), causing measurable beam losses and heating. Beyond this beam-loss sense, pC polarimetry also relies on thin carbon-strip target technology with finite thermal, mechanical, and lifetime margins: a recent study of the thermal and electromechanical response of these targets in relativistic bunched beams indicates that the technology has operational limits and may become inadequate under the more demanding EIC beam conditions, strengthening the motivation for complementary noninvasive diagnostics~\cite{Rathmann2026CarbonTarget}.

The EIC physics program demands a relative polarization uncertainty $\delta P/P \leq 1\%$ (where $P$ is the magnitude of the beam polarization vector and $\delta P$ is its absolute uncertainty), bunch-by-bunch measurement capability, and full characterization of the polarization vector $\vec{P} = (P_x, P_y, P_z)$~\cite{eic_global_req}. Existing storage-ring polarimetry at RHIC measures the transverse projections $(P_x, P_y)$ bunch by bunch via pC azimuthal sampling, with the absolute $P_y$ scale calibrated by the HJET polarimeter operated with a vertical target polarization~\cite{Rathmann2026PRAB}. Dedicated procedures at RHIC combined the transverse pC/STAR measurements with controlled spin rotations and offline transport to recover a fill-averaged three-dimensional stable spin direction~\cite{Schoefer:2024TECH}, but cannot follow $\vec P(t)$ bunch by bunch or continuously across a store. The upgraded HJET planned for the EIC will additionally measure $P_x$ and $P_z$ via target-polarization rotation, using techniques established for $pp$ spin-correlation measurements~\cite{PhysRevC.58.658}; even with this upgrade, however, no current or planned scattering polarimeter reconstructs the full polarization vector bunch by bunch and continuously throughout the fill. Given the tenfold reduction in bunch spacing relative to RHIC (\SI{11.0}{ns} vs.\ \SI{106.6}{ns}) and the threefold increase in average beam current, noninvasive alternatives to scattering-based polarimetry are of particular interest. The three-channel SQUID polarimeter proposed in this paper addresses this gap, supplementing the established scattering-based capabilities  with continuous, noninvasive, bunch-resolved measurement of the full polarization vector $\vec P(t)$.

Earlier noninvasive beam-polarimetry concepts pursued two related routes for accessing the collective spin signal electromagnetically. One approach used RF-resonant Stern-Gerlach interactions, in which the magnetic moment of a polarized beam couples to selected cavity modes; this led to proposals for spin-state separation and absolute polarimetry in storage rings~\cite{Derbenev1995RFResonancePolarimeter, Conte2000SternGerlachInteraction}. A passive RF cavity for such a Stern-Gerlach polarimeter was carried as far as prototype fabrication and bench characterization for an eventual test with a transversely polarized electron beam at the MIT-Bates storage ring~\cite{Cameron2005SternGerlachCavity}, and the relativistic Stern-Gerlach deflection of polarized electron beams has continued to be developed theoretically and in proposed experimental configurations~\cite{Talman2016RelativisticSternGerlach, Mane2016SternGerlachHamiltonian}; to date, however, none of these electromagnetic Stern-Gerlach schemes has been demonstrated on a beam. A second approach, closer in spirit to the present work, proposed direct SQUID detection of the magnetic moment of polarized RHIC beams using a pickup loop coupled to a SQUID readout, favoring a room-temperature loop for practical accelerator installation; that proposal already identified the central difficulty, namely that the ordinary beam-current field exceeds the spin field by many orders of magnitude and must be separated by pickup geometry and frequency-domain methods~\cite{Cameron1997SquidsSnakesPolarimeters}. The present study revisits this direct magnetic-moment approach for the EIC HSR, using the six-snake $\nu_s=1/2$ spin lattice, the EIC multi-bunch fill pattern, and matched-filter summation of the bunch-resolved SQUID signal.

A closely related and concurrent proposal also treats the stored-beam polarization as a continuous, phase-coherent observable read out noninvasively from the spin-dependent electromagnetic field of the beam~\cite{Kim2026CoherentPolarimetry}. That work couples symmetric pickup electrodes to a high-$Q$ radio-frequency resonator tuned to a MHz bunch-train harmonic, and extracts the electric dipole moment of counter-propagating beams in a dedicated frozen-spin ring from the clockwise-minus-counter-clockwise difference of a controlled spin-wheel precession frequency. The present scheme differs in its detector, frequency regime, observable, and machine. We read the bunch magnetic moment directly with a SQUID flux pickup rather than an electrode-coupled resonator; we operate at the spin-precession sideband $f_s = f_\mathrm{rev}/2 \simeq \SI{39}{\kilo\hertz}$ set by the six-snake $\nu_s = 1/2$ lattice rather than at a MHz bunch harmonic; we reconstruct the full polarization vector $(P_x, P_y, P_z)$ bunch by bunch as a routine beam diagnostic rather than a single electric-dipole-moment frequency difference; and we target the EIC HSR collider with a single circulating beam and matched-filter summation across the EIC fill, rather than a dedicated counter-rotating ring. The two approaches are complementary: the resonant electrode readout is favored at the MHz harmonics of that proposal, where SQUID magnetometry is less natural, whereas the SQUID is the natural sensor at the low-frequency $\nu_s = 1/2$ sideband exploited here. Ref.~\cite{Kim2026CoherentPolarimetry} also identifies polarized-proton and light-ion rings at the EIC as a target for its architecture; the present work develops the complementary SQUID-based, bunch-resolved, full-vector realization specific to the EIC HSR spin lattice.

A SQUID magnetometer can detect the net macroscopic magnetic dipole moment of a polarized bunch without any material interaction with the beam. This paper develops a quantitative design and feasibility analysis of this approach for the EIC HSR, taking full account of the spin dynamics imposed by the six Siberian snakes, the multi-bunch fill structure with its alternating spin-sign pattern, and the matched-filter reconstruction that recovers the full multi-bunch coherence from a SQUID time-stream sampled at the bunch rate. The detector concept uses a three-channel local pickup basis: a saddle cosine-$\theta$ loop sensitive to the horizontal transverse dipole component, a saddle sine-$\theta$ loop sensitive to the vertical component, and a coaxial axial gradiometer sensitive to the longitudinal component. The three channels are read out simultaneously by independent low-noise dc-SQUID front ends and reconstruct $(P_x, P_y, P_z)$. The polarimeter is designed to operate in two complementary modes. In \emph{static} mode (no active spin manipulation), the stored vertical polarization $P_y$ appears at the revolution frequency and any residual transverse and longitudinal components precess at $\nu_s f_\mathrm{rev}$; all three are read continuously and bunch by bunch throughout the fill, giving direct access to the longitudinal projection $P_z$ that is inaccessible to single-spin scattering polarimetry by parity conservation. In \emph{dynamic} mode, a longitudinal kicker tips the stored polarization by a controlled small angle to produce a free-induction-decay (FID) signal, and many phase-locked tip-$\pi$-echo-restore cycles are accumulated coherently via matched-filter summation across all bunches; the closing restore pulse returns the tipped in-plane component to the stable spin axis, so the cumulative cost in beam polarization across a full $\delta P/P = 1\%$ measurement remains negligible. The full engineering implementation including saddle-coil topology, mechanical mounting, magnetic shielding, and cryogenic readout chain is the subject of a dedicated companion paper; the present work establishes the principle, the sensitivity budget, and the operational concept. The same architecture extends to deuteron and $^3$He beams via species-specific spin-magnetic factors, with storage-ring electric-dipole-moment applications treated in a dedicated follow-up paper.

The concept is related to the free-precession spin-tune measurement demonstrated by the JEDI collaboration at COSY~\cite{Eversmann:2015}, where the spin vector of a coherently precessing ensemble was sampled turn-by-turn using an internal scattering polarimeter. The JEDI program has produced a series of advances directly relevant to the present work: the establishment of long in-plane spin coherence times through chromaticity correction and sextupole tuning~\cite{guidoboni2018}, spin-tune mapping as a precision diagnostic~\cite{saleev2017}, the demonstration of phase-locking the spin precession~\cite{hempelmann2018}, and a comprehensive framework for spin dynamics in storage rings~\cite{rathmann2020}. The deuteron EDM program has further developed the tools for precision spin manipulation and noninvasive polarimetry~\cite{CPEDM:2019nwp,PhysRevResearch.7.023257,10.1063/1.4967465}. Without dedicated machine development effort analogous to that at COSY, in particular sextupole tuning to correct chromaticity-driven spin decoherence, achieving long spin coherence times in the HSR cannot be assumed. The SQUID approach replaces the destructive internal polarimeter used at COSY with a passive magnetic pickup, making the measurement fully noninvasive.

Beyond the absolute calibration use case, the static-mode SQUID readout feeds back to the spin lattice and the operating point. With per-bunch matched filtering on the saddle and axial channels, drifts and localized anomalies in $(P_x, P_y, P_z)$ can be resolved bunch by bunch with few-percent precision in measurement windows of order one minute, providing real-time feedback on the machine spin state. The EIC HSR will require tuning of the spin lattice, snake configuration, cooling state, emittance, momentum spread, and bunch intensity to preserve polarization over stores lasting several hours. RHIC experience shows that spin-tune spread is an operationally important and tunable quantity: matching
the dispersion-function derivatives at the Siberian snakes reduced the spin-tune spread and substantially improved spin-flip efficiency~\cite{Liu2019SpinTuneSpread}. At the same time, polarization transmission during acceleration was not determined by spin-tune spread alone, but also depended on deviations from $\nu_s=1/2$, snake resonances, and interference between intrinsic and imperfection resonances~\cite{Liu2019SpinTuneSpread}. High-intensity operation adds further dependence on bunch population, emittance growth, momentum tails, beam-beam tune spread, and intrabeam scattering, as observed in RHIC polarized-proton stores~\cite{LuoFischer2014BeamBeam}. A noninvasive pickup that directly observes the free spin precession after a controlled tipping pulse would therefore provide a new diagnostic observable: the spin-precession frequency, linewidth, coherence envelope, and response to spin-echo or restore-pulse sequences. Such a measurement would be valuable not only for polarimetry, but also for
tuning the HSR spin lattice and operating point.

The paper is organized as follows. Section~\ref{sec:measurement_concept} sets the measurement concept and the HSR six-snake spin lattice. Section~\ref{sec:spin-dynamics-and-signal-geometry} develops the spin dynamics, the bunch magnetic dipole moment, and the FID protocol. Section~\ref{sec:fid_spin_tune_phase_lock} treats the initial spin-tune diagnostic via spectral search and phase lock. Section~\ref{sec:fid-echo-sequence} presents the spin-echo rephasing that closes the tip-$\pi$-restore cycle and makes the measurement noninvasive. Section~\ref{sec:flux-and-matched-filter} establishes the matched-filter reconstruction and the resulting per-bunch sensitivity. Section~\ref{sec:design-overview} overviews the three-channel SQUID polarimeter pickup. Sections~\ref{sec:static_vector_extension} and~\ref{sec:pz_axial_channel} extend the treatment to static transverse vector polarimetry and to the longitudinal $P_z$ via the axial gradiometer. Section~\ref{sec:bunch-resolved-vec-polarimetry} develops
bunch-resolved $(P_x, P_y, P_z)$ reconstruction over the full fill as the operational deliverable. An appendix covers spin manipulation with an RF Wien filter, with an emphasis on the two-branch resonance degeneracy that arises at $\nu_s = 1/2$ in the HSR.

\section{Measurement concept and HSR spin lattice}
\label{sec:measurement_concept}

The SQUID polarimeter considered here measures the magnetic field produced by the collective magnetic dipole moment of the stored polarized bunches. It therefore does not require an internal target or a scattering reaction. The relevant signal is not the ordinary beam-current field, but the much smaller field associated with the aligned proton magnetic moments. The measurement concept uses the known HSR spin lattice to place this spin signal at a well-defined frequency and then uses bunch-resolved phase information to recover it coherently.

\subsection{Six-snake spin lattice}
\label{subsec:six_snake_spin_structure}

In the EIC HSR, six full Siberian snakes are used to enforce the synchronous-particle spin tune
\begin{equation}
	\nu_s = \frac{1}{2} \, .
	\label{eq:hsr_spin_tune_half_sec2}
\end{equation}
For the Lee-Courant configuration considered here, the stable spin axis in suitable arc regions is approximately vertical,
\begin{equation}
	\vec n_0 \simeq \ey \, .
	\label{eq:arc_stable_axis_vertical_sec2}
\end{equation}
A spin component perpendicular to $\vec n_0$ therefore changes sign from one turn to the next. This turn-by-turn sign reversal is the central feature used by the SQUID measurement. It shifts the transverse spin signal to half the revolution frequency, while the ordinary beam-current field remains tied to the usual revolution harmonics and bunch pattern.

The detailed stable-spin-axis calculation and the corresponding six-snake layout are discussed below. In this section we only use the qualitative consequence of Eq.~\eqref{eq:hsr_spin_tune_half_sec2}: after a controlled tipping pulse, the transverse component of the bunch magnetic moment oscillates with the spin phase and reverses sign on successive turns.

\subsection{Transverse FID measurement}
\label{subsec:transverse_fid_measurement}

\begin{figure}[htb]
	\centering
	\includegraphics[width=0.5\textwidth]{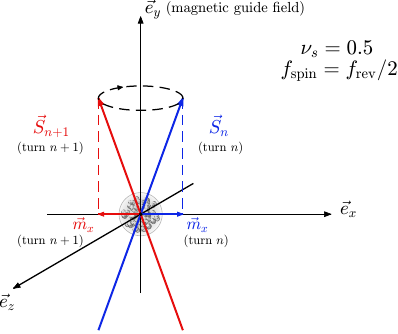}
	\caption{Schematic of the spin-precession geometry for a guide field along the vertical direction $\ey$. A tipping pulse creates a transverse magnetic moment $\vec m$ of the bunch. For $\nu_s=0.5$, the transverse projection advances by $\pi$ per turn, so the $\ex$ component changes sign between turns $n$ and $n+1$. This produces the spin signal at $f_s=f_{\mathrm{rev}}/2$.}
	\label{fig:spin_precession}
\end{figure}

A short tipping pulse (produced by a longitudinal magnetic field $B_z$ applied for a fraction of a microsecond, as detailed in Sec.~\ref{sec:tipping_intro}; a rotation about $B_z$ is a  feasible option here, since it does not produce a transverse kick on the closed orbit) applied in a region with $\vec n_0\simeq\ey$ rotates a small fraction of the stored vertical polarization into the ring plane. Immediately after this pulse, the bunch carries a coherent transverse magnetic moment. As illustrated in Fig.~\ref{fig:spin_precession}, the transverse projection changes sign on successive turns for $\nu_s=0.5$. This produces a free-induction-decay (FID) signal in the SQUID pickup at
\begin{equation}
	f_s = \nu_s f_{\mathrm{rev}} \simeq \frac{1}{2}f_{\mathrm{rev}} \, .
	\label{eq:spin_signal_frequency_sec2}
\end{equation}
For the EIC HSR this is near \SI{39}{kHz}, in the useful bandwidth range of wideband DC SQUID readout.

The measurement sequence is therefore a tip-read-restore cycle. The tipping pulse creates the transverse spin component, the SQUID pickup records the FID signal over the available spin-coherence window, and a phase-locked restore pulse returns the in-plane component close to the stable-spin direction. The numerical baseline used later is a tipping angle of $\alpha = \SI{30}{mrad}$ and an effective rms spin-tune spread $\sigma_{\nu_s}^{\mathrm{eff}}=10^{-3}$ (Table~\ref{tab:params}), with these values entering only when the coherence time and signal-to-noise ratio are estimated.

\subsection{Local vector-pickup basis}
\label{subsec:local_vector_pickup_basis}

The transverse FID signal can be detected with two orthogonal pickup channels. A cosine-$\theta$ pickup couples to the transverse magnetic-field pattern associated with the local $P_x$ component, while a sine-$\theta$ pickup couples to the corresponding $P_y$ component. In this basis, the two transverse channels provide local sensitivity to
\begin{equation}
	P_x
	\quad\mathrm{and}\quad
	P_y \, .
	\label{eq:px_py_channels_sec2}
\end{equation}

\begin{figure}[htb]
	\centering
	\includegraphics[width=0.55\textwidth]{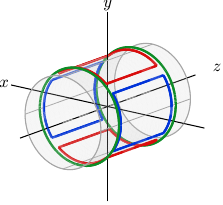}
	\caption{Conceptual schematic of the three-channel SQUID pickup basis on a common cylindrical support around the beam pipe. The beam direction is along the cylinder axis. The blue saddle-loop pair represents the cosine-$\theta$ transverse pickup, which is sensitive to the local $P_x$ component. The red saddle-loop pair represents the sine-$\theta$ transverse pickup, rotated by $90^\circ$ around the cylinder and sensitive to the local $P_y$ component. The green pair of coaxial loops forms an axial gradiometer, sensitive to the longitudinal component $P_z$. The drawing is schematic and illustrates the pickup symmetries and common geometric center, not the detailed winding implementation.}
	\label{fig:pickup_system_schematic}
\end{figure}

The same pickup platform can be extended with an axial loop-gradiometer channel. This channel is sensitive to the longitudinal component
\begin{equation}
	P_z \, ,
	\label{eq:pz_channel_sec2}
\end{equation}
and is geometrically insensitive to the ideal azimuthal magnetic field of a centered beam current. Figure~\ref{fig:pickup_system_schematic} shows this three-channel pickup basis schematically on a common cylindrical support around the beam pipe.

In the arc locations emphasized here, the stored polarization is expected to be close to vertical before the tipping pulse, so the transverse FID channels are the primary measurement mode. The axial channel is nevertheless useful as part of a local vector-polarimetry basis and as a relative diagnostic for longitudinal spin components or spin leakage.

The three-channel pickup basis therefore groups the polarization measurement into two related parts. The cosine-$\theta$ and sine-$\theta$ channels measure the transverse FID signal and provide the main sensitivity to $P_x$ and $P_y$. The axial gradiometer provides the corresponding $P_z$ sensitivity. All three channels are gradiometric in the sense that they reject the dominant azimuthal beam-current field, but the rejection mechanism differs: the saddle pickups use their first-harmonic ($\cos\theta$ or $\sin\theta$) winding pattern to suppress the azimuthally symmetric current field, while the coaxial axial loop measures axial flux and is therefore intrinsically blind to that field by geometry. The detailed pickup geometry, field coupling, and common-mode rejection requirements are analyzed in later sections.

\subsection{Two-stage measurement strategy}
\label{subsec:two_stage_measurement}

The SQUID FID signal supports two operating modes that address different measurement goals. In the initial mode (Sec.~\ref{sec:fid_spin_tune_phase_lock}), a tipping pulse with a sufficiently large tipping angle is followed by a short observation within one coherence window, providing a signal-to-noise ratio of order unity in a single record before the spin coherence is known to last longer. This first record determines the spin-precession frequency $f_s = \nu_s f_\mathrm{rev}$ and the spin-tune spread $\sigma_{\nu_s}$ from the spectral peak and its linewidth, respectively.

In the precision mode (Sec.~\ref{sec:coherent-integration}), $f_s$ and the available coherence window are taken as known from the initial mode. With the tipping angle $\alpha = \SI{30}{\milli\radian}$, the closing restore pulse returns the in-plane component to the stable spin axis with a per-cycle loss of only $\mathcal{O}(\alpha^2/\pi^2) \sim 10^{-4}$, and many tip-read-restore cycles are accumulated coherently using matched-filter summation. Throughout each cycle the spin-precession phase is tracked continuously from a reference oscillator phase-locked to $f_s$~\cite{hempelmann2018}, so that the closing restore pulse is applied at the precession phase required to rotate the in-plane component back to the stable spin axis. The relative statistical precision on the polarization magnitude $P$ then improves as $T^{-1/2}$ in the total integration time $T$, with the beam polarization preserved across the measurement.

\section{Spin dynamics and signal geometry}
\label{sec:spin-dynamics-and-signal-geometry}

\subsection{Numerical input parameters}

We use the convention $\ez$ = beam direction (longitudinal), $\ey$ = vertical (along the arc dipole guide field), and $\ex$ = horizontal transverse throughout. The relevant EIC Hadron Storage Ring parameters are summarised in Table~\ref{tab:params}. All numerical values are taken from Table~II of Ref.~\cite{Rathmann2026PRAB}. The HSR reuses the RHIC Yellow ring tunnel (circumference $L = 3833.85$\,m). The revolution frequency at injection is slightly lower than at flattop because the proton velocity at \SI{23.5}{GeV} ($\gamma = 25.05$) is marginally below $c$.

At injection the HSR is filled with 290 bunches at \SI{44.1}{ns} spacing. After the energy ramp, RF bunch splitting increases the population to 1160 bunches at \SI{11.0}{ns} spacing. Each parent bunch produces four daughter bunches that inherit its polarisation sign. There are no plans to change the spin pattern after injection. Although noninvasive spin-pattern manipulation using a radio-frequency Wien filter appears technically feasible, as demonstrated at COSY and discussed in the context of future collider experiments~\cite{PhysRevResearch.7.023257},  such a device is not part of the baseline HSR design.

\begin{table}[htb]
	\centering
	\caption{Primary EIC HSR proton beam and restored-FID input parameters used in this analysis. Beam parameters are based on Table~II of Ref.~\cite{Rathmann2026PRAB}. Derived quantities such as revolution frequency, spin-precession frequency, coherence time, sideband linewidth, and SNR integration times are computed from these inputs.}
	\label{tab:params}
	\renewcommand{\arraystretch}{1.15}
	\sisetup{table-number-alignment=center}
	\begin{tabular}{l c c S[table-format=4.6] S[table-format=4.6]}
		\hline\hline
		{Parameter} & {Symbol} & {Unit} & {EIC injection} & {EIC flattop} \\
		\hline
		Ring circumference            & $C$                                  & \si{\meter}                & 3833.85 & {same} \\
		Total beam energy             & $E$                                  & \si{\giga\electronvolt}    &   23.5  & 275 \\
		Lorentz factor                & $\gamma$                             & {--}                       &   25.05 & 293.1 \\
		Relativistic factor           & $\beta$                              & {--}                       & 0.9992 & 0.999994 \\
		Protons per bunch             & $N_p$                                & $\times \SI{1e10}{}$       &   27.6  &   6.9 \\
		Number of bunches             & $N_b$                                & {--}                       &  290    & 1160 \\
		Bunch spacing                 & $T_b$                                & \si{\nano\second}          &   44.1  &  11.0 \\
		RMS bunch width               & $\sigma_t$                           & \si{\nano\second}          &    0.801 &   0.200 \\
		RMS bunch length              & $\sigma_L$                           & \si{\meter}                &    0.24 &   0.06 \\
		Average beam current          & $I_\mathrm{avg}$                     & \si{\ampere}               &    1.002 &   1.003 \\
		Beam polarization             & $P$                                  & {--}                       &    0.70 & {same} \\
		FID tipping angle             & $\alpha$                             & \si{\milli\radian}         &   30    & {same} \\
		Effective RMS spin-tune spread & $\sigma_{\nu_s}^{\mathrm{eff}}$     & {--}                       & {$\num{1.0e-3}$} & {same} \\
		Spin tune                     & $\nu_s$                              & {--}                       &    0.5  & {same} \\
		\hline\hline
	\end{tabular}
\end{table}

The effective value $\sigma_{\nu_s}^{\mathrm{eff}}=\num{1.0e-3}$ is used here as a working rms spin-tune-spread input for the FID coherence and matched-filter estimates. It should not be interpreted as a fixed machine constant: spin-tune spread and polarization transmission depend on the detailed HSR lattice, snake matching, momentum spread, and operational tuning state, as emphasized in HSR spin-tracking and snake-matching studies~\cite{Hamwi2024SnakeMatching,polar_trans}.

\subsection{Snake locations and spin tune}

\begin{figure}[htb]
	\centering
	\includegraphics[width=0.85\linewidth]{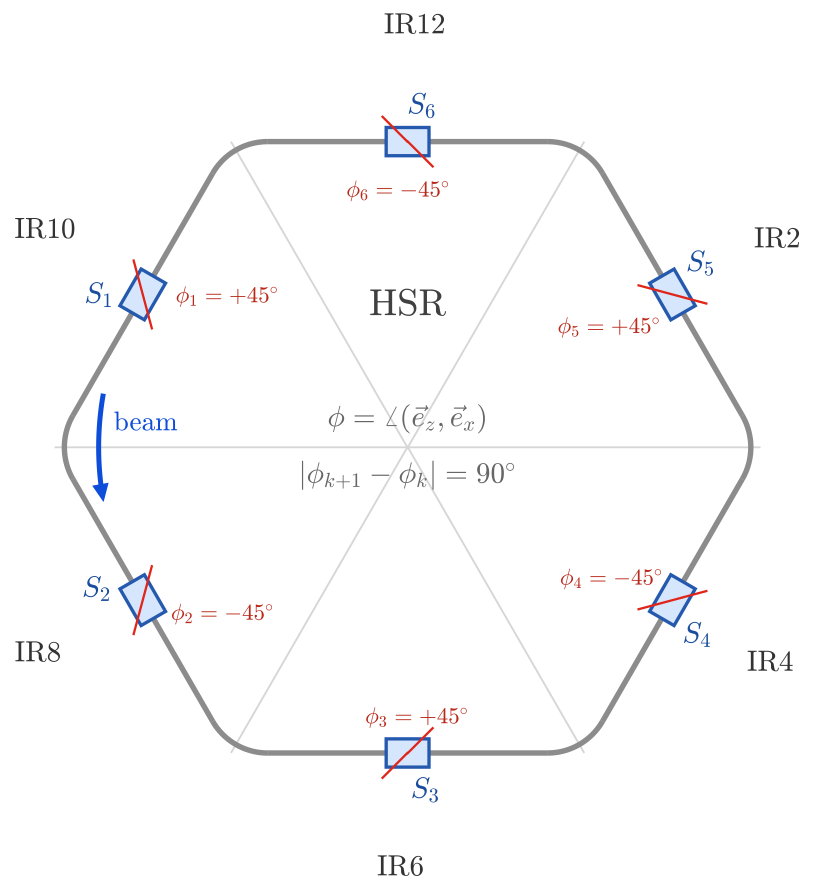}
	\caption{Schematic layout of the EIC HSR in hexagonal representation. The six interaction regions IR2--IR12 are located at the midpoints of the six straight sections. Each straight section hosts one Siberian snake. The red line through each snake box indicates the Lee--Courant rotation axis.}
	\label{fig:hsr_layout}
\end{figure}

The EIC HSR has six interaction regions and six helical Siberian snakes distributed around the ring, as shown in Fig.~\ref{fig:hsr_layout}. The figure shows the simplified hexagonal representation used in this note: six equal arc sectors separated by localized $180^\circ$ snake rotations. The red lines indicate the Lee--Courant snake-axis pattern used for the analytical spin model.

\subsection{Siberian snakes and the spin tune}
\label{sec:sib-snakes-and-spin-tune}

The HSR employs six helical superconducting Siberian snakes, each producing a spin rotation of $180^\circ$ about a horizontal axis. The six snakes are arranged as three pairs. Within each pair, the rotation axes are chosen to be perpendicular, so that each pair contributes an effective spin rotation by $\pi$. The product of the three pair rotations is again an odd multiple of $\pi$, which fixes the one-turn spin tune of the synchronous particle at
\begin{equation}
	\nu_s = \frac{1}{2}
\end{equation}
independently of $G\gamma$~\cite{polar_trans}. This removes the direct crossing of first-order imperfection resonances (at integer values of $G\gamma$) and of intrinsic spin resonances during acceleration.

The increase from two snakes in RHIC to six snakes in the HSR is driven by the stronger spin-resonance structure of the EIC lattice and by the need to accelerate additional polarized ion species. In particular, polarized $^3$He has an anomalous magnetic moment $G=-4.19$ (Table~\ref{tab:g_factors}), more than twice as large in magnitude as the proton value. Spin resonances are therefore correspondingly stronger for $^3$He, which further motivates the six-snake scheme.

Because each snake produces a full $180^\circ$ spin rotation, the six-snake configuration fixes $\nu_s=1/2$ independently of energy, so the half-integer spin tune can be maintained throughout acceleration rather than only at collision energy. When the spin rotators reach their nominal values at collision energy, the snake currents are adjusted simultaneously to compensate for the resulting spin-tune shift~\cite{Ptitsyn2022SpinRotators}. The detector solenoid does not change the spin tune, but it can perturb the local stable spin axis $\vec n_0(s)$. Matching the injected polarization direction from the AGS to the HSR stable spin direction is therefore still required.

It is important to distinguish the spin tune from the stable spin axis. The six-snake configuration fixes $\nu_s=1/2$ for the synchronous particle, but the local stable spin axis $\vec n_0(s)$ is determined by the full sequence of snake axes, arc spin rotations, spin rotators, detector solenoids, and local lattice fields. Thus $\vec n_0(s)$ is a local lattice quantity, not simply a consequence of the value $\nu_s=1/2$.

The snake-axis orientations $\{\phi_k\}$ are therefore free design parameters with two separate roles. They preserve the energy-independent spin tune, but they also determine the direction of $\vec n_0(s)$ and the suppression of higher-order spin resonances during the energy ramp. Of particular importance for the HSR is the Doubly Lee-Courant (DLC) configuration~\cite{polar_trans}, in which the snake-axis pattern is chosen to provide local cancellation of spin-phase errors and robust polarization transmission. In the idealized DLC picture, the stable spin axis remains close to vertical in suitable arc regions,
\begin{equation}
	\vec n_0(s) \simeq \ey \, .
\end{equation}
This vertical-axis approximation is the working assumption used in the simplified signal estimates below.

For the actual HSR implementation, however, the relevant quantity is the local $\vec n_0(s)$ map from the machine spin-tracking model. The placement and orientation of any spin-manipulation element, including the FID tipping element, must be chosen so that its spin-kick axis is transverse to the local $\vec n_0(s)$ at that location. In an arc region where $\vec n_0\simeq\ey$, a longitudinal field $B_z$ provides such a transverse spin kick that rotates a small fraction of the vertical polarization into the ring plane, where it precesses and flips the sign of the transverse moment $m_x$ on successive turns at $\tfrac{1}{2}f_\mathrm{rev}$, as shown in Fig.~\ref{fig:spin_precession}. If $\vec n_0(s)$ has a significant horizontal component at a candidate location, the required field orientation must be adjusted accordingly.

Table~\ref{tab:g_factors} also indicates how the SQUID signal generalizes to other polarized ion species. The SQUID couples to the bunch net magnetic dipole moment, which is by construction a rank-1 (vector) observable; tensor polarizations of any rank are not accessible. For spin-$1/2$ particles such as the proton and $^3$He, the polarization state is purely vectorial and the SQUID signal is proportional to $P$. For spin-1 species (deuteron, $^6$Li), the polarization state contains both a vector polarization $p_z$ and a rank-2 tensor polarization $p_{zz}$~\cite{Stone2016}; the SQUID measures $p_z$ and is blind to $p_{zz}$. For spin-$3/2$ ($^7$Li), ranks 0--3 are present; the SQUID still picks up only the rank-1 component. Rank-$k$ tensor components precess at $k\omega_s$, so higher ranks would alias to harmonics of the spin-precession sideband, but they produce no magnetic flux and therefore no SQUID signal at any harmonic. For $^3$He, the analysis is identical to the proton case with $G=-4.19$ substituted throughout. For deuterons and $^6$Li, $|G|\approx 0.14$--$0.18$, so fewer spin resonances are crossed during acceleration, but the SQUID spin signal per nucleon is also reduced.

\begin{table}[ht]
	\centering
	\caption{Spin and magnetic properties of ion species relevant for the EIC,
	using the BMT definition $g = A\mu/(ZI\mu_N)$ and $G=(g-2)/2$.
	Magnetic moments from Stone~\cite{Stone2016}; proton and deuteron values
	from CODATA~\cite{codata}.}
	\label{tab:g_factors}
	\sisetup{table-number-alignment=center,table-format=+1.4}
	\begin{tabular}{l c c c S[table-format=+1.4] S[table-format=+1.4] S[table-format=+1.4]}
		\toprule
		{Nucleus} & {$Z$} & {$A$} & {$I$} & {$\mu\;(\mu_N)$} & {$g = A\mu/(ZI\mu_N)$} & {$G=(g-2)/2$} \\
		\midrule
		{$p$}      & 1 & 1 & {$1/2$} & +2.7928 & +5.5856 & +1.7928 \\
		{$^{3}$He} & 2 & 3 & {$1/2$} & -2.1276 & -6.3828 & -4.1914 \\
		{$d$}      & 1 & 2 & {$1$}   & +0.8574 & +1.7148 & -0.1426 \\
		{$^{6}$Li} & 3 & 6 & {$1$}   & +0.8220 & +1.6440 & -0.1780 \\
		{$^{7}$Li} & 3 & 7 & {$3/2$} & +3.2564 & +5.0655 & +1.5328 \\
		\bottomrule
	\end{tabular}
\end{table}

\subsection{Spin dynamics, stable spin axis, and spin tune}
\label{nzero:sec:spin_dynamics}

\subsubsection{Thomas-BMT equation}

The spin vector $\vec{S}$ of a relativistic particle with charge $q$, mass $m$, and anomalous magnetic moment $G=(g-2)/2$ obeys the Thomas-Bargmann-Michel-Telegdi (T-BMT) equation in the laboratory frame~\cite{tmt, doi:10.1142/S0217751X13501479}
\begin{equation}
	\frac{\dd\vec{S}}{\dd t} = \vec{\Omega}_\mathrm{BMT} \times \vec{S},
	\label{nzero:eq:tbmt_full}
\end{equation}
with precession vector
\begin{equation}
		\vec{\Omega}_\mathrm{BMT} = -\frac{q}{m}\left[
		\left(G+\frac{1}{\gamma}\right)\vec{B}
		- \frac{G\gamma}{\gamma+1}(\vec{\beta}\cdot\vec{B})\vec{\beta}
		- \left(G+\frac{1}{\gamma+1}\right)\frac{\vec{\beta}\times\vec{E}}{c}
		\right],
		\label{nzero:eq:Omega_BMT}
\end{equation}
where $\vec{\beta}=\vec{v}/c$, $\gamma=(1-\beta^2)^{-1/2}$, and $\vec{B}$, 	$\vec{E}$ are the laboratory magnetic and electric fields.

For the HSR the following simplifications apply. The beam is ultrarelativistic ($\gamma\gg1$), so $1/\gamma\approx0$ and $\gamma/(\gamma+1)\approx1$. On the design orbit there are no electric fields ($\vec{E}=0$) and the bending field is purely transverse ($\vec{\beta}\cdot\vec{B}=0$). This $\vec\beta\cdot\vec B = 0$ simplification is exact only on the design closed orbit; for particles with finite longitudinal momentum spread or transverse motion, and in regions where the equilibrium $\vec\beta$ acquires a transverse component (snake and spin-rotator interiors, low-$\beta^*$ insertions near the interaction points), $\vec\beta\cdot\vec B$ is generally nonzero and contributes small additional spin rotations beyond the closed-orbit precession. These contributions are the dominant source of the spread of the stable spin direction across phase space, and hence of the residual spin-tune spread $\sigma_{\nu_s}$ that drives the coherence-time analysis of Sec.~\ref{sec:coherent-integration} and the working value adopted in Table~\ref{nzero:tab:matlab_params}. Transforming from $t$ to arc length $s$ via $dt=ds/v$ and using $p=q B\rho$ (magnetic rigidity), Eq.~(\ref{nzero:eq:Omega_BMT}) reduces to
\begin{equation}
	\vec{\Omega}(s) = -\frac{q}{p}(G\gamma+1)\,B_\perp(s)\,\vec{e}_\perp,
	\label{nzero:eq:Omega_HSR}
\end{equation}
where $B_\perp$ is the field component perpendicular to the design orbit. In a vertical dipole field $\vec{B}=B_y\ey$ the precession is purely around $\ey$ at rate $G\gamma/\rho$ per unit arc length, so the spin accumulates a total precession angle $G\gamma\theta$ over a sector of orbital bending angle $\theta$, giving the arc rotation matrix of Eq.\,\eqref{nzero:eq:R_arc}. The quantity $G=1.7928$ for the proton. The T-BMT equation in arc-length form is
\begin{equation}
	\frac{\dd\vec{S}}{\dd s} = \vec{\Omega}(s) \times \vec{S}.
	\label{nzero:eq:tbmt}
\end{equation}

\subsubsection{Arc sector and snake rotation matrices}

In a pure vertical dipole field the spin precesses around $\ey$ by $G\gamma\theta$ for an arc sector of orbital bending angle $\theta$, leading to
\begin{equation}
	\Rmat_\mathrm{arc}(\theta) = \Rmat_y(G\gamma\theta),
	\qquad
	\Rmat_y(\varphi) =
	\begin{pmatrix}
		\cos\varphi & 0 & \sin\varphi \\
		0           & 1 & 0           \\
		-\sin\varphi & 0 & \cos\varphi
\end{pmatrix}.
	\label{nzero:eq:R_arc}
\end{equation}
Snake $k$ rotates the spin by $180^\circ$ around the horizontal axis $\vec{n}_k = \sin\phi_k\,\ex + \cos\phi_k\,\ez$, where $\phi_k$ is measured from the local beam direction $\ez$~\cite{ptitsyn2022}. Applying the Householder reflection formula~\cite{10.1145/320941.320947}, $\vec{S}' = 2(\vec{S}\cdot\vec{n}_k)\vec{n}_k - \vec{S}$, where $\vec{S}'$ is the spin after the snake rotation, yields
\begin{equation}
		\Rmat_\mathrm{snake}(\phi_k) =
		\begin{pmatrix}
			-\cos 2\phi_k  & 0  & \sin 2\phi_k \\
			0              & -1 & 0            \\
			\sin 2\phi_k   & 0  & \cos 2\phi_k
		\end{pmatrix}.
		\label{nzero:eq:R_snake}
	\end{equation}
	The action on the spin components is most transparent for three special cases:
	\begin{center}
		\renewcommand{\arraystretch}{1.3}
		\begin{tabular}{lccccc}
			\hline\hline
			Axis & $\phi_k$ & $S_x'$ & $S_y'$ & $S_z'$ \\
			\hline
			$\ez$ (longitudinal) & $0^\circ$  & $-S_x$ & $-S_y$ & $+S_z$ \\
			diagonal             & $45^\circ$ & $+S_z$ & $-S_y$ & $+S_x$ \\
			$\ex$ (radial)       & $90^\circ$ & $+S_x$ & $-S_y$ & $-S_z$ \\
			\hline\hline
		\end{tabular}
	\end{center}

In every case $S_y \to -S_y$ because the snake axis is always horizontal. The component along $\vec{n}_k$ is preserved; the component perpendicular to $\vec{n}_k$ in the $xz$ plane is flipped. The arc rotations act only in the $xz$ plane and leave $S_y$ unchanged.

\subsubsection{One-turn spin transfer matrix}

The one-turn matrix $\mathbf{M}(s_0)$ is the ordered product of arc and snake matrices for one full revolution. For the analytical six-snake model (six equal arcs, $\theta = \pi/3$, six point-like snakes), one obtains
\begin{equation}
	\mathbf{M}(s_0) =
	\prod_{k=6}^{1}
	\Rmat_\mathrm{snake}(\phi_k)\,\Rmat_\mathrm{arc}(\pi/3),
	\label{nzero:eq:M_oneturn}
\end{equation}
evaluated in the order of traversal (indices modulo~6).

\subsubsection{Stable spin axis $\vec{n}_0$ and spin tune $\nu_s$}
\label{nzero:sec:spin_formalism}

The one-turn matrix $\mathbf{M}(s)$ belongs to $SO(3)$, the group of proper rotations in $\mathbb{R}^3$, defined by $\mathbf{M}^T\mathbf{M}=\mathbf{I}$ and $\det(\mathbf{M})=+1$. Its three eigenvalues therefore take the form $\{+1,\,e^{+2\pi i\nu_s},\,e^{-2\pi i\nu_s}\}$, where the real eigenvalue $+1$ always exists because every $SO(3)$ rotation has a fixed axis. The stable spin axis $\vec{n}_0(s)$ is the eigenvector belonging to $\lambda=+1$: it is the direction around which all other spin components precess, and it returns to itself after every full revolution:
\begin{equation}
	\mathbf{M}(s)\,\vec{n}_0(s) = +\vec{n}_0(s).
	\label{nzero:eq:n0_def}
\end{equation}
The complex pair $e^{\pm 2\pi i\nu_s}$ describes the precession of spin components perpendicular to $\vec{n}_0$: they rotate by angle $2\pi\nu_s$ per turn. Since $|e^{\pm 2\pi i\nu_s}|=1$ the precession is length-preserving, and the spin vector does not grow or decay. The spin tune $\nu_s$ is extracted from the matrix trace via
\begin{equation}
	\mathrm{tr}(\mathbf{M}) = 1 + 2\cos(2\pi\nu_s).
	\label{nzero:eq:nu_from_trace}
\end{equation}

\subsubsection{Phasor derivation of $\nu_s = \tfrac{1}{2}$}

Since every arc leaves $S_y$ unchanged and every snake flips $S_y \to -S_y$, after six snakes $S_y \to (-1)^6 S_y = +S_y$. Hence $\ey$ is always an eigenvector of $\mathbf{M}$ with eigenvalue $+1$: $\vec{n}_0 = \ey$ in the arc regions, independently of energy.

To find $\nu_s$ we work in the $xz$ plane using the complex variable $w = S_x + iS_z$, in which a rotation by angle $\varphi$ becomes multiplication by $e^{i\varphi}$ and a reflection becomes complex conjugation. The spin precession angle accumulated per arc sector is
\begin{equation}
		\alpha = G\gamma\,\frac{\pi}{3},
		\label{nzero:eq:alpha_arc}
\end{equation}
which depends on the beam energy through $G\gamma$. The two building blocks in the $xz$ plane are:
\begin{align}
		\text{Arc:} \quad     & w \;\to\; e^{-i\alpha}\,w, \label{nzero:eq:arc_complex}\\
		\text{Snake }k:\quad  & w \;\to\; -e^{-2i\phi_k}\,\bar{w}. \label{nzero:eq:snake_complex}
\end{align}
The conjugation $\bar{w} = S_x - iS_z$ in the snake operation flips the sign of $S_z$, equivalent to a reflection in the $xz$ plane; the prefactor $-e^{-2i\phi_k}$ rotates this reflection to the actual snake axis orientation.

The key observation is that the conjugation in each snake turns the arc phase $e^{-i\alpha}$ accumulated before it into $e^{+i\alpha}$, which then cancels exactly with the $e^{-i\alpha}$ of the following arc. Tracing through two consecutive arc--snake pairs:
	\begin{equation}
		w
		\;\xrightarrow{A_1}\; e^{-i\alpha}w
		\;\xrightarrow{S_1}\; {-}e^{-2i\phi_1}e^{+i\alpha}\bar{w}
		\;\xrightarrow{A_2}\; {-}e^{-2i\phi_1}\bar{w}
		\;\xrightarrow{S_2}\; e^{2i(\phi_1-\phi_2)}w.
		\label{nzero:eq:two_pairs}
	\end{equation}
The arc phase $\alpha$ has cancelled completely after each pair. After all three pairs of arc--snake steps the one-turn map in the $xz$ plane is a pure rotation by the total spin phase $\Phi$:
	\begin{equation}
		w_\mathrm{final} = e^{i\Phi}\,w_0,
		\qquad
		\Phi = -2\sum_{k=1}^{3}(\phi_{2k} - \phi_{2k-1}),
		\label{nzero:eq:one_turn_complex}
	\end{equation}
	independent of $G\gamma$.
	Comparing with $e^{\pm 2\pi i\nu_s}$ from Eq.~(\ref{nzero:eq:nu_from_trace}),
	$\nu_s = 1/2$ requires $\Phi = \pi$.

	With the LC axes $\phi_k = (-1)^{k+1}\times 45^\circ$, each pair contributes
	$\phi_{2k} - \phi_{2k-1} = -90^\circ$, giving:
	\begin{equation}
		\Phi = -2\times 3\times(-90^\circ) = +540^\circ \equiv \pi \pmod{2\pi}.
		\label{nzero:eq:Phi_lc}
	\end{equation}
	Hence $e^{i\Phi} = e^{i\pi} = -1$, and the one-turn trace is
	$\mathrm{tr}(\mathbf{M}) = 1 + 2\cos(\pi) = -1$,
	giving $\nu_s = 1/2$ at every beam energy.
	This is verified numerically in Section~\ref{nzero:sec:numerical-results}.

\subsubsection{Lee--Courant and Doubly Lee--Courant configurations for the HSR}
\label{sec:lc_dlc_config}

The HSR employs six full Siberian snakes, distributed approximately uniformly around the ring. Each snake rotates the spin by $180^\circ$ around a horizontal axis with angle $\phi_k$. For an even number of full snakes, the closed-orbit spin tune is fixed at
\begin{equation}
	\nu_s = \frac{1}{2},
	\label{nzero:eq:snake_spin_tune_half}
\end{equation}
independently of energy. The snake-axis pattern therefore does not primarily determine the closed-orbit spin tune. Instead, it determines the closed-orbit stable spin direction $\vec n_0(s)$ and the degree to which higher-order spin-orbit resonances, and therefore amplitude-dependent spin-tune spread, are suppressed during the ramp.

The Lee--Courant (LC) condition imposes local cancellation of spin motion across selected snake pairs~\cite{PhysRevD.41.292}. In the simplified six-snake model with equal $60^\circ$ arc sectors, this corresponds to successive snake axes differing by $90^\circ$ in absolute value,
\begin{equation}
	|\phi_{k+1}-\phi_k| = 90^\circ.
	\label{nzero:eq:lc_condition}
\end{equation}
A convenient representative is
\begin{equation}
	\phi_k = (-1)^{k+1}\times 45^\circ, \quad k=1,\ldots,6,
	\label{nzero:eq:lc_hsr}
\end{equation}
i.e., $\phi=[+45^\circ,-45^\circ,+45^\circ,-45^\circ,+45^\circ,-45^\circ]$. These are the axes shown in Fig.~\ref{fig:hsr_layout}. In the idealized analytical model this configuration gives $\vec n_0=\ey$ throughout the arc regions, as confirmed numerically in Section~\ref{nzero:sec:numerical-results}.

A stronger symmetry is obtained in the Doubly Lee--Courant (DLC) scheme.
The DLC condition enforces the local $\pi$ spin-phase cancellation across every
consecutive pair of snakes, so that each snake participates in two overlapping
cancellation segments. For six snakes this reduces the allowed axis pattern to
one continuous overall rotation angle $\psi$,
\begin{equation}
	\phi_k = \psi + (k-1)\times45^\circ,
	\quad k=1,\ldots,6.
	\label{nzero:eq:dlc_hsr}
\end{equation}
Equivalently, neighboring axes step by $45^\circ$ around the ring. Tracking
studies of the realistic HSR lattice~\cite{Hamwi2024SnakeMatching} find an optimized DLC pattern with
$\psi\simeq22.5^\circ$, giving
\begin{equation}
	\phi \simeq
	[22.5^\circ,\ 67.5^\circ,\ 112.5^\circ,\ 157.5^\circ,\ 202.5^\circ,\ 247.5^\circ].
	\label{nzero:eq:dlc_opt_axes}
\end{equation}
The important point for the present SQUID estimate is that the DLC condition
suppresses spin-orbit resonance driving terms over the full energy ramp without
requiring dynamic adjustment of the snake axes.

Recent HSR spin-tracking studies by Hamwi provide useful context for the expected scale of spin-tune spread in different six-snake configurations, especially the snake-axis comparison in Sec.~4.4.4 and Figs.~4.12--4.13 of Ref.~\cite{HamwiThesis2025}. In that work, Lee--Courant schemes are shown to improve the polarization-preserving properties of the HSR snake lattice, while the Doubly Lee--Courant (DLC) scheme imposes a stronger overlapping local-cancellation condition in which each snake participates in two cancellation segments.

Hamwi quantifies the residual spin spread through the amplitude-dependent spin tune (ADST). The ADST is the dependence of the spin tune on the orbital actions of a particle, caused by spin-orbit coupling in the real lattice. Particles at different transverse amplitudes therefore do not all precess with exactly the closed-orbit spin tune $\nu_s=1/2$, but acquire slightly different spin tunes. This spread is precisely the quantity that limits the coherent transverse-spin signal. Hamwi reports the maximum ADST spread over the tracked ensemble and energy range. In the present discussion we denote the corresponding conservative spin-tune-spread scale by $\sigma_{\nu_s}$, to keep the notation consistent with the coherence-time estimate. The resulting hierarchy is
\begin{equation}
	\begin{split}
		\sigma_{\nu_s}^{\mathrm{DLC}} & \lesssim 10^{-2}, \\
		\sigma_{\nu_s}^{\mathrm{LC}}  & \sim 0.02\text{--}0.04,\\
		\sigma_{\nu_s}^{\mathrm{gen}} & \sim 0.04\text{--}0.08 .
	\end{split}
	\label{nzero:eq:spin_tune_spread_hierarchy}
\end{equation}
These values should be understood as configuration-dependent spread scales inferred from maximum ADST measures, not as final rms values for the stored beam. Nevertheless, they show that the snake-axis configuration is a central control knob for $\sigma_{\nu_s}$ and provide a scale against which the coherent integration assumption should be tested.

\begin{table}[htb]
	\centering
	\caption{Input parameters and spin-tune-spread scales used for the analytical six-snake model and SQUID coherence estimates. The working value $\sigma_{\nu_s}^{\mathrm{eff}}=\num{1.0e-3}$ is used for the baseline FID coherence and matched-filter estimates. The LC, DLC, and general-axis entries are conservative spin-tune-spread scales inferred from Hamwi's maximum ADST measures~\cite{HamwiThesis2025} and are used as test scales rather than fixed rms beam parameters.}
	\label{nzero:tab:matlab_params}
	\renewcommand{\arraystretch}{1.05}
	\sisetup{table-number-alignment=center}
\begin{tabular}{l c c S[table-format=3.4]}
	\hline\hline
	{Description} & {Parameter} & {Unit} & {Value} \\\hline
	Proton anomalous magnetic moment & $G$ & {--} & 1.7928 \\
	Lorentz factor (injection) & $\gamma_\mathrm{inj}$ & {--} & 25.05 \\
	Lorentz factor (flattop) & $\gamma_\mathrm{flat}$ & {--} & 293.1 \\
	Number of snakes & $N_s$ & {--} & 6 \\
	Arc bending angle per sector & $\theta$ & \si{\degree} & 60 \\
	LC axis angles & $\phi_k$ & \si{\degree} & {alternating $\pm\num{45}$} \\
	DLC axis offset & $\psi$ & \si{\degree} & {$\simeq 22.5$} \\
	FID tipping angle & $\alpha$ & \si{\milli\radian} & 30 \\
	Working effective RMS spin-tune spread & $\sigma_{\nu_s}^{\mathrm{eff}}$ & {--} & {$\num{1.0e-3}$} \\
	DLC ADST spin-tune-spread scale & $\sigma_{\nu_s}^{\mathrm{DLC}}$ & {--} & {$\lesssim\num{1.0e-2}$} \\
	LC ADST spin-tune-spread scale & $\sigma_{\nu_s}^{\mathrm{LC}}$ & {--} & {$\num{2.0e-2}$--$\num{4.0e-2}$} \\
	General-axis ADST spin-tune-spread scale & $\sigma_{\nu_s}^{\mathrm{gen}}$ & {--} & {$\num{4.0e-2}$--$\num{8.0e-2}$} \\
	Revolution frequency (injection) & $f_\text{rev,inj}$ & \si{\kilo\hertz} & 78.1339 \\
	Revolution frequency (flattop) & $f_\mathrm{rev,flat}$ & \si{\kilo\hertz} & 78.1957 \\\hline\hline
	\end{tabular}
\end{table}

\subsubsection{Numerical results}
\label{nzero:sec:numerical-results}

To illustrate the spin motion in the LC configuration introduced in Sec.~\ref{sec:lc_dlc_config}, we evaluate the transport for the hexagonal six-snake model shown in Fig.~\ref{fig:hsr_layout}. The calculation uses the one-turn map of Eq.~(\ref{nzero:eq:M_oneturn}) together with the sector-wise rotations of Eqs.~(\ref{nzero:eq:R_snake}) and (\ref{nzero:eq:R_arc}). In this analytical model, the lattice is represented by six identical arc sectors separated by localized snake rotations at the positions $S_k$, $k=1,\ldots,6$. The orbit is discretized into straight sections and curved arc segments so that the continuous evolution in the arcs and the localized snake rotations can be resolved explicitly.

The invariant spin axis $\vec{n}_0$ is obtained from the eigenvalue condition of Eq.~(\ref{nzero:eq:n0_def}) and transported along the ring using the same sequence of spin maps, thereby defining the local stable spin direction $\vec{n}_0(s)$ along the closed orbit. Due to the localized action of the snakes, $\vec{n}_0(s)$ undergoes finite rotations at the snake locations and evolves continuously in the arcs. A physical spin vector $\vec{S}(s)$ is initialized at the reference point with a fixed opening angle $\psi$ relative to $\vec{n}_0$ and propagated with the same spin-transfer sequence, ensuring that the angle between $\vec{S}(s)$ and $\vec{n}_0(s)$ remains constant while the phase around $\vec{n}_0(s)$ evolves.

\begin{figure}[tb]
	\centering
	\includegraphics[width=0.85\linewidth]{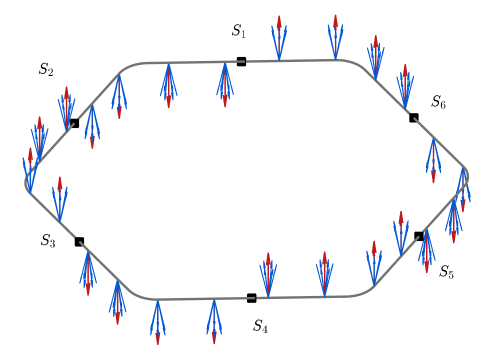}
	\caption{Numerical spin lattice for the hexagonal six-snake model introduced in Fig.~\ref{fig:hsr_layout}. The black squares mark the localized Siberian snakes $S_k$, $k=1,\ldots,6$, and the gray curve represents the simplified closed orbit. The red vectors show the transported stable spin axis $\vec{n}_0(s)$ at selected locations, while the blue vectors show spin vectors $\vec{S}(s)$ with a fixed opening angle $\psi=\SI{30}{\degree}$ around $\vec{n}_0(s)$. The multiple blue vectors at a given location represent different phase angles on the local precession cone; they do not represent successive turns of one particle.}
	\label{nzero:fig:spin_precession_hexagon}
\end{figure}

The resulting spin lattice is shown in Fig.~\ref{nzero:fig:spin_precession_hexagon}. It corresponds to the same hexagonal layout and snake labeling shown in Fig.~\ref{fig:hsr_layout}. The piecewise behavior of $\vec{n}_0(s)$ reflects the symmetry of the lattice, with finite rotations at the localized snakes and smooth evolution in the intervening arc sectors. The spin vectors $\vec{S}(s)$ populate the corresponding precession cone with fixed opening angle $\psi$. The different blue vectors at each selected location represent distinct phase angles around $\vec{n}_0(s)$ and are included to visualize the local precession cone; they do not represent subsequent turns of a single particle. All parameters used in the calculation correspond to the LC configuration defined in Eq.~(\ref{nzero:eq:lc_hsr}) and are summarized in Table~\ref{nzero:tab:matlab_params}.

\subsection{Net magnetic dipole moment of a polarized bunch}
\label{sec:bunch_magnetic_moment}

A proton has magnetic moment
\begin{equation}
	\mu_p
	=
	(1+G)\mu_N
	=
	\SI{1.410583e-26}{J\,T^{-1}},
	\label{eq:proton_magnetic_moment}
\end{equation}
where $\mu_N=e\hbar/(2m_p)$ is the nuclear magneton and $G=\num{1.7928}$ is the proton anomalous magnetic moment~\cite{codata}. For a fully polarized bunch of $N_p$ protons with spins aligned along the stable spin axis $\vec{n}_0 \approx \ey$, the net magnetic dipole moment is
\begin{equation}
	\vec{m}
	=
	N_p\,\mu_p\,\ey.
	\label{eq:moment_def}
\end{equation}
For a partially polarized beam with polarization $P$, this becomes
\begin{equation}
	\vec{m}
	=
	P\,N_p\,\mu_p\,\ey.
	\label{eq:moment_def_polarized}
\end{equation}
In steady state the moment is static along $\ey$; the cosine-$\theta$ transverse SQUID pickup, which is sensitive to the local $m_x$ component, therefore sees no FID signal before a tipping pulse is applied. How an oscillating in-plane signal is generated via a tipping pulse is discussed in Sec.~\ref{sec:tipping_intro}.

Since $\nu_s = 0.5$, once tipped, the in-plane component reverses sign every turn, as shown in Fig.~\ref{fig:spin_precession}. The moment sampled by a fixed cosine-$\theta$ pickup after $n$ turns is
\begin{equation}
	m_x(n) = P\,N_p\,\mu_p\,\sin\alpha\,(-1)^n,
	\label{eq:mxn}
\end{equation}
where $\alpha$ is the tipping angle. This is the same spin-precession signal introduced in Eq.~\eqref{eq:spin_signal_frequency_sec2}, with
\begin{equation}
	f_s = \nu_s f_\mathrm{rev} = \frac{1}{2} f_\mathrm{rev}.
	\label{eq:fspin_sec36}
\end{equation}
Numerically, this gives $f_s=\SI{39.0669}{kHz}$ at injection and $f_s=\SI{39.0979}{kHz}$ at flattop.

For a single fully polarized bunch at injection and flattop (Table\,\ref{tab:params},  Eq.\,\eqref{eq:proton_magnetic_moment}),  the magnetic-moment scale is
\begin{align}
	|m|^\mathrm{inj}
	=
	N_p^\mathrm{inj}\,\mu_p
	&\approx
	\SI{3.893e-15}{J\,T^{-1}},
	\label{eq:mx_inj}
	\\[0.5ex]
	|m|^\mathrm{flat}
	=
	N_p^\mathrm{flat}\,\mu_p
	&\approx
	\SI{9.733e-16}{J\,T^{-1}}.
	\label{eq:mx_flat}
\end{align}
The fourfold reduction at flattop reflects RF bunch splitting. Signal estimates in subsequent sections use $P=\num{0.7}$ and include the $\sin\alpha$ factor from Eq.~\eqref{eq:mxn}. For the operating tipping angle $\alpha=\SI{30}{mrad}$ (Table~\ref{tab:params}), the corresponding FID magnetic moments are
\begin{align}
	|m_x|^\mathrm{inj}_\mathrm{FID}
	&=
	P\,N_p^\mathrm{inj}\,\mu_p\,\sin\alpha
	\approx
	\SI{8.174e-17}{J\,T^{-1}},
	\\
	|m_x|^\mathrm{flat}_\mathrm{FID}
	&=
	P\,N_p^\mathrm{flat}\,\mu_p\,\sin\alpha
	\approx
	\SI{2.044e-17}{J\,T^{-1}}.
\end{align}

\subsection{Tipping the spin ensemble: the FID protocol}
\label{sec:tipping_intro}

In steady state the spins are aligned with the local stable spin axis, taken here as $\vec{n}_0\simeq\ey$. The bunch magnetic moment is then static and points approximately along $\ey$, as in Eq.~\eqref{eq:moment_def_polarized}. The cosine-$\theta$ pickup, which measures the local $m_x$ component, therefore sees no FID signal before the spin ensemble is tipped.

To initiate a measurement, a dedicated spin-kick element tips the spin ensemble away from $\ey$ by a small angle $\alpha$ into the ring plane. For the nominal arc geometry with $\vec{n}_0\simeq\ey$, a longitudinal magnetic field $B_z$ provides a spin-kick axis perpendicular to $\vec{n}_0$. The tipping angle is related to the integrated longitudinal field by the T-BMT equation~\cite{tmt},
\begin{equation}
	\alpha = \frac{(1+G)}{B\rho} \int B_z\,dl,
	\label{eq:alpha}
\end{equation}
where $B\rho=p/e$ is the magnetic rigidity and $G$ is the proton anomalous magnetic moment (Table~\ref{nzero:tab:matlab_params}). The factor $(1+G)$ appears because the spin rotation in a longitudinal field combines the anomalous-moment contribution with the rotation along the velocity direction; see, e.g., \cite{Lee1997SpinDynamics,tmt}. A longitudinal field is attractive because $B_z\parallel\vec v$ gives $\vec v\times\vec B=0$ for the reference particle. It therefore produces a magnetic spin kick without a transverse Lorentz-force kick on the closed orbit.

After the pulse, the tipped in-plane component precesses freely at the spin-precession frequency introduced in Eq.~\eqref{eq:spin_signal_frequency_sec2}. For $\nu_s=0.5$, this component changes sign on every turn. The moment sampled by a fixed cosine-$\theta$ pickup after $n$ turns is therefore
\begin{equation}
	m_x^\mathrm{FID}(n)
	=
	P\,N_p\,\mu_p\,\sin\alpha\,(-1)^n .
	\label{eq:mxfid}
\end{equation}
This freely precessing transverse signal is the free-induction decay (FID), in direct analogy with nuclear magnetic resonance (NMR).

The kick element must create the transverse spin component without significantly perturbing the orbit. A purely longitudinal magnetic kick satisfies this condition geometrically, but a conventional high-inductance solenoid may not be the most practical implementation for a fast, well-defined pulse. An alternative is a transverse RF spin-kick element operated in Wien-filter mode, where the electric and magnetic fields satisfy
\begin{equation}
	\vec E+\vec v\times\vec B=0
	\label{eq:wien_condition_spin_kick}
\end{equation}
for the reference orbit. This cancels the orbital Lorentz force while retaining the spin rotation. RF Wien filters for spin manipulation have been developed and operated in the COSY/JEDI program~\cite{SLIM2016116,PhysRevResearch.7.023257}. Phase-locking such a transverse Wien-filter field to a longitudinal RF field, $90^\circ$ out of phase, selects a single rotating field component and drives a coherent reversal of a bunch's stored polarization at $\nu_s=1/2$; gated bunch by bunch, this provides per-bunch polarization reversal, which is developed in Appendix~\ref{app:spin-manip-rf-WF}.

For the kick to be treated as impulsive in the spin dynamics, its duration should be short compared with the spin-precession period,
\begin{equation}
	T_s
	=
	\frac{1}{f_s}
	\simeq
	\SI{25.6}{\micro\second}.
	\label{eq:spin_precession_period}
\end{equation}
A sub-microsecond to few-microsecond spin-kick element is therefore the natural target. Slower or resonant implementations are not excluded, but then the finite time profile of the kick must be included explicitly in the spin map rather than approximated as an instantaneous rotation.

The tipping angle $\alpha$ sets the size of the observed FID signal. If the spin ensemble were tipped and then simply left unrestored, the stored projection would be reduced by $1-\cos\alpha\simeq\alpha^2/2$. In the intended tip-read-restore protocol, however, a restore pulse is applied after the readout. The residual operational polarization loss is therefore determined by the accuracy and reproducibility of the restore operation, not by the tipping angle alone.

If the local stable spin axis differs appreciably from $\ey$, the spin-kick field must be oriented perpendicular to the actual local $\vec n_0(s)$. This is a placement and hardware-orientation constraint, not a change to the basic FID concept.

%

\section{Spin-tune measurement: spectral search and phase lock}
\label{sec:fid_spin_tune_phase_lock}


The first operational task of the SQUID system is not absolute
polarimetry but the direct, on-line measurement of the ensemble spin
tune $\nu_s$ and of the spin-tune spread $\sigma_{\nu_s}$. Both must
be determined from the SQUID signal itself, before any phase-tracking
or coherent-accumulation strategy of
Sec.~\ref{sec:coherent-integration} can be invoked. The FID protocol
of Sec.~\ref{sec:tipping_intro}, which produces a freely precessing
transverse magnetic moment after a controlled spin tip, supplies the
required time-domain signal; what is needed in addition is a
data-analysis strategy that can extract the precession frequency
from this signal without prior knowledge of either the frequency
itself or the coherence window.

The principal difficulty is that the SQUID polarimeter must commission
itself in a machine state whose spin-coherence time is essentially
unknown and likely short. A first measurement concept that assumed an
observation time longer than the untuned coherence time would be
circular: it would require the machine to be tuned for long coherence
before the diagnostic capable of guiding that tuning is operational.
The initial measurement strategy must therefore work in the
short-coherence-time regime expected from conservative start-up
estimates ($\sigma_{\nu_s}\sim 10^{-2}$--$10^{-3}$, corresponding to
$\tau_\mathrm{coh}\sim\SI{0.2}{}$--$\SI{2}{\milli\second}$) and must
extract both $f_s$ and $\sigma_{\nu_s}$ from the SQUID signal alone.
The technique introduced in this section meets this constraint with a
noninvasive spectral search performed entirely with the SQUID
polarimeter.

\subsection{Spectral-search strategy}
\label{subsec:spectral_search_strategy}

One measurement cycle consists of (i) a longitudinal kicker pulse, defined in Sec.~\ref{sec:tipping_intro}, that tips the stored polarization by the angle $\alpha=\SI{30}{\milli\radian}$; (ii) a SQUID recording of the resulting FID for a duration $T_\mathrm{meas}\!\le\!\tau_\mathrm{coh}^{\mathrm{untuned}}$; (iii) a fast Fourier transform of the recorded record to produce a single power spectrum; and (iv) a closing restore pulse that returns the polarization to vertical. This cycle is repeated $N$ times. The $N$ single-record power spectra are then averaged incoherently.

\begin{figure*}[!tb]
	\centering
	\includegraphics[width=\textwidth]{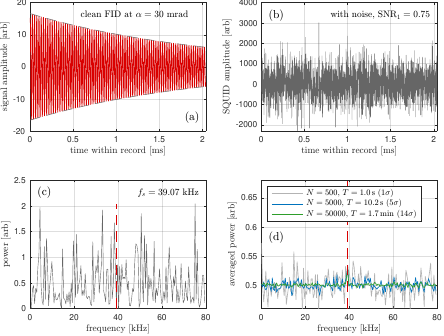}
	\caption{%
		Spectral search with the SQUID polarimeter at injection energy: the precession signal emerges from SQUID noise through incoherent power-spectrum averaging. All panels use the locked operating point $\alpha=\SI{30}{\milli\radian}$, $P=0.5$, $\sigma_{\nu_s}=10^{-3}$, yielding a coherence time $\tau_\mathrm{coh}=\SI{2.04}{\milli\second}$ and an expected precession frequency $f_s=\SI{39.07}{\kilo\hertz}$. 
		(a) Clean reference FID over one record of duration $T_\mathrm{meas}=\tau_\mathrm{coh}$; the red trace is the underlying physical signal, the dashed gray curves mark its exponential envelope. 
		(b) Same record plus realistic SQUID noise; the per-record signal-to-noise ratio $\mathrm{SNR}_1\!\approx\!0.75$ leaves the signal entirely buried in noise. 
		(c) Power spectrum of one such record; the underlying peak at $f_s$ is indistinguishable from random spectral spikes. 
		(d) Power spectrum averaged over $N$ records; the peak at $f_s$ emerges with $N$ and reaches $\sim 14\sigma$ above the noise floor in $T_\mathrm{tot}=N\,T_\mathrm{meas}=\SI{1.7}{\minute}$ of integration, well inside a routine machine-development session. The detection significance grows as $\sqrt{N}=\sqrt{T_\mathrm{tot}/\tau_\mathrm{coh}}$.}
	\label{fig:spectral_search_intro}
\end{figure*}

The qualitative behaviour expected from this averaging is the content of Fig.~\ref{fig:spectral_search_intro}. The underlying physical signal is a damped cosine at the precession frequency $f_s$, with envelope $\exp(-t/\tau_\mathrm{coh})$ set by the spin-tune spread (panel~a). In any single record this signal is buried in SQUID noise,  the per-record signal-to-noise ratio at $\sigma_{\nu_s}=10^{-3}$ is only $\mathrm{SNR}_1\!\approx\!0.75$ (panel~b), so the single-record power spectrum carries no unambiguous peak (panel~c). The key feature of incoherent power-spectrum averaging is that the deterministic signal contribution at $f_s$ adds in phase from record to record (because power is a positive quantity), while the random noise contributions reduce as $1/\sqrt{N}$ in the averaged spectrum. After enough records the peak at $f_s$ therefore emerges above the noise floor at a significance that grows as $\sqrt{N}=\sqrt{T_\mathrm{tot}/\tau_\mathrm{coh}}$ (panel~d). Importantly, this strategy
requires no a~priori knowledge of the precession phase between records; what is preserved across the average is the precession frequency, which is the quantity to be measured.

Two pieces of information are extracted from the averaged spectrum. The peak position determines the spin tune, $\nu_s = f_s/f_\mathrm{rev}$, with statistical uncertainty $\delta\nu_s$ that decreases as $1/\sqrt{T_\mathrm{tot}}$ once the peak is resolved. The peak width determines the spin-tune spread $\sigma_{\nu_s}$ through the relation $\tau_\mathrm{coh}=1/(2\pi f_\mathrm{rev}\sigma_{\nu_s})$, with narrower peaks corresponding to better-tuned machine states. A single spectral search therefore returns simultaneously the spin tune that will be used to set up phase-locked operation (Sec.~\ref{subsec:two_stage_measurement}) and the coherence-time diagnostic that will guide subsequent lattice tuning.

\subsection{Detection feasibility and operating-point numbers}
\label{subsec:spin_tune_detection_feasibility}

Figure~\ref{fig:spectral_search_analysis} extends the single-energy demonstration of Fig.~\ref{fig:spectral_search_intro} to both storage energies and quantifies the detection and precision scaling. Panels~(a)~and~(b) show averaged spectra at injection ($\SI{1.7}{\minute}$ of integration) and at flattop ($\SI{27.1}{\minute}$). The peak at $f_s$ is resolved at both energies. The slower emergence at flattop is not a consequence of fewer total stored protons (RF bunch-splitting quadruples the bunch count and conserves the total integrated proton current) but of the SQUID pickup responding to the magnetic dipole moment of each \emph{individual} bunch as it passes. Each flattop bunch carries one quarter of the protons of an injection bunch, hence one quarter of the per-bunch FID amplitude. Since detection significance scales linearly in this amplitude and only as $\sqrt{T_\mathrm{tot}}$ in time, the integration required for matched significance grows by a factor of about sixteen relative to injection.  Panel~(c) plots the universal $\sqrt{T_\mathrm{tot}}$ scaling of the detection significance against both energies, with the $5\sigma$ and $10\sigma$ detection thresholds shown for orientation. At injection, the spectral peak is detected at $10\sigma$ after $T_\mathrm{tot}\!\lesssim\!\SI{1}{\second}$; at flattop, after $T_\mathrm{tot}\!\sim\!\SI{6}{\second}$. Table~\ref{tab:first_spin_tune_kicker_requirements} collects the corresponding numerical timing scenarios for three machine-tuning states characterized by $\sigma_{\nu_s}$. The canonical value $\sigma_{\nu_s}=10^{-3}$ reflects conservative start-up expectations from COSY-class lattices~\cite{Eversmann:2015,Guidoboni2016}; $\sigma_{\nu_s}=10^{-2}$ represents a poorly tuned commissioning state, while $\sigma_{\nu_s}=10^{-4}$ represents the level projected after lattice optimization informed by an initial SQUID measurement (still two orders of magnitude above the best COSY deuteron benchmark of $\sigma_{\nu_s}\sim\num{2e-10}$ achieved with deliberate spin-coherence tuning~\cite{Guidoboni2016}). The required integration times to reach $\delta\nu_s=10^{-5}$ at injection range from $\SI{1.0}{\hour}$ in the poorly-tuned case down to $\SI{0.36}{\second}$ in the ML/AI-tuned case. The corresponding integrated longitudinal kicker field, fixed by $\alpha$ and the magnetic rigidity, is $\int B_z\,dl=\SI{0.84}{\tesla\meter}$ at injection and $\int B_z\,dl=\SI{9.85}{\tesla\meter}$ at flattop. 

\begin{table*}[!tb]
	\centering
	\caption{%
		Spin-tune-search timing scenarios as a function of the assumed spin-tune spread $\sigma_{\nu_s}$. All entries assume the locked operating point $\alpha=\SI{30}{\milli\radian}$, $P=0.5$ (a conservative start-up polarization). The per-record duration is matched to the coherence time $T_\mathrm{meas}=\tau_\mathrm{coh}= 1/(2\pi f_\mathrm{rev}\sigma_{\nu_s})$ set by the assumed $\sigma_{\nu_s}$. $\mathrm{SNR}_1$ is the spectral signal-to-noise ratio of the peak at $f_s$ after one such record, computed from the SQUID polarimeter and beam parameters in Table~\ref{tab:params}. $T_\mathrm{tot}$ is the total integration time required to reach $\delta\nu_s=10^{-5}$, corresponding to a $1\%$ relative measurement of $\sigma_{\nu_s}$ in the canonical scenario ($\sigma_{\nu_s}=10^{-3}$). The required integrated longitudinal kicker field is $\int B_z\,dl=\SI{0.84}{\tesla\meter}$ at injection and $\int B_z\,dl=\SI{9.85}{\tesla\meter}$ at flattop, independent of $\sigma_{\nu_s}$. }
	\label{tab:first_spin_tune_kicker_requirements}
	\sisetup{table-number-alignment=center}
	\begin{tabular}{c c S[table-format=1.3] S[table-format=1.3] c c}
		\toprule
		{$\sigma_{\nu_s}$} & {$\tau_\mathrm{coh}$} &
		{$\mathrm{SNR}_1$ (inj)} & {$\mathrm{SNR}_1$ (flat)} &
		{$T_\mathrm{tot}$ (inj)} & {$T_\mathrm{tot}$ (flat)} \\
		\midrule
		$10^{-2}$ & \SI{0.20}{\milli\second}  & 0.237 & 0.059 &
		\SI{1.0}{\hour}    & \SI{16.1}{\hour}  \\
		$10^{-3}$ & \SI{2.04}{\milli\second}  & 0.750 & 0.187 &
		\SI{36.2}{\second} & \SI{9.7}{\minute}  \\
		$10^{-4}$ & \SI{20.37}{\milli\second} & 2.372 & 0.593 &
		\SI{0.36}{\second} & \SI{5.8}{\second}  \\
		\bottomrule
	\end{tabular}
\end{table*}

\subsection{Iterative improvement loop}
\label{subsec:ml_ai_iterative_loop}

\begin{figure*}[!tb]
	\centering
	\includegraphics[width=\textwidth]{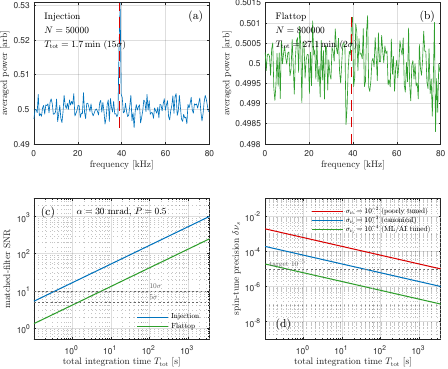}
	\caption{%
		Operational analysis of the spin-tune spectral search. \textbf{(a)}~Averaged SQUID power spectrum at injection ($N=5\times 10^{4}$ records, $T_\mathrm{tot}=\SI{1.7}{\minute}$): the peak at $f_s=\SI{39.07}{\kilo\hertz}$ stands $\sim 15\sigma$ above the noise floor. (b) Same at flattop ($N=8\times 10^{5}$ records, $T_\mathrm{tot}=\SI{27.1}{\minute}$); RF bunch-splitting reduces the proton number per bunch by a factor of four, so the same detection significance requires a factor of sixteen more 		integration time than at injection, but the peak is still resolved within one hour of machine-development time. (c) Detection significance vs total integration time $T_\mathrm{tot}$ at $\alpha=\SI{30}{\milli\radian}$, $P=0.5$, for both energies; the universal $\sqrt{T_\mathrm{tot}}$ scaling shows that $10\sigma$ detection is reached in $T_\mathrm{tot}\!\lesssim\!\SI{1}{\second}$ at injection and $T_\mathrm{tot}\!\sim\!\SI{6}{\second}$ at flattop. Horizontal dotted lines mark $5\sigma$ and $10\sigma$ thresholds; vertical dotted lines indicate \SI{1}{\minute} and \SI{1}{\hour} time budgets. d) Spin-tune precision $\delta\nu_s$ at injection vs $T_\mathrm{tot}$ for three machine-tuning states characterized by the spin-tune spread $\sigma_{\nu_s}$. The horizontal dotted line marks the representative target $\delta\nu_s=10^{-5}$, corresponding to a $1\%$ relative measurement of $\sigma_{\nu_s}$ in the canonical scenario ($\sigma_{\nu_s}=10^{-3}$). Reducing $\sigma_{\nu_s}$ by one order of magnitude (e.g., through ML/AI-guided lattice tuning informed by an initial SQUID measurement) compresses the integration time required to reach this target by two orders of magnitude.}
	\label{fig:spectral_search_analysis}
\end{figure*}

Panel~(d) of Fig.~\ref{fig:spectral_search_analysis} makes explicit the operational consequence of measuring $f_s$ and $\sigma_{\nu_s}$ simultaneously: for a fixed target precision $\delta\nu_s$, the required integration time scales as $\sigma_{\nu_s}^{2}$. A factor-of-ten reduction in the spin-tune spread therefore compresses the measurement time by two orders of magnitude. This is the qualitative advantage of noninvasive SQUID polarimetry over destructive carbon-target spin-tune measurement~\cite{Eversmann:2015}: each measurement returns not only $\nu_s$ but also a fresh, non-destructive estimate of $\sigma_{\nu_s}$ from the peak linewidth, providing the diagnostic feedback needed to guide sextupole, chromaticity, and bunch-cooling adjustments without heating the beam. The next measurement is then both more precise at fixed time and faster at fixed precision. A realistic factor-of-ten reduction in $\sigma_{\nu_s}$, obtainable from, e.g., ML/AI-guided lattice tuning, compresses an initial $\sim$twenty-minute spectral search at $\sigma_{\nu_s}\!\sim\!10^{-2}$ to a few-second update at $\sigma_{\nu_s}\!\sim\!10^{-4}$ (Table~\ref{tab:first_spin_tune_kicker_requirements}), once the lattice has been optimized for long spin coherence.

\subsection{Transition to phase-locked operation}
\label{subsec:phase_lock_transition}

Once the spectral search has located $f_s$ to a small fraction of the coherence-broadened peak width, the SQUID reference oscillator is phase-locked to $f_s$~\cite{hempelmann2018} and the system transitions to a precision operating mode that reaches $\delta P/P = 1\%$ within seconds at injection and minutes at flattop. Two ingredients enable this transition; both are developed in the next two sections, and the present subsection sketches them in enough detail to explain the operational picture before the reader meets them in full.

The first is the spin echo, developed in Sec.~\ref{sec:fid-echo-sequence}. A phase-coherent $\pi$ rephasing pulse applied at time $t = \tau$ after the initial tipping pulse reflects the phase fan in the transverse precession plane, so that the dephasing accumulated during $0 \le t \le \tau$ closes again at $t = 2\tau$. The echo extends the usable in-plane coherence window from the reversible $T_2^\ast = \tau_\mathrm{coh}$ set by the spin-tune spread (Eq.~\ref{eq:t2_star_definition}) to the much longer irreversible $T_2$ (Eq.~\ref{eq:t2_definition}), so a single measurement record is no longer bounded by $\tau_\mathrm{coh}$. The full four-pulse cycle (tip, $\pi$ rephase, echo readout, restore) is closed by a phase-coherent restore pulse that returns the tipped in-plane component to the stable spin axis with a per-cycle polarization cost of $\mathcal{O}(\alpha^2/\pi^2) \sim 10^{-4}$, so many cycles can be accumulated within a single fill with negligible cumulative depletion.

The second is the matched-filter coherent summation across all bunches, developed in Sec.~\ref{sec:matched-filter-coherent-summation}. The injected spin-sign pattern and the deterministic lattice phase advance from bunch to bunch are both known, so each per-bunch contribution can be rotated into a common phase before summation; the result is a per-record signal-to-noise ratio that scales as $\sqrt{N_\mathrm{fill}}$ rather than the order-unity gain of an uncorrected sum. Combined with the phase-locked tip-$\pi$-echo-restore cycle, the relative statistical precision on the polarization magnitude $P$ scales as $1/\sqrt{T}$ in the total integration time $T$, with the stored beam polarization preserved across the measurement. A slow feedback on the fitted $f_s$ from rolling spectral windows tracks drift in the lattice, the closed orbit, and the snake matching during the fill~\cite{Eversmann:2015}, maintaining phase lock across hours of running.

A complementary precision tool for the coherent ring-magnetic-field environment, of interest to BSM observables such as charged-particle electric dipole moments, is the radio-frequency Wien filter operated at $\nu_s=1/2$. Its physics, the corresponding waveguide implementation, and its connection to the spin-tune measurement established here are discussed in Appendix~\ref{app:spin-manip-rf-WF}.

\section{Spin-echo rephasing and the noninvasive measurement cycle}
\label{sec:fid-echo-sequence}

The spectral search of Sec.~\ref{sec:fid_spin_tune_phase_lock} returns the ensemble spin tune $\nu_s$ from the location of the spectral peak, the spin-tune spread $\sigma_{\nu_s}$ from its width, and concludes by phase-locking the SQUID reference oscillator to the measured precession at $f_s = \nu_s f_\mathrm{rev}$. With phase lock in hand, the next step is to extend the measurement window beyond the reversible-dephasing time $\tau_\mathrm{coh} = (2\pi f_\mathrm{rev}\sigma_{\nu_s})^{-1}$ set by the spin-tune spread. The instrument for this extension is a Hahn spin-echo sequence~\cite{PhysRev.80.580}: a phase-coherent $\pi$ pulse applied at $t = \tau$ rephases the dephased ensemble at $t = 2\tau$, cancelling the reversible $\sigma_{\nu_s}$-driven dephasing and exposing the irreversible decoherence due to processes that change individual spin tunes during the measurement.

Equally important, when the echo sequence is closed by a phase-coherent restore pulse the entire cycle (tip, $\pi$ rephase, echo readout, restore) returns the beam polarization to the stable spin direction with negligible loss per cycle. This is the operational property that makes the SQUID polarimeter noninvasive: many measurement cycles can be repeated within the same fill without measurable depletion of the stored polarization. By contrast, the fast carbon-target polarimeters used at RHIC~\cite{Bravar2005ProtonPolarimetryRHIC} sampled the stored beam destructively, with each measurement depleting beam intensity and degrading the emittance. Operationally this restricted RHIC pC polarimetry to a small number of measurements at the beginning, middle, and end of each store, rather than continuous monitoring throughout the fill.

The hardware requirement is no more demanding than what the spin-tune search of Sec.~\ref{sec:fid_spin_tune_phase_lock} already needs. The longitudinal kicker that delivers the tipping pulse also delivers, when synchronized to the SQUID reference oscillator, both the $\pi$ rephasing pulse and the restore pulse. The same lattice element therefore performs all three roles, and the phase lock established at the end of Sec.~\ref{sec:fid_spin_tune_phase_lock} is the prerequisite that makes the timed sequence close.

\subsection{Reversible and irreversible decoherence}
\label{sec:fid-echo-t2-star-t2}

Two physically distinct mechanisms reduce the coherent transverse magnetic moment of the stored ensemble. The first is reversible dephasing: particles with different spin tunes $\nu_{s,i} = \nu_s + \delta\nu_{s,i}$ accumulate transverse phase differences $\Delta\phi_i(t) = 2\pi f_\mathrm{rev}\,\delta\nu_{s,i}\,t$ relative to the ensemble mean, so the coherent vector sum decays with the rms spin-tune spread $\sigma_{\nu_s}$. The individual spin vectors are not lost; only their phase-coherent sum is reduced. This is the FID envelope of Sec.~\ref{sec:fid_spin_tune_phase_lock}, characterized by the reversible coherence time
\begin{equation}
	T_2^\ast \;=\; \tau_\mathrm{coh}
	\;=\; \frac{1}{2\pi f_\mathrm{rev}\sigma_{\nu_s}}.
	\label{eq:t2_star_definition}
\end{equation}
At canonical start-up ($\sigma_{\nu_s} = 10^{-3}$), $T_2^\ast = \SI{2.04}{\milli\second}$ (Table~\ref{tab:first_spin_tune_kicker_requirements}). Reverting the relative phases by an external operation can in principle recover the lost coherence, because the dephasing is deterministic in the individual spin frequencies.

The second mechanism is irreversible. Processes that change individual spin tunes during the measurement window, intra-bunch Coulomb scattering, slow drifts of the closed orbit and the snake matching, ionization of residual gas anywhere in the ring vacuum, and beam losses, add a random walk to each particle's transverse phase. These mechanisms cannot be undone by an external pulse and define the irreversible (homogeneous) coherence time
\begin{equation}
	T_2 \;\gg\; T_2^\ast,
	\label{eq:t2_definition}
\end{equation}
which sets the ultimate envelope of the polarimeter signal regardless of any rephasing strategy. The spin-echo sequence developed in this section is the instrument by which $T_2$ is measured separately from $T_2^\ast$.

\subsection{Echo rephasing and the $\pi$-pulse axis}
\label{sec:fid-echo-rephasing}

\begin{figure}[htbp]
	\centering
	\includegraphics[width=0.45\linewidth]{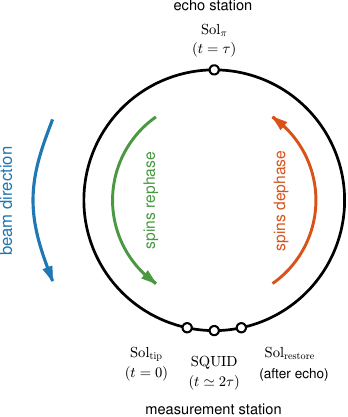}
	\caption{Functional layout of the tip-$\pi$-echo-restore sequence. The measurement station contains the longitudinal kicker (which delivers the tip, $\pi$, and restore pulses) and the SQUID pickup. The tipping pulse at $t=0$ creates the transverse spin component; the $\pi$ pulse at $t=\tau$ reflects the phase fan; the echo maximum is sampled near $t\simeq 2\tau$; and the restore pulse, fired after sampling, returns the polarization to the stored direction. All four pulses are synchronized to the SQUID reference oscillator phase-locked at $f_s$.}
	\label{fig:fid_echo_layout}
\end{figure}

Figure~\ref{fig:fid_echo_layout} shows the functional layout of the four-pulse sequence around the ring: the longitudinal kicker delivers the tipping pulse at $t=0$, the $\pi$ rephasing pulse at $t=\tau$, and the closing restore pulse, while the SQUID pickup samples the echo maximum near $t\simeq 2\tau$. The remainder of this section develops the spin-dynamics content of each step in turn.

\begin{figure}[htb]
	\centering
	\includegraphics[width=\linewidth]{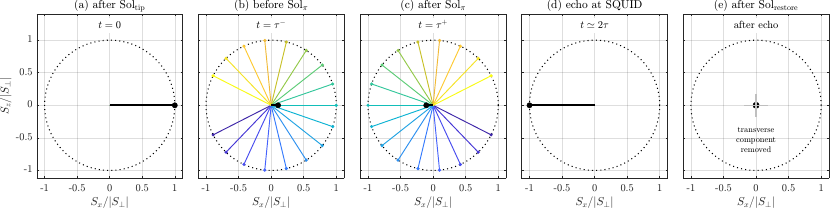}
	\caption{Simplified rotating-frame phase-fan picture of the FID and spin-echo sequence. Each colored vector represents a representative particle or beam slice with a slightly different spin tune, and the black vector denotes the ensemble-averaged transverse component. (a)~Immediately after the tipping pulse, the transverse phases are aligned. (b)~During $0<t<\tau$, the phases fan out because of the spin-tune spread. (c)~The $\pi$-pulse at $t=\tau$ reflects the phase fan across the longitudinal axis $\vec e_z$. (d)~During $\tau<t<2\tau$, the reflected phases re-align and the echo envelope reaches its maximum near $t\simeq2\tau$. (e)~The restore pulse, applied after the echo signal has been sampled, returns the ensemble to the stored-spin direction.}
	\label{fig:fid_echo_phase_fan}
\end{figure}

The spin-echo sequence applies a phase-coherent $\pi$ rotation pulse at time $t = \tau$ after the initial tipping pulse, reflecting the phase fan in the transverse precession plane (Fig.~\ref{fig:fid_echo_phase_fan}). Particles that had advanced furthest in phase before the pulse are repositioned to lag the ensemble mean afterwards, while those that had lagged behind move ahead. Continued free precession then drives the ensemble toward rephasing, with the coherent transverse vector reaching an echo maximum at $t = 2\tau$.

The $\pi$ pulse is applied about the longitudinal axis $\vec e_z$ (the beam direction). This axis lies in the transverse precession plane, perpendicular to the local stable spin axis $\vec n_0 \simeq \vec e_y$, and is therefore one of the two possible choices for an echo rephasing pulse (a rotation about $\ex$ would also rephase). The longitudinal axis is the preferred choice because the tipping kicker of Sec.~\ref{sec:tipping_intro} already delivers rotations about $\vec e_z$: the same element provides the $\pi$ pulse and the restore pulse without modification. In spin components,
\begin{equation}
	R_z(\pi)\,:\;(S_x,S_y,S_z) \;\to\; (-S_x,\,-S_y,\,S_z),
	\label{eq:r_z_pi}
\end{equation}
so the transverse components $(S_x,S_z)$ are reflected across $\vec e_z$ while the stored polarization along $\vec n_0$ is reversed in sign. The $S_y$ sign reversal is the operational price of using $\vec e_z$ as the rotation axis; it is undone by the restore pulse in Sec.~\ref{sec:fid-echo-restore} below.

The rephasing can be verified directly. Consider a particle with spin tune $\nu_{s,i} = \nu_s + \delta\nu_{s,i}$, angular precession rate $\Omega_i = 2\pi(\nu_s+\delta\nu_{s,i})f_\mathrm{rev}$, and an initial tipping pulse $R_z(\alpha)$ applied at $t=0$ to the stored state $(0,1,0)$. The post-tip state is $(-\sin\alpha,\cos\alpha,0)$. Free precession about $\vec n_0 = \vec e_y$ to $t = \tau^-$ gives
\begin{equation}
	\bigl(S_x,S_y,S_z\bigr)_{\tau^-}
	= \bigl(-\sin\alpha\cos\Omega_i\tau,\;\cos\alpha,\;
	\sin\alpha\sin\Omega_i\tau\bigr).
	\label{eq:state_before_pi}
\end{equation}
Application of $R_z(\pi)$ from Eq.~\eqref{eq:r_z_pi} flips $S_x$ and $S_y$:
\begin{equation}
	\bigl(S_x,S_y,S_z\bigr)_{\tau^+}
	= \bigl(\sin\alpha\cos\Omega_i\tau,\;-\cos\alpha,\;
	\sin\alpha\sin\Omega_i\tau\bigr).
	\label{eq:state_after_pi}
\end{equation}
Free precession from $\tau$ to $2\tau$ rotates the transverse components by an additional $\Omega_i\tau$. At $t = 2\tau$,
\begin{equation}
	\bigl(S_x,S_y,S_z\bigr)_{2\tau}
	= \bigl(\sin\alpha,\;-\cos\alpha,\;0\bigr),
	\label{eq:echo_state}
\end{equation}
independent of $\delta\nu_{s,i}$. The transverse phase fan that grew during $[0,\tau]$ closes exactly during $[\tau,2\tau]$; every particle, regardless of its individual spin tune, lands in the same transverse orientation at $t=2\tau$. This is the spin echo.

The irreversible decoherence on the time scale $T_2$ is not cancelled by this construction. Including it, the echo envelope decays from its tip-pulse value as
\begin{equation}
	M_\perp(2\tau) \;=\; M_\perp(0)\,
	\exp\!\left(-\frac{2\tau}{T_2}\right),
	\label{eq:echo_amplitude}
\end{equation}
with the $T_2^\ast$ contribution having cancelled at $t = 2\tau$ by construction. The echo amplitude as a function of $\tau$ thus measures $T_2$ directly, separately from $T_2^\ast$.

\subsection{Hardware reuse, phase-locked timing, and field budget}
\label{sec:fid-echo-hardware}

The tipping pulse of Sec.~\ref{sec:tipping_intro} produces a spin rotation about the longitudinal axis whose magnitude is set by the integrated longitudinal field,
\begin{equation}
	\alpha \;=\; \frac{(1+G)}{B\rho}\,\int B_z\,dl.
	\label{eq:alpha_recap}
\end{equation}
The $\pi$ rephasing pulse and the restore pulse rotate the spin about the same longitudinal axis, so the relation Eq.~\eqref{eq:alpha_recap} applies to all three pulses, with only the rotation angle changing. From Table~\ref{tab:first_spin_tune_kicker_requirements}, the tipping rotation of $\alpha = \SI{30}{\milli\radian}$ already requires $\int B_z\,dl = \SI{0.84}{\tesla\meter}$ at injection. For the $\pi$ rephasing pulse the rotation angle is larger by a factor $\pi/\alpha \simeq 105$, and the restore rotation $R_z(\pi-\alpha)$ demands almost the same. If delivered in a single passage of the beam through the kicker, the $\pi$ pulse would therefore call for $\int B_z\,dl = \SI{88}{\tesla\meter}$ at injection and $\SI{1032}{\tesla\meter}$ at flattop, two orders of magnitude above the tipping requirement and well beyond what a single short pulse from any practical solenoid can deliver.

The way around the single-shot limit is to spread the $\pi$ rotation over $N$ machine turns, with the kicker fired synchronously with the SQUID reference oscillator at the precession frequency $f_s$. The synchronization is essential: at $\nu_s = 1/2$ the transverse spin projection in the $(x,z)$ plane reverses sign on every machine turn, so a static field would give per-turn contributions that average to zero. A pulsed waveform at $f_s$ instead produces per-turn contributions of the same sign, which accumulate constructively. Each pulse then delivers a per-pass rotation $\pi/N$ requiring per-pass integrated field
\begin{equation}
	\int B_z\,dl\Big|_{\pi,\text{per-pass}}
	\;=\; \frac{1}{N}\,\cdot\,\frac{\pi\,B\rho}{1+G}.
	\label{eq:per_pass_pi_field}
\end{equation}
The maximum allowed $N$ is set by the requirement that the entire $\pi$ pulse complete in much less than the coherence time, so that particles with different $\delta\nu_{s,i}$ have not yet drifted significantly during the pulse itself. Taking a safety factor of two,
\begin{equation}
	N \;\le\; \frac{f_\mathrm{rev}\,\tau_\mathrm{coh}}{2}
	\;=\; \frac{1}{4\pi\sigma_{\nu_s}}.
	\label{eq:N_bound}
\end{equation}

Table~\ref{tab:pi_pulse_field_requirements} translates Eqs.~\eqref{eq:per_pass_pi_field}--\eqref{eq:N_bound} into per-pass integrated-field requirements for the three machine-tuning scenarios of Sec.~\ref{sec:fid_spin_tune_phase_lock} at both energies.

\begin{table*}[!tb]
	\centering
	\caption{%
		Per-pass integrated-field requirement for the synchronized $\pi$ rephasing pulse as a function of the spin-tune spread $\sigma_{\nu_s}$. The number of passes $N$ is set to the safety-factor-2 maximum allowed by Eq.~\eqref{eq:N_bound}. For reference, the tipping kicker of Table~\ref{tab:first_spin_tune_kicker_requirements} delivers $\int B_z\,dl = \SI{0.84}{\tesla\meter}$ at injection and $\SI{9.85}{\tesla\meter}$ at flattop in a single pass.}
	\label{tab:pi_pulse_field_requirements}
	\sisetup{table-number-alignment=center}
	\begin{tabular}{c c S[table-format=3.0] c c}
		\toprule
		{$\sigma_{\nu_s}$} & {$\tau_\mathrm{coh}$} & {$N$ (max)} &
		{$\int B_z\,dl|_{\pi,\text{per-pass}}$ (inj)} &
		{$\int B_z\,dl|_{\pi,\text{per-pass}}$ (flat)} \\
		\midrule
		$10^{-2}$ & \SI{0.20}{\milli\second}  &   8 &
		\SI{11}{\tesla\meter}    & \SI{129}{\tesla\meter}  \\
		$10^{-3}$ & \SI{2.04}{\milli\second}  &  80 &
		\SI{1.1}{\tesla\meter}   & \SI{12.9}{\tesla\meter}  \\
		$10^{-4}$ & \SI{20.4}{\milli\second}  & 796 &
		\SI{0.111}{\tesla\meter} & \SI{1.30}{\tesla\meter}  \\
		\bottomrule
	\end{tabular}
\end{table*}

At injection, all three machine-tuning scenarios in Table~\ref{tab:pi_pulse_field_requirements} sit at or below the tipping-kicker requirement of Table~\ref{tab:first_spin_tune_kicker_requirements}; the same longitudinal kicker delivers tip, $\pi$ rephase, and restore without modification. At flattop the same is true for the canonical and ML/AI-tuned cases ($\sigma_{\nu_s} \le 10^{-3}$): the per-pass $\int B_z\,dl$ for the $\pi$ pulse is between $\SI{1.3}{\tesla\meter}$ and $\SI{13}{\tesla\meter}$, comparable to or below the flattop tipping requirement of $\SI{9.85}{\tesla\meter}$. Only the worst-case flattop scenario ($\sigma_{\nu_s} = 10^{-2}$) is a demanding condition, requiring an order of magnitude larger per-pass field than tipping ($\SI{132}{\tesla\meter}$). This single design point motivates the spin-coherence improvement loop of Sec.~\ref{subsec:ml_ai_iterative_loop} as a flattop commissioning priority: bringing $\sigma_{\nu_s}$ down to the canonical $10^{-3}$ level reduces the
per-pass $\pi$-pulse field requirement by a factor of ten and places it comfortably within the existing hardware envelope.

Three phase-locked timings follow from the SQUID reference oscillator of Sec.~\ref{sec:fid_spin_tune_phase_lock}. The tipping pulse, fired at $t=0$, has no phase requirement: it acts on a fresh stored $S_y$. The $\pi$ rephasing pulse must be fired at a precession phase that cleanly reflects $(S_x,S_z)$ across $\vec e_z$, at a fixed offset from the SQUID reference, calibrated once per fill. The restore pulse must be timed so that the spin is again at the precession phase it occupied immediately after the tipping pulse, modulo the $S_y$ flip introduced by the $\pi$ pulse. All three timings reduce operationally to fixed offsets from the SQUID reference oscillator.

\subsection{Restore pulse and the closed measurement cycle}
\label{sec:fid-echo-restore}

The state of the ensemble at the echo maximum, Eq.~\eqref{eq:echo_state}, is $(\sin\alpha,-\cos\alpha,0)$. After the echo signal has been sampled by the SQUID, a phase-coherent restore pulse rotates the ensemble back to the stored-spin direction. The required rotation is $R_z(\pi-\alpha)$:
\begin{equation}
	R_z(\pi-\alpha)\,(\sin\alpha,-\cos\alpha,0) \;=\; (0,1,0),
	\label{eq:restore_action}
\end{equation}
which simultaneously undoes the $S_y$ sign reversal introduced by the $\pi$ pulse and the small tipping rotation $\alpha$. The integrated-field requirement is therefore $(1-\alpha/\pi)\simeq 99\%$ of the $\pi$-pulse field requirement already established in Eqs.~\eqref{eq:alpha_recap}--\eqref{eq:per_pass_pi_field}. Figure~\ref{fig:fid_echo_phase_fan}(e) shows the action of the restore pulse in the rotating-frame phase-fan picture.

The full four-pulse cycle (tip, $\pi$ rephase, echo readout, restore) thus returns the beam polarization to within $\mathcal{O}(\alpha^2/\pi^2)\sim 10^{-4}$ of its starting value per cycle, modulo irreversible $T_2$ loss accumulated during the $2\tau$ free-evolution interval. The cycle can be repeated continuously throughout a fill at a duty cycle determined by $\tau$. This is the operational property that makes the SQUID polarimeter noninvasive: the stored polarization is preserved across the measurement, and many cycles can be averaged to improve the statistical precision on the echo amplitude and on the spin tune extracted from the SQUID signal between pulses.

Cumulative drift from imperfect calibration of the restore angle can be eliminated entirely by a Carr--Purcell--Meiboom--Gill (CPMG) sequence~\cite{PhysRev.94.630,10.1063/1.1716296}, in which a second $\pi$ pulse at $t = 3\tau$ produces a second echo at $t = 4\tau$, after which the spin state coincides with the configuration immediately following the tipping pulse, so that the stored polarization sign is restored automatically by the second $\pi$ flip. A single $R_z(-\alpha)$ tip then closes the cycle, with the restore-pulse field requirement reduced to that of the original tipping pulse.

The closed cycle enables a qualitatively new operating mode at the HSR. With a fill lasting several hours and the tip-$\pi$-echo-restore cycle returning the beam polarization to its starting value each time, the spin-precession frequency $f_s$, the spin-tune spread $\sigma_{\nu_s}$, and the in-plane beam polarization can be monitored essentially continuously throughout the entire $\sim 8$-hour store, with negligible cumulative impact on the beam intensity or polarization. A slow feedback on the fitted $f_s$ from rolling spectral windows tracks the spin tune as it drifts during the store, maintaining phase lock across hours of running. The drift itself carries physics information: the spin tune is acutely sensitive to lattice imperfections, closed-orbit distortions, and snake-matching errors, as demonstrated in dedicated spin-tune mapping experiments at COSY~\cite{saleev2017}, so continuous spin-tune monitoring with the SQUID polarimeter functions in parallel as a high-precision diagnostic of HSR magnetic imperfections and operational stability. The precision achievable by this class of free-precession measurement is high: at COSY the JEDI free-precession method determined the spin tune to order $10^{-8}$ in \SI{2.6}{\second} and to $1\times10^{-10}$ over a continuous \SI{100}{\second} cycle~\cite{Eversmann:2015}, among the most precisely measured quantities in any accelerator, and spin-tune mapping fixed the angular orientation of the stable spin axis $\vec{n}_0$ to better than \SI{2.8}{\micro\radian}~\cite{saleev2017}. These benchmarks were reached in a regime far removed from HSR operating conditions: in-plane spin-coherence times near \SI{1000}{\second} obtained through dedicated bunching, electron cooling, and sextupole correction~\cite{Guidoboni2016}, with the stored beam current deliberately limited to below $10^{9}$ deuterons per fill to suppress collective effects~\cite{guidoboni2018}. They indicate the ultimate reach of the free-precession method rather than the precision expected at HSR start-up, where the coherence time is initially of order milliseconds and the stored proton intensity is far higher. Continuous polarimetry of this kind is beyond the reach of the polarimetry methods deployed at RHIC and required at the EIC. The fast carbon-target (pC) polarimeter~\cite{Bravar2005ProtonPolarimetryRHIC} samples the stored beam destructively, with each measurement depleting beam intensity and degrading the emittance, restricting pC polarimetry operationally to a small number of measurements at the beginning, middle, and end of each store rather than continuous monitoring. The noninvasive atomic-hydrogen-jet (HJET)  polarimeter, also operated at RHIC and required at the EIC for absolute proton beam polarization calibration, removes the destructive aspect but provides only marginal statistical precision per unit time relative to what the SQUID concept offers at HSR conditions~\cite{Rathmann2026PRAB}.

\subsection{The echo amplitude $\tau$-scan and what is measured}
\label{sec:fid-echo-tau-scan}

\begin{figure}[!tb]
	\centering
	\includegraphics[width=0.9\linewidth]{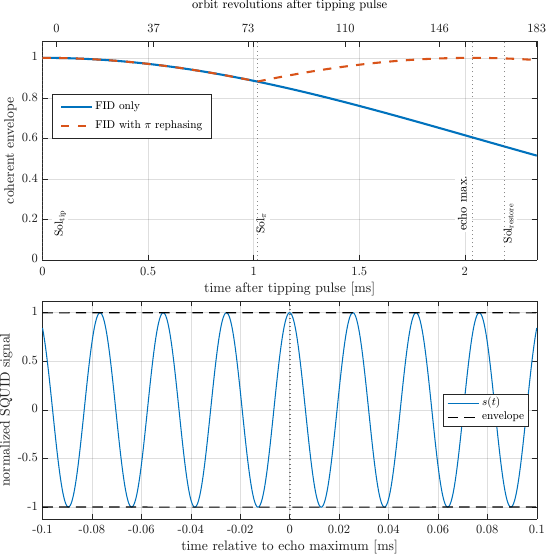}
	\caption{Continuous SQUID readout for the full tip-$\pi$-echo-restore sequence. The upper panel shows the coherent transverse-spin envelope: after the initial FID decay on the reversible time scale $T_2^\ast=\tau_\mathrm{coh}$, the $\pi$ pulse at $t=\tau$ reverses the phase fan and produces an echo maximum near $t\simeq 2\tau$, reduced from the tip-pulse value by $\exp(-2\tau/T_2)$ from irreversible decoherence on the time scale $T_2$. The lower panel zooms on the SQUID signal near the echo maximum: the precession at $f_s = \nu_s f_\mathrm{rev}\simeq\SI{39.1}{\kilo\hertz}$ continues uninterrupted throughout the sequence, and the echo restores the slowly varying coherent envelope, not the underlying precession.}
	\label{fig:fid_echo_continuous_signal}
\end{figure}

The deliverable of the spin-echo sequence is the echo amplitude as a function of the operator-chosen delay $\tau$. From Eq.~\eqref{eq:echo_amplitude},
\begin{equation}
	\frac{M_\perp(2\tau)}{M_\perp(0)} \;=\;
	\exp\!\left(-\frac{2\tau}{T_2}\right),
	\label{eq:echo_tau_scan}
\end{equation}
so a logarithmic-slope fit of the SQUID echo amplitude versus $\tau$ extracts $T_2$ directly. The FID linewidth measured from the same SQUID data within each cycle continues to yield $T_2^\ast$ via Eq.~\eqref{eq:t2_star_definition} as in Sec.~\ref{sec:fid_spin_tune_phase_lock}. The two coherence times can therefore be measured simultaneously in the same fill, and the ratio $T_2/T_2^\ast$ quantifies how much of the apparent $\tau_\mathrm{coh}$ is recoverable dephasing versus genuine irreversible loss. Figure~\ref{fig:fid_echo_continuous_signal} illustrates the appearance of the continuous SQUID signal across one such cycle.

The HSR-specific contributions to $T_2$ include intra-bunch Coulomb scattering, slow drifts of the closed orbit and the snake matching, residual-gas scattering throughout the ring, and beam losses. Quantitative estimates of these contributions require dedicated HSR spin tracking and are beyond the scope of this section. The operational point is that $T_2$ is a measurable, noninvasive diagnostic of these processes via Eq.~\eqref{eq:echo_tau_scan}, complementing the $\sigma_{\nu_s}$ feedback used by the lattice-tuning loop of Sec.~\ref{subsec:ml_ai_iterative_loop}.

\section{Magnetic flux signal and matched-filter reconstruction}
\label{sec:flux-and-matched-filter}

\begin{figure}[htb]
	\centering
	\includegraphics[width=0.90\linewidth]{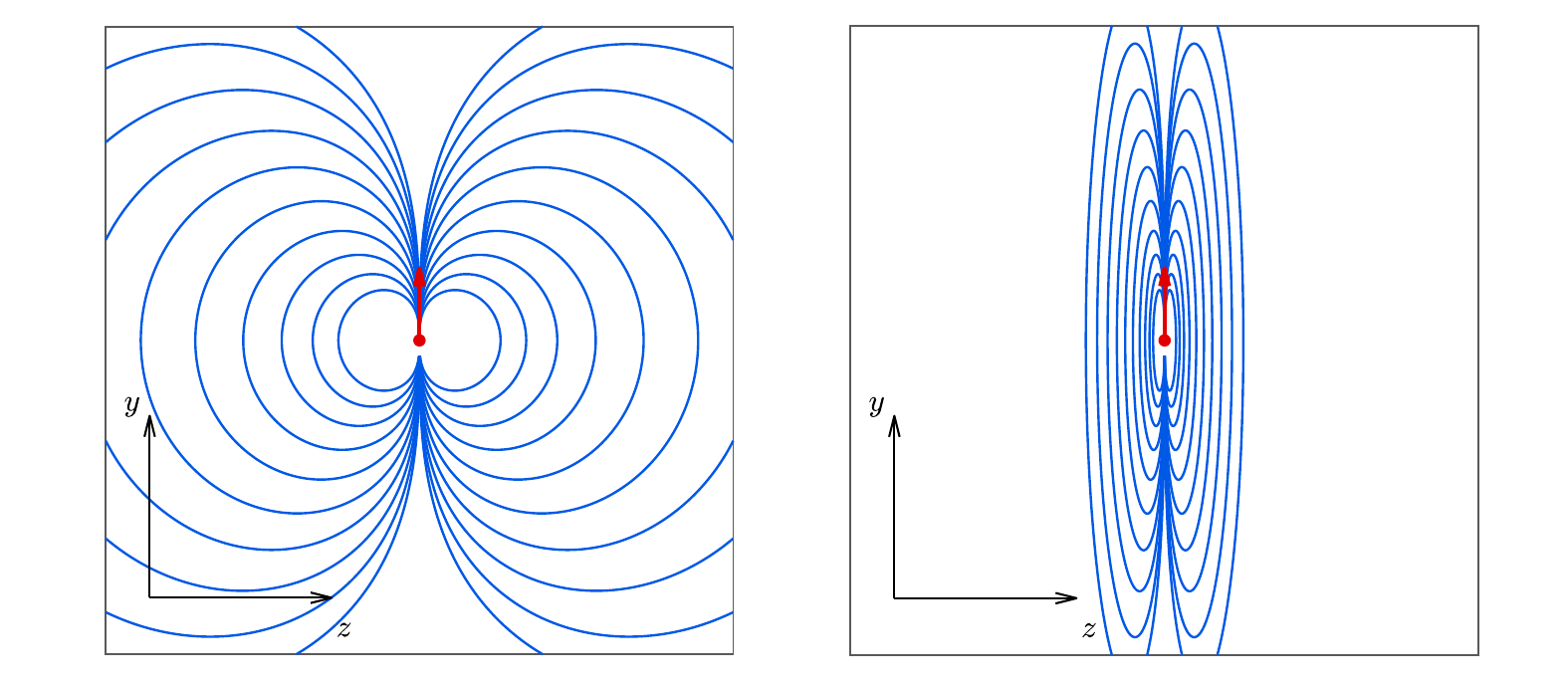}
	\caption{Magnetic-field pattern of a transverse magnetic dipole before and after a Lorentz boost. Both panels are shown in the $y$--$z$ plane, with $y$ vertical and $z$ horizontal. Left: magnetic field lines of a dipole moment $\vec m=m_y\,\ey$, shown in red, at rest. Right: the corresponding Lorentz-boosted field pattern for the same dipole boosted along $+\ez$, shown for $\beta=0.98$. The boost compresses the field pattern in the longitudinal $z$ direction, so that a fixed pickup samples the field over a short time interval set by the bunch length.}
	\label{fig:moving_dipole}
\end{figure}
	
\subsection{Relativistic field of a moving transverse magnetic dipole}
\label{sec:moving-transverse-dipole-field}

The SQUID pickup coil is placed at position $\rcoil\,\ex$ (distance $\rcoil$ from the beam axis along $\ex$), with its effective area normal along $\ex$, outside the beam pipe. For a magnetic dipole $\vec{m} = m_x\,\ex$ moving along $\ez$ at $v \approx c$, the Lorentz transformation of the rest-frame magnetic field gives, at the moment of closest approach of the bunch centre to the coil (i.e.\ when the bunch centre passes the transverse plane $z = 0$ containing the coil)
	\begin{equation}
		B_x^\mathrm{lab} = -\frac{\mu_0}{4\pi}\frac{m_x}{\rcoil^3}.
		\label{eq:Blab}
	\end{equation}
The transverse dipole field at a transverse observation point is not boosted by $\gamma$: the orbit excursion and the spin precession contributions scale in ways that largely cancel for the helical dipole geometry.
	
The bunch has a lab-frame rms length $\sigma_L$ (listed in Table~\ref{tab:params}), related to the rms temporal width by $\sigma_t = \sigma_L / (\beta c)$; both are already lab-frame quantities and require no Lorentz correction. The time for the bunch centre to traverse the coil is negligible; what matters is the duration of the magnetic pulse at the coil, which is set by the bunch length divided by $c$:
	\begin{equation}
		\Delta t \approx 2\sigma_t,
		\label{eq:dt}
	\end{equation}
where the factor of 2 accounts for the full $\pm 1\sigma$ passage time and $\sigma_t$ is the lab-frame rms temporal width from Table~\ref{tab:params}. At EIC flattop ($\sigma_t = \SI{0.200}{ns}$): $\Delta t \approx \SI{0.40}{ns}$. The magnetic pulse is far too brief for time-domain detection; the signal is therefore extracted in the frequency domain via the matched-filter coherent summation described in Sec.~\ref{sec:matched-filter-coherent-summation}.

\subsection{Flux per bunch pass}
\label{sec:flux-per-bunch-pass}

The single-turn flux at the pickup for a bunch carrying transverse
magnetic moment $m_x$ is
\begin{equation}
	\Phi_\mathrm{bunch}
	=
	\frac{\mu_0}{4\pi}\,
	\frac{|m_x|}{\rcoil^3}\,
	A\,
	f_\mathrm{geom},
	\label{eq:flux}
\end{equation}
with $\rcoil$, $A$, and $f_\mathrm{geom}$ from
Table~\ref{tab:three_channel_pickup_inputs}. The longitudinal form
factor $f_\mathrm{geom}\sim 1$--$3$ accounts for the finite saddle
extent along $\ez$ relative to the dipole field profile.

For unit polarization ($P=1$) and full transverse projection
($\sin\alpha=1$), the bunch magnetic moment is
\begin{equation}
	|m_x| = N_p\,\mu_p ,
	\label{eq:mx_unit_projection}
\end{equation}
with $N_p$ from Table~\ref{tab:params} and
$\mu_p=\SI{1.4106e-26}{\joule\per\tesla}$ the proton magnetic moment.
Evaluating Eq.~\eqref{eq:flux},
\begin{equation}
	\Phi_\mathrm{bunch}^{\mathrm{inj}}
	\approx
	\SI{17.6}{\micro\Phi_0},
	\qquad
	\Phi_\mathrm{bunch}^{\mathrm{flat}}
	\approx
	\SI{4.41}{\micro\Phi_0},
	\label{eq:flux_per_loop_numeric}
\end{equation}
where $\Phi_0=h/(2e)=\SI{2.07e-15}{Wb}$ is the flux quantum. These
values represent the flux per single saddle turn that would be
delivered by a fully polarized ensemble with the spin vector lying
entirely in the transverse plane, and serve as the geometric flux
template for the matched-filter sensitivity that follows.

The pickup is wound as a multi-turn cosine-$\theta$ saddle connected
to the SQUID input through a superconducting flux transformer. With
winding multiplicity $N_\mathrm{turns}$ from
Table~\ref{tab:three_channel_pickup_inputs} and flux-transformer
coupling efficiency $\eta$ from Table~\ref{tab:squid_readout}, the
flux delivered to the SQUID input per bunch passage is
\begin{equation}
	\Phi_\mathrm{pickup}
	=
	N_\mathrm{turns}\,\eta\,\Phi_\mathrm{bunch}.
	\label{eq:flux_pickup}
\end{equation}
For unit transverse polarization,
\begin{equation}
	\Phi_\mathrm{pickup}^{\mathrm{inj}}
	\approx
	\SI{1236}{\micro\Phi_0},
	\qquad
	\Phi_\mathrm{pickup}^{\mathrm{flat}}
	\approx
	\SI{309}{\micro\Phi_0}.
	\label{eq:flux_pickup_numeric_full}
\end{equation}
The operating polarization $P$ and tipping angle $\alpha$
(Table~\ref{tab:params}) enter as a single transverse-polarization
factor $P_\perp = P\sin\alpha$ at the matched-filter sensitivity stage
in Sec.~\ref{sec:coherent-integration}, not in the per-bunch flux
template here. The SQUID array architecture with $N_\mathrm{squids}$
independent channels (Table~\ref{tab:squid_readout}) contributes an
additional factor $\sqrt{N_\mathrm{squids}}$ to the matched-filter
signal-to-noise ratio, also derived in
Sec.~\ref{sec:coherent-integration}.
	
\begin{table}[htb]
	\centering
	\caption{SQUID readout parameters. Flux noise is the conservative published value for the Magnicon-class two-stage current sensor at $\SI{4.2}{\kelvin}$ read out by the XXF-1 electronics~\cite{4277368}, anchoring the canonical $\SI{4}{\kelvin}$ operating temperature assumed throughout this paper.}
	\label{tab:squid_readout}
	\renewcommand{\arraystretch}{1.15}
	\sisetup{table-number-alignment=center}
	\begin{tabular}{l c c S[table-format=1.2]}
		\hline\hline
		{Parameter} & {Symbol} & {Unit} & {Value} \\
		\hline
		Flux-transformer coupling efficiency & $\eta$                                  & 1                                       & 0.7 \\
		SQUID array channels                 & $N_\mathrm{squids}$                     & 1                                       & 4 \\
		SQUID white flux noise               & $S_\Phi^{1/2}$                          & \si{\micro\Phi_0\per\sqrt{\hertz}} & 0.4 \\
		SQUID operating temperature          & $T$                                      & \si{\kelvin}                           & 4 \\
		\hline\hline
	\end{tabular}
\end{table}

The SQUID sensors and pickup coils are operated at a base temperature of $T = \SI{4}{\kelvin}$, anchored to the published noise specification of the Magnicon-class two-stage current sensor at that temperature~\cite{4277368}. The cold mass is contained in a vacuum envelope with two beam-tube apertures of \SI{100}{\milli\meter} diameter (combined aperture area $\approx \SI{157}{\centi\meter\squared}$) for the upstream and downstream beam pipes. On the beam axis, the transverse dipole field of a single polarized bunch falls as $1/z^3$ outside the active pickup region, dropping to $\sim\SI{e-19}{\tesla}$ at $z = \SI{10}{\centi\meter}$. For comparison, the single-bunch SQUID noise floor on a 40-mm-radius single-turn pickup is $\sim S_\Phi^{1/2}\sqrt{\mathrm{BW}}\,/(\pi r_\mathrm{coil}^2) \sim \SI{e-14}{\tesla}$ for the per-bunch bandwidth $\mathrm{BW}\sim 1/\sigma_t \sim \SI{1}{\giga\hertz}$ at injection, i.e.\ five orders of magnitude above the on-axis dipole field at $z = \SI{10}{\centi\meter}$; the ceramic former that carries the pickup coils can therefore be extended axially upstream and downstream of the active region at no signal cost. We assume such an extension is configured as a graded thermal shield, with the warm end at $\sim\SI{300}{\kelvin}$ at the beam-tube interface, an intermediate stage at $\sim\SI{40}{\kelvin}$, and the active region at $\SI{4}{\kelvin}$, so that the room-temperature radiative load through the apertures is intercepted before reaching the active pickup. With this geometry the residual load at the $\SI{4}{\kelvin}$ stage is well within the $\sim\SI{1}{\watt}$ capacity of a standard pulse-tube cryocooler. Operation at sub-K temperatures with a multi-stage refrigerator would further reduce the flux noise but is not assumed for the baseline performance estimates of this paper.

\subsection{Bunch spin pattern at injection}
\label{sec:bunch-spin-pattern}
	
The bunch spin pattern at the EIC HSR has not yet been settled. Two operationally realistic options are the RHIC-style alternating pattern $s_j = (-1)^j$ for $j = 0, \ldots, 289$, which is the canonical fill at RHIC for systematic cancellation of spin asymmetry errors, and a random pattern, which may be preferred at the EIC for added systematic robustness in spin-correlation measurements. Whatever pattern is loaded at injection is not modified during the store. At flattop, each parent bunch $j$ produces four daughters at positions $4j, \ldots, 4j+3$, all inheriting the parent sign $s_j$. The matched-filter formulation developed below in Sec.~\ref{sec:matched-filter-coherent-summation} accommodates any known spin-sign pattern $\{s_j\}$ (Eq.~\eqref{eq:mf_result}); the alternating pattern is used throughout this paper as the worked example because it is concrete, operationally familiar from RHIC, and is the most demanding case for an uncorrected sum (Sec.~\ref{sec:naive-every-other-turn-summation}), providing a worst-case anchor for the matched-filter gain.
	
The important point is not the visual pattern itself, but the cancellation it produces in an unweighted sum. 
For the alternating injection pattern, a naive sum over bunches gives only
\begin{equation}
	C_\mathrm{naive}
	=
	\sum_{j=0}^{N_\mathrm{fill}-1}
	s_j \cos\psi_j
	\simeq 1
	\quad
	\text{(injection)} ,
	\label{eq:cnaive_inj}
\end{equation}
and, after fourfold RF splitting,
\begin{equation}
	C_\mathrm{naive}
	\simeq 4
	\quad
	\text{(flattop)} .
	\label{eq:cnaive_flat}
\end{equation}
Thus an every-other-turn or uncorrected bunch sum discards essentially the entire multi-bunch gain. The matched filter avoids this cancellation by applying the known spin sign and lattice phase correction to each bunch before summation.

\subsection{Lattice spin phase offsets}
\label{sec:lattice-spin-phase-offsets}

The bunch index also carries a spin-phase offset. Bunch $j$ is displaced from bunch 0 by the fraction $j/N_\mathrm{fill}$ of one ring circumference. Since the spin advances by $2\pi\nu_s$ per full revolution, the same longitudinal displacement corresponds to the spin phase advance
\begin{equation}
	\psi_j
	=
	2\pi\nu_s\,\frac{j}{N_\mathrm{fill}} .
	\label{eq:psi_general}
\end{equation}
For the HSR operating point $\nu_s=1/2$, this becomes
\begin{equation}
	\psi_j
	=
	\frac{\pi j}{N_\mathrm{fill}},
	\label{eq:psi}
\end{equation}
where $N_\mathrm{fill}=290$ at injection and $N_\mathrm{fill}=1160$ at flattop.

This phase is not an additional unknown spin rotation. It is the deterministic phase offset that follows from the position of bunch $j$ around the ring. Even if all bunches had the same polarization magnitude, their signals at a fixed pickup would not arrive with identical spin phase, because they sample different azimuthal positions of the same spin-precession wave around the ring.

The signal from bunch $j$ on turn $n$ can therefore be written as
\begin{equation}
	x_{j,n}
	=
	\Phi_\mathrm{pickup}\,
	s_j\,P_j\,\sin\alpha\,
	\cos(\pi n+\psi_j)
	+
	\mathrm{noise}_{j,n}.
	\label{eq:xjn}
\end{equation}
Here $\Phi_\mathrm{pickup}$ is the per-bunch flux at the SQUID input for unit polarization and full transverse projection [Eq.~\eqref{eq:flux_pickup_numeric_full}], $s_j=\pm1$ is the injected spin-sign pattern, $P_j$ is the polarization magnitude of bunch $j$, $\sin\alpha$ is the transverse projection set by the operating tipping angle $\alpha$ (Table~\ref{tab:params}), and the factor $\cos(\pi n+\psi_j)$ describes the spin-precession phase at the pickup. The term $\pi n$ gives the turn-by-turn sign reversal from $\nu_s=1/2$, while $\psi_j$ is the fixed bunch-dependent phase offset.

The matched filter uses both known pieces of information, the injected sign $s_j$ and the deterministic phase $\psi_j$, to rotate each bunch contribution into a common phase before summation. This correction chain is illustrated schematically in Fig.~\ref{fig:matched_filter}.
	
\subsection{Naive every-other-turn summation}
\label{sec:naive-every-other-turn-summation}
	
Summing all bunch signals on even turns yields an effective pattern-weighted sum $C_\mathrm{naive} = \sum_j s_j \cos\psi_j$. For $s_j = (-1)^j$ with $\psi_j = \pi j/N_\mathrm{fill}$ this is a geometric series evaluating to $C_\mathrm{naive} = 1$: the alternating pattern nearly cancels, leaving only one equivalent bunch at injection and four at flattop.

\subsection{Matched filter: phase-corrected coherent summation}
\label{sec:matched-filter-coherent-summation}

\begin{figure}[!tb]
	\centering
	\includegraphics[width=\linewidth]{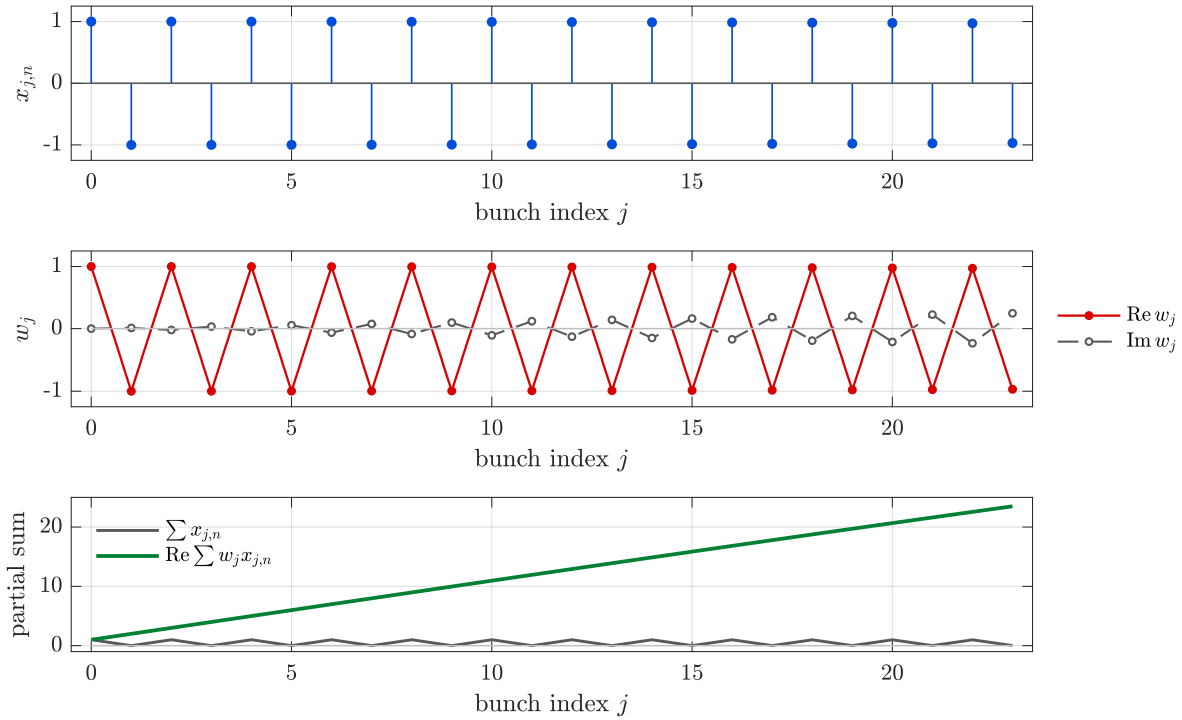}
	\caption{Schematic illustration of the matched-filter correction for a representative subset of bunches. Top: raw bunch samples $x_{j,n}$ contain the injected spin sign $s_j$, the lattice spin-phase offset $\psi_j$, and the turn-by-turn factor $(-1)^n$ from $\nu_s=1/2$ [Eq.~\eqref{eq:xjn}]. Middle: real and imaginary parts of the known per-bunch correction factor, as defined in Eqs.~\eqref{eq:matched_filter_weight} and \eqref{eq:matched_filter_weight_re_im}. Bottom: running sum over the displayed bunch subset. The uncorrected partial sum corresponds to the naive pattern-weighted sum $C_\mathrm{naive}$, which remains small because adjacent bunch contributions cancel [Sec.~\ref{sec:naive-every-other-turn-summation}]. The phase-corrected sum corresponds to $\mathrm{Re}[\tilde S_n]$ after applying the matched-filter weights [Eq.~\eqref{eq:mf}], and grows coherently as in Eq.~\eqref{eq:mf_result}. The full analysis applies the same correction to all $N_\mathrm{fill}$ bunches and then accumulates over turns.}
	\label{fig:matched_filter}
\end{figure}

The purpose of the matched filter is to remove the two known bunch-dependent factors before the bunches are summed: the injected spin sign $s_j$ and the lattice spin-phase offset $\psi_j$. We therefore define the complex correction weight
\begin{equation}
	w_j
	=
	s_j e^{-i\psi_j}
	=
	s_j\left(\cos\psi_j-i\sin\psi_j\right).
	\label{eq:matched_filter_weight}
\end{equation}
The real and imaginary parts of this weight are
\begin{equation}
	\mathrm{Re}\,w_j
	=
	s_j\cos\psi_j,
	\qquad
	\mathrm{Im}\,w_j
	=
	-s_j\sin\psi_j .
	\label{eq:matched_filter_weight_re_im}
\end{equation}
These are the two correction components shown in the middle panel of Fig.~\ref{fig:matched_filter}.

The phase-corrected bunch sum on turn $n$ is then
\begin{equation}
	\tilde{S}_n
	=
	\sum_{j=0}^{N_\mathrm{fill}-1}
	w_j\,x_{j,n}
	=
	\sum_{j=0}^{N_\mathrm{fill}-1}
	s_j e^{-i\psi_j} x_{j,n}.
	\label{eq:mf}
\end{equation}
Substituting Eq.~\eqref{eq:xjn}, and taking $P_j=P$ for the coherent-sum estimate, gives
\begin{equation}
	\tilde{S}_n
	=
	\Phi_\mathrm{pickup}\,P_\perp
	\sum_{j=0}^{N_\mathrm{fill}-1}
	s_j e^{-i\psi_j}s_j
	\cos(\pi n+\psi_j)
	+
	\tilde{n}_n ,
	\label{eq:mf_substituted}
\end{equation}
where $\tilde{n}_n$ denotes the filtered noise contribution. Using $s_j^2=1$ and writing the cosine as the real part of a complex exponential, the coherent contribution becomes
\begin{equation}
	\mathrm{Re}\,\tilde{S}_n
	=
	\Phi_\mathrm{pickup}\,P_\perp\,(-1)^n
	\sum_{j=0}^{N_\mathrm{fill}-1} 1
	=
	\Phi_\mathrm{pickup}\,P_\perp\,(-1)^n N_\mathrm{fill}.
	\label{eq:mf_result}
\end{equation}
Thus the matched filter rotates all bunch contributions into the same phase before summation. The uncorrected partial sum cancels for the alternating pattern, while the corrected sum grows coherently, as illustrated in the bottom panel of Fig.~\ref{fig:matched_filter}.

The optimal pattern sum is therefore
\begin{equation}
	C_\mathrm{opt}
	=
	N_\mathrm{fill}
	=
	290\;(\text{injection}),
	\qquad
	1160\;(\text{flattop}).
	\label{eq:copt}
\end{equation}
This result is independent of the particular spin-sign pattern $\{s_j\}$, provided that the pattern is known and included in $w_j$.

The matched-filter output of Eq.~\eqref{eq:mf_result} is a turn-by-turn
signal at the spin-precession frequency
$f_s = \nu_s\,f_\mathrm{rev}$ defined in
Eq.~\eqref{eq:spin_signal_frequency_sec2}, which evaluates to
$\SI{39.07}{\kilo\hertz}$ at injection and $\SI{39.10}{\kilo\hertz}$ at
flattop for $\nu_s = 1/2$.

A real-time implementation of the matched-filter sum, Eq.~\eqref{eq:mf}, does not require sampling each bunch individually and applying the weights in software. The general firmware implementation mixes the SQUID time-stream with a digital per-bunch reference encoding the known pattern $\{s_j\}$ and the lattice phase offsets $\psi_j$, generated by a digital signal processor locked to the machine RF clock. For the alternating-pattern worked example $s_j = (-1)^j$ this digital reference reduces to a single sinusoidal oscillator at the bunch-pattern alternation rate $f_\mathrm{ref} = N_\mathrm{fill}\,f_s$, where $f_s$ is the spin-precession frequency, defined in Eq.\,\eqref{eq:spin_signal_frequency_sec2}
\begin{align}
	f_\mathrm{ref}^\mathrm{inj}
	&=
	N_\mathrm{fill}^\mathrm{inj}\,f_s^\mathrm{inj}
	=
	290 \times \SI{39.07}{\kilo\hertz}
	=
	\SI{11.33}{MHz},
	\label{eq:fref_inj}
	\\
	f_\mathrm{ref}^\mathrm{flat}
	&=
	N_\mathrm{fill}^\mathrm{flat}\,f_s^\mathrm{flat}
	=
	1160 \times \SI{39.10}{\kilo\hertz}
	=
	\SI{45.35}{MHz}.
	\label{eq:fref_flat}
\end{align}
The relation $f_\mathrm{ref} = N_\mathrm{fill}\,f_s$ makes the physical content of the alternating-pattern case explicit: $f_\mathrm{ref}$ is $N_\mathrm{fill}$ times the spin-precession frequency, equivalently one half of the bunch-passage rate $N_\mathrm{fill}\,f_\mathrm{rev}$ at $\nu_s = 1/2$. It is the frequency at which the alternating pattern $s_j = (-1)^j$ deposits a spectral line in the SQUID time-stream. The polarization signal appears not as a line at $f_\mathrm{ref}$ itself but as a pair of sidebands at $f_\mathrm{ref} \pm f_s$ around it. Mixing the SQUID output with the sinusoidal reference at $f_\mathrm{ref}$ and applying a low-pass filter at a bandwidth slightly above $f_s$ demodulates these sidebands down to baseband, leaving a complex oscillator at $\pm f_s$ that carries the bunch-pattern-corrected polarization signal. For a non-alternating pattern (e.g.\ a random pattern) the spin signal is distributed over multiple spectral lines determined by the Fourier content of $\{s_j\}$, and the firmware reference is the corresponding digital per-bunch sequence rather than a single sinusoid; the matched-filter principle and the per-bunch phase rotation by $e^{-i\psi_j}$ are unchanged.

The lattice spin-phase offsets $\psi_j$ enter as a fixed per-bunch phase that the firmware-side matched filter applies as a static phase rotation to the demodulated samples; this is the hardware analog of the complex weight $w_j$ in Eq.~\eqref{eq:matched_filter_weight}. Phase-locking the reference oscillator to the RF clock ensures that the bunch timing, the spin-sign pattern, and the phase correction share the same temporal reference, so that the matched filter remains coherent over the full integration time.

\subsection{In-plane coherence time and spin tune spread}
\label{sec:coherent_integration_spin_tune_spread}

The matched-filter sum assumes that the transverse spin phase remains
coherent over the integration time. The synchronous particle has
$\nu_s=0.5$, but off-momentum particles and particles at different
betatron amplitudes acquire spin-tune offsets,
\begin{equation}
	\nu_{s,i}
	=
	\nu_{s,0}
	+
	\delta\nu_{s,i},
	\label{eq:matched_filter_spin_tune_offsets}
\end{equation}
with $\nu_{s,0}=1/2$ for the ideal HSR spin tune. The rms spread
$\sigma_{\nu_s}$ of these offsets sets the FID envelope decay rate,
with characteristic coherence time given by
Eq.~\eqref{eq:t2_star_definition}.

The working value adopted throughout this analysis is the effective rms
spin-tune spread $\sigma_{\nu_s}^\mathrm{eff} = 10^{-3}$
(Table~\ref{nzero:tab:matlab_params}), which corresponds to
\begin{equation}
	\tau_{\mathrm{coh}}^\mathrm{eff}
	\simeq
	\SI{2.04}{\milli\second}
	\,.
	\label{eq:tau_coh_canonical}
\end{equation}
This is the value used for the matched-filter sensitivity estimate in Sec.~\ref{sec:coherent-integration} and for the multi-cycle accumulation that follows. Two larger reference scales are tabulated for context: the DLC ADST scale $\sigma_{\nu_s}^\mathrm{DLC} \lesssim 10^{-2}$ (Table~\ref{nzero:tab:matlab_params}) gives $\tau_{\mathrm{coh}}\simeq\SI{0.20}{\milli\second}$, and the LC ADST scale $\sigma_{\nu_s}^\mathrm{LC} \simeq 2\times 10^{-2}$ gives $\tau_{\mathrm{coh}}\simeq\SI{0.10}{\milli\second}$. These scales are used in Sec.~\ref{sec:fid_spin_tune_phase_lock} for sizing the initial longitudinal kicker, not for the matched-filter sensitivity estimate.

At the operating tipping angle $\alpha$ (Table~\ref{tab:params}), the matched-filter integration time required to reach $\delta P/P = 1\%$ exceeds $\tau_{\mathrm{coh}}^\mathrm{eff}$. The measurement therefore proceeds as a sequence of $N$ tip-read-restore cycles, each within one coherence window, and is averaged statistically; the corresponding cycle counts and integration times are given in Sec.~\ref{sec:coherent-integration}. For the first spin-tune determination, before the spin-coherence environment has been characterized, a larger tipping angle and a single-record (short-window) detection are used, as discussed in Sec.~\ref{sec:fid_spin_tune_phase_lock}.

The corresponding sideband linewidth around the spin-precession
frequency $f_s$ [Eq.~\eqref{eq:spin_signal_frequency_sec2}] is set by
the spin-tune spread itself,
\begin{equation}
	\sigma_f
	\simeq
	f_{\mathrm{rev}}\,\sigma_{\nu_s} ,
	\label{eq:sigma_f_def}
\end{equation}
which evaluates to $\sigma_f^\mathrm{eff} \simeq \SI{78}{\hertz}$ for
the working value $\sigma_{\nu_s}^\mathrm{eff} = 10^{-3}$, and to
$\SI{0.78}{\kilo\hertz}$ and $\SI{1.56}{\kilo\hertz}$ for the DLC and
LC reference scales, respectively. Measuring this linewidth provides a
direct noninvasive diagnostic of the spin-tune spread and of
subsequent machine tuning.

\subsection{Coherent integration}
\label{sec:coherent-integration}

The SQUID readout measures a beam-synchronous sequence of single-pass flux waveforms. Let $\Phi_1(t)$ denote the flux waveform produced at the SQUID input by one bunch passage for unit transverse polarization. This waveform includes the magnetic coupling to the pickup, the finite bunch length, the pickup and flux-transformer response, and the SQUID/electronics transfer function. Its Fourier component at angular frequency $\omega$ is
\begin{equation}
	\widetilde{\Phi}_1(\omega)
	=
	\int_{-\infty}^{+\infty}
	\Phi_1(t)\,e^{-i\omega t}\,dt .
	\label{eq:single_pass_flux_spectrum}
\end{equation}
The quantity $\widetilde{\Phi}_1(\omega)$ has units of flux times time.

For the transverse FID measurement, the relevant frequency is the spin-precession angular frequency
\begin{equation}
	\omega_s
	=
	2\pi \nu_s f_{\mathrm{rev}} .
	\label{eq:spin_angular_frequency}
\end{equation}
The bunch train converts the single-pass response into a coherent spectral line at this spin sideband. After applying the known injected spin sign and the lattice phase correction described in Sec.~\ref{sec:matched-filter-coherent-summation}, the coherent line amplitude is
\begin{equation}
	\Phi_{\mathrm{line}}
	=
	f_{\mathrm{rev}}\,
	C_{\mathrm{opt}}\,
	P_\perp\,
	\left|
	\widetilde{\Phi}_1(\omega_s)
	\right| ,
	\label{eq:spin_sideband_line_amplitude}
\end{equation}
where $P_\perp = P\sin\alpha$ is the transverse polarization component, with $P$ the beam polarization and $\alpha$ the operating tipping angle (Table~\ref{tab:params}), and $C_{\mathrm{opt}}=N_{\mathrm{fill}}$ for the matched-filter sum.

For white SQUID flux noise with amplitude spectral density $S_\Phi^{1/2}$, coherent integration for a time $T$ gives
\begin{equation}
	\mathrm{SNR}(T)
	=
	\frac{
		\Phi_{\mathrm{line}}
	}{
		S_\Phi^{1/2}
	}
	\sqrt{T}
	=
	\frac{
		f_{\mathrm{rev}}\,
		C_{\mathrm{opt}}\,
		P_\perp\,
		\left|
		\widetilde{\Phi}_1(\omega_s)
		\right|
	}{
		S_\Phi^{1/2}
	}
	\sqrt{T} \, .
	\label{eq:matched_filter_snr_spectral}
\end{equation}
This expression is the normalization used for the numerical sensitivity estimates. The peak single-pass flux scale in Sec.~\ref{sec:flux-per-bunch-pass} fixes the magnetic coupling strength, while the matched-filter sensitivity is set by the spectral response $\left|\widetilde{\Phi}_1(\omega_s)\right|$ of the full pickup and readout chain.

An alternative time-domain template form of the matched-filter
signal-to-noise ratio exposes the hardware and ensemble inputs
separately and is the form used to evaluate the numerical sensitivity
in Sec.~\ref{sec:numerical-sensitivity-estimates}. For a Gaussian
bunch profile $\Phi_1(t)=\Phi_\mathrm{pickup}\exp(-t^2/2\sigma_t^2)$
with peak $\Phi_\mathrm{pickup}$ at the SQUID input
[Eq.~\eqref{eq:flux_pickup_numeric_full}] and rms width $\sigma_t$
(Table~\ref{tab:params}), define the matched-filter effective pulse
width
\begin{equation}
	\tau_h
	=
	\sqrt{\pi}\,\sigma_t ,
	\label{eq:matched_filter_window}
\end{equation}
chosen so that $\int|\Phi_1(t)|^2\,dt = \Phi_\mathrm{pickup}^2\,\tau_h$.
Accounting for an array of $N_\mathrm{squids}$ independent SQUID
channels reading the same pickup (Table~\ref{tab:squid_readout}), the
matched-filter signal-to-noise ratio after coherent integration over
time $T$ is
\begin{equation}
	\mathrm{SNR}(T)
	=
	K_\mathrm{template}\,P_\perp\sqrt{T},
	\label{eq:matched_filter_snr_template}
\end{equation}
with
\begin{equation}
	K_\mathrm{template}
	=
	\frac{\Phi_\mathrm{pickup}}{S_\Phi^{1/2}}
	\sqrt{N_\mathrm{fill}\,f_\mathrm{rev}\,\tau_h\,N_\mathrm{squids}} ,
	\label{eq:K_template}
\end{equation}
where $N_\mathrm{fill}$ and $f_\mathrm{rev}$ are the fill bunch count
and revolution frequency
(Tables~\ref{tab:params} and~\ref{nzero:tab:matlab_params}), and
$S_\Phi^{1/2}$ is the SQUID white flux noise
(Table~\ref{tab:squid_readout}).

The template form factorizes the matched-filter sensitivity
$K_\mathrm{template}$ into four physical contributions:
$\Phi_\mathrm{pickup}/S_\Phi^{1/2}$ is the single-channel
flux-to-noise ratio set by the pickup geometry and the SQUID
hardware; $\sqrt{N_\mathrm{fill}\,f_\mathrm{rev}}$ is the statistical
gain from $N_\mathrm{fill}\,f_\mathrm{rev}$ bunch passages per unit
time; $\sqrt{\tau_h}$ is the matched-filter window per pulse, set by
the bunch length; and $\sqrt{N_\mathrm{squids}}$ is the
parallel-channel gain from the SQUID array. The transverse
polarization factor $P_\perp = P\sin\alpha$ multiplies this template
linearly, separating the operating-point choice of polarization and
tipping angle from the hardware-fixed sensitivity scale
$K_\mathrm{template}$.

\subsection{Numerical sensitivity estimates}
\label{sec:numerical-sensitivity-estimates}

Inserting the values from Tables~\ref{tab:params},
\ref{nzero:tab:matlab_params},
\ref{tab:three_channel_pickup_inputs}, and \ref{tab:squid_readout}
into Eq.~\eqref{eq:K_template} gives the matched-filter sensitivity at
the two operating energies,
\begin{align}
	K_\mathrm{template}^\mathrm{inj}
	&\approx \SI{1108}{\per\sqrt{\second}},
	\label{eq:K_template_inj_numeric}
	\\
	K_\mathrm{template}^\mathrm{flat}
	&\approx \SI{277}{\per\sqrt{\second}}.
	\label{eq:K_template_flat_numeric}
\end{align}
The injection sensitivity exceeds the flattop sensitivity by a factor
of four: the four times larger protons-per-bunch
$N_p^\mathrm{inj}/N_p^\mathrm{flat} = 4$ enters $\Phi_\mathrm{pickup}$
linearly, while the four times larger fill count
$N_\mathrm{fill}^\mathrm{flat}/N_\mathrm{fill}^\mathrm{inj} = 4$ enters
$K_\mathrm{template}$ only as $\sqrt{N_\mathrm{fill}}$. The remaining factor of two comes from the matched-filter window $\sqrt{\tau_h}\propto\sqrt{\sigma_t}$: the injection bunch is four times longer ($\sigma_t = \SI{0.801}{ns}$ versus $\SI{0.200}{ns}$), so $\sqrt{\tau_h^\mathrm{inj}/\tau_h^\mathrm{flat}} = 2$.

At the operating tipping angle $\alpha = \SI{30}{mrad}$ and beam
polarization $P = 0.7$ (Table~\ref{tab:params}), the transverse
polarization factor is $P_\perp = P\sin\alpha \approx 0.021$. The
integration time required to reach $\delta P/P = 1\%$
($\mathrm{SNR} = 100$) follows from
Eq.~\eqref{eq:matched_filter_snr_template} as
\begin{equation}
	T_{1\%}
	=
	\left(
	\frac{100}{K_\mathrm{template}\,P_\perp}
	\right)^{\!2},
	\label{eq:T1pct_general}
\end{equation}
giving
\begin{align}
	T_{1\%}^\mathrm{inj}
	&\approx \SI{18.5}{\second},
	\\
	T_{1\%}^\mathrm{flat}
	&\approx \SI{4.9}{\minute}.
	\label{eq:T1pct_numeric}
\end{align}
Both times exceed the coherence time $\tau_\mathrm{coh}^\mathrm{eff} \simeq \SI{2.04}{\milli\second}$ [Eq.~\eqref{eq:tau_coh_canonical}], so the measurement proceeds as a sequence of tip-read-restore cycles, each within one coherence window, accumulated statistically over $N \sim T_{1\%}/T_\mathrm{cycle}$.

For reference, the integration time at full transverse projection
($P_\perp = 0.7$, corresponding to $\sin\alpha = 1$) would be
$T_{1\%}^\mathrm{inj} \approx \SI{16.6}{\milli\second}$ at injection and
$T_{1\%}^\mathrm{flat} \approx \SI{266}{\milli\second}$ at flattop. The
$1/\sin^2\alpha$ scaling of $T_{1\%}$ at fixed precision accounts for
the three-orders-of-magnitude increase between full projection and the
operating tipping angle.

Table~\ref{tab:snr} summarizes the three analysis strategies discussed
in Sec.~\ref{sec:naive-every-other-turn-summation} and
Sec.~\ref{sec:matched-filter-coherent-summation}. The near-cancellation
of the alternating injection pattern in the naive every-other-turn sum
leaves an effective sample of one bunch at injection and four at
flattop, and the corresponding $1/N$ degradation makes the integration
time impractically long. Selecting only the same-sign bunches recovers
approximately half the fill at injection, but with a factor-of-two
reduction in usable samples. The matched filter recovers the full
$N_\mathrm{fill}$ samples and is the strategy used throughout this
analysis.

\begin{table}[ht]
	\centering
	\caption{Effective pattern sum $C$ and integration time $T_{1\%}$ for $\delta P / P = 1\%$ at the operating tipping angle $\alpha = \SI{30}{mrad}$ and beam polarization $P = 0.7$ (Table~\ref{tab:params}), with $K_\mathrm{template}$ from Eq.~\eqref{eq:K_template} evaluated at injection and flattop and the alternative pattern sums discussed in Sec.~\ref{sec:naive-every-other-turn-summation}.}
	\label{tab:snr}
	\renewcommand{\arraystretch}{1.15}
	\sisetup{table-number-alignment=center}
	\begin{tabular}{l l S[table-format=4.0] c}
		\hline\hline
		{Condition} & {Analysis} & {$C$} & {$T_{1\%}$} \\
		\hline
		Injection, 290 bunches & Naive every-other-turn &    1 & 17.98~days \\
		Injection, 290 bunches & Select same-sign only  &  145 & \SI{73.9}{\second} \\
		Injection, 290 bunches & Matched filter         &  290 & \SI{18.47}{\second} \\
		\hline
		Flattop, 1160 bunches  & Naive every-other-turn &    4 & 287.9~days \\
		Flattop, 1160 bunches  & Matched filter         & 1160 & \SI{4.929}{\minute} \\
		\hline\hline
	\end{tabular}
\end{table}

\begin{figure}[htb]
	\centering
	\includegraphics[width=0.85\linewidth]{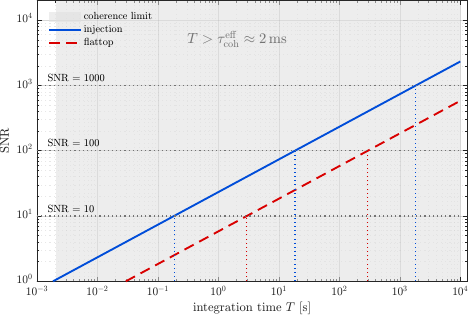}
	\caption{SNR as a function of integration time $T$ for the matched filter at injection and flattop, for the operating-point transverse polarization $P_\perp = P\sin\alpha = 0.021$ (Table~\ref{tab:params}) and the beam parameters of Tables~\ref{tab:params}, \ref{nzero:tab:matlab_params}, and \ref{tab:three_channel_pickup_inputs}. Horizontal dotted lines indicate $\mathrm{SNR}=10$, $100$, and $1000$, corresponding to relative statistical uncertainties of $10\%$, $1\%$, and $0.1\%$, respectively. Vertical dotted lines mark the corresponding integration times. The grey shaded region indicates $T > \tau_\mathrm{coh}^\mathrm{eff}$, using the canonical spin-tune spread $\sigma_{\nu_s}^\mathrm{eff} = 10^{-3}$ and $\tau_\mathrm{coh}^\mathrm{eff} \simeq \SI{2.04}{\milli\second}$ from Eq.~\eqref{eq:tau_coh_canonical}.}
	\label{fig:snr_time}
\end{figure}

\begin{figure}[tbp]
	\centering
	\begin{subfigure}[t]{0.32\textwidth}
		\centering
		\includegraphics[width=\linewidth]{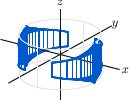}
		\caption{Cosine-$\theta$ pickup.}
		\label{fig:three_channel_pickup_basis_cos}
	\end{subfigure}
	\hfill
	\begin{subfigure}[t]{0.32\textwidth}
		\centering
		\includegraphics[width=\linewidth]{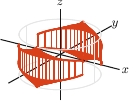}
		\caption{Sine-$\theta$ pickup.}
		\label{fig:three_channel_pickup_basis_sin}
	\end{subfigure}
	\hfill
	\begin{subfigure}[t]{0.32\textwidth}
		\centering
		\includegraphics[width=\linewidth]{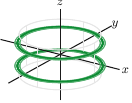}
		\caption{Axial gradiometer.}
		\label{fig:three_channel_pickup_basis_axial}
	\end{subfigure}
	\caption{
		Schematic pickup-basis geometries used to illustrate the three magnetic-field channels considered in this work.
		The cosine-$\theta$ and sine-$\theta$ pickups form an orthogonal transverse basis and are shown as saddle-coil patterns rotated by $90^{\circ}$ with respect to each other.
		The axial gradiometer consists of two longitudinally separated loops with differential readout for common-mode rejection.
		The former is drawn as a transparent guide to show the winding geometry.
		For visual clarity, the transverse saddle-coil panels are shown with the reduced value $n_{\mathrm{turns}}=10$; this value is used only for the schematic visualization.
		The geometric input parameters used for the drawings are summarized in Table~\ref{tab:three_channel_pickup_inputs}.
	}
	\label{fig:three_channel_pickup_basis}
\end{figure}

\begin{table}[tbp]
	\centering
	\caption{
		Pickup-coil parameters for the three-channel SQUID polarimeter. Top block: saddle parameters for the cosine-$\theta$ and sine-$\theta$ transverse channels, following the construction shown schematically in Fig.~\ref{fig:three_channel_pickup_basis}; the schematic in that figure is drawn with reduced turn count $n_\mathrm{turns}=10$ for visual clarity, the operational value is $n_\mathrm{turns}=N_\mathrm{turns}=100$. Bottom block: axial-gradiometer parameters for the longitudinal channel (Sec.~\ref{sec:pz_axial_channel}).
	}
	\label{tab:three_channel_pickup_inputs}
	\renewcommand{\arraystretch}{1.15}
	\sisetup{table-number-alignment=center}
	\begin{tabular}{l l c S[table-format=4.1]}
		\hline\hline
		{Parameter} & {Symbol} & {Unit} & {Value} \\
		\hline
		\multicolumn{3}{l}{\textit{Cosine-$\theta$ and sine-$\theta$ transverse saddle pickup}} & \\
		Saddle former radius                    & $r_\mathrm{coil}$    & \si{\milli\meter}         &  40 \\
		Base half-length of saddle loop         & $L_{\mathrm{base}}$  & \si{\milli\meter}         &   7.5 \\
		Length increment between turns          & $\Delta L$           & \si{\milli\meter}         &   1.0 \\
		End-loop height parameter               & $H_{\mathrm{end}}$   & \si{\milli\meter}         &   4.5 \\
		Maximum saddle angle                    & $\theta_{\max}$      & \si{\degree}              &  75 \\
		Winding multiplicity                    & $N_\mathrm{turns}$   & 1                         & 100 \\
		Total length along $z$                  & {\small$L_z = 2\left[L_{\mathrm{base}} + (N_{\mathrm{turns}}-1)\Delta L + H_{\mathrm{end}}\right]$} & \si{\milli\meter} & 222 \\
		Single-turn effective coupling area     & $A$                  & \si{\centi\meter\squared} &  40 \\
		Longitudinal form factor                & $f_\mathrm{geom}$    & 1                         &   1.5 \\
		\hline
		\multicolumn{3}{l}{\textit{Axial gradiometer (longitudinal $P_z$ pickup)}} & \\
		Axial-loop radius                       & $r_\mathrm{coil}$    & \si{\milli\meter}         &  40 \\
		Gradiometer baseline (centre-to-centre) & $\Delta z$           & \si{\milli\meter}         & 300 \\
		\hline\hline
	\end{tabular}
\end{table}

\section{Design overview: three-channel SQUID polarimeter}
\label{sec:design-overview}

The single-channel transverse-pickup operation developed in Secs.~\ref{sec:fid_spin_tune_phase_lock}--\ref{sec:fid-echo-sequence} for spin-tune determination and the matched-filter precision-mode analysis of Sec.~\ref{sec:coherent-integration} measure one in-plane component of the polarization vector at a time. The full proton polarimetry programme at the HSR, including absolute determination of the stored vertical polarization $P_y$ and on-line monitoring of the longitudinal component $P_z$ at the interaction region, requires resolving all three orthogonal components of the polarization vector $\vec P = (P_x, P_y, P_z)$. This is achieved by a three-channel SQUID pickup that shares a single cylindrical former around the beam pipe and a single \SI{4}{\kelvin} cryogenic envelope.

Figure~\ref{fig:three_channel_pickup_basis} shows the three pickup channels schematically. A cosine-$\theta$ saddle-loop pair, with winding density proportional to $\cos\phi$ around the cylinder azimuth, picks up the radial in-plane magnetic-moment component and is therefore sensitive to $P_x$. A sine-$\theta$ saddle-loop pair, identical in geometry but rotated by $90^\circ$ about the cylinder axis, picks up the vertical component and is therefore sensitive to $P_y$. A coaxial pair of azimuthally symmetric loops separated along the beam direction forms an axial gradiometer sensitive to the longitudinal component $P_z$. The three winding patterns are orthogonal in the sense that each one projects onto a single component of the bunch magnetic moment as the bunch passes through, with negligible cross-talk for an ideally-wound common-cylinder geometry.

The three channels couple to the three orthogonal polarization components in distinct temporal patterns set by the spin dynamics. The cosine-$\theta$ channel sees an oscillating signal at $f_s = \nu_s f_\mathrm{rev} = f_\mathrm{rev}/2$ during free-induction decay, because the $P_x$ component reverses sign every machine turn as the spin precesses about the stable spin axis $\vec n_0 \simeq \vec e_y$; this is the channel implicitly used in Secs.~\ref{sec:fid_spin_tune_phase_lock}--\ref{sec:fid-echo-sequence}. The axial gradiometer sees a similar oscillating signal at $f_s$ for any non-zero $P_z$ component, whether deliberately excited by a tipping pulse applied about $\ex$ or already present in the stored beam as a residual longitudinal polarization. In a non-ideal lattice such residual $P_z$, and equivalently residual $P_x$ on the cosine-$\theta$ channel, arises from imperfect injection matching, snake mismatch, rotator transients, or other lattice imperfections, and produces a
continuous signal at $f_s$ in the absence of any tipping pulse. The two transverse channels therefore serve a dual role: they support deliberate FID and spin-echo measurements after a tipping pulse, and they monitor the residual non-vertical polarization components of the stored beam in steady state. The sine-$\theta$ channel, in contrast, picks up the stored vertical polarization $P_y$ that lies along $\vec n_0$ and is therefore static on the spin-precession time scale, producing a signal at the bunch repetition frequency $f_\mathrm{rev}$ that requires no tipping pulse to read out. All three channels share the same beam pass, the same mechanical support, and the same cryostat, so their relative geometric calibration is fixed by the mechanical reference frame and does not drift between channels.

The remainder of the paper develops the polarimetric reconstruction for each channel class in turn. Section~\ref{sec:static_vector_extension} treats the transverse polarimetry of $P_x$ and $P_y$ via the two saddle-loop channels, including the static bunch magnetic moment, the three-channel single-passage response, the finite-bunch-length convolution, the response matrix, and the cross-talk hierarchy. Section~\ref{sec:pz_axial_channel} treats the longitudinal polarimetry of $P_z$ via the axial gradiometer, with the charge-current rejection geometry that distinguishes it from the saddle-loop pickups, the gradiometer-specific signal-to-noise treatment, and the operational regime in which the longitudinal channel adds information beyond what the transverse channels already provide.
	
\section{Full spin-vector extension and static vector polarimetry}
\label{sec:static_vector_extension}

The design overview of Sec.~\ref{sec:design-overview} introduced the three-channel pickup basis on a common cylindrical former. This section develops the polarimetric reconstruction of the two transverse polarization components $P_x$ and $P_y$ from the saddle-loop channels (cosine-$\theta$ and sine-$\theta$). The longitudinal component $P_z$, picked up by the axial gradiometer, is treated separately in Sec.~\ref{sec:pz_axial_channel} because its winding geometry and charge-current rejection mechanism differ qualitatively from the saddle-loop case. The cross-talk between channels in a realistic apparatus, and the broader systematic-effect program required to convert per-channel sensitivities into absolute polarimetric precision, are framed in Sec.~\ref{sec:static_vector_systematics_subsec} but their quantitative evaluation is deferred to a dedicated engineering paper.

The starting point is the local polarization vector
\begin{equation}
	\vec P = (P_x,P_y,P_z)
\end{equation}
at the pickup location. The FID, spin-echo, and matched-filter modes of Secs.~\ref{sec:fid_spin_tune_phase_lock}--\ref{sec:coherent-integration} are special cases of this framework in which a controlled tipping pulse generates a deliberately oscillating in-plane component and the cosine-$\theta$ channel reads it out. The static-vector framework developed in this section is the more general case: in steady state, each saddle-loop channel reads out a definite component of the bunch magnetic moment $\vec m = N_p\mu_p\vec P$, with the time structure imposed by the bunch passage rather than by any deliberate spin manipulation.

At a pickup location in an ordinary HSR arc or straight section where the local stable spin axis is close to vertical, $\vec n_0\simeq \ey$, a stored vertical polarization component $P_y$ produces a magnetic moment component $m_y$ that is constant on the spin-precession time scale and pulses at $f_\mathrm{rev}$ as each bunch passes the pickup. Residual in-plane and longitudinal components $P_x$ and $P_z$, arising from imperfect injection matching, snake mismatch, rotator transients, or other lattice imperfections, produce moment components that oscillate at $f_s = f_\mathrm{rev}/2$ in the spin-precession frame and therefore appear as alternating bunch-to-bunch signals at the pickup. The three symmetry classes of the pickup basis separate these contributions cleanly.

\subsection{Static bunch magnetic moment}

For bunch $b$, the total magnetic moment is written as
\begin{equation}
	\vec m_b = N_{p,b}\mu_p\,\vec P_b,
	\label{eq:static_vector_moment}
\end{equation}
where $N_{p,b}$ is the bunch population, $\mu_p$ is the proton magnetic moment, and $\vec P_b=(P_{x,b},P_{y,b},P_{z,b})$ is the local bunch polarization vector. Equation~\eqref{eq:static_vector_moment} is the common starting point for both the FID-based transverse signal and the static vector measurement. The distinction is not the magnetic moment itself, but the time dependence imposed on it: in the FID mode the transverse component is deliberately generated and allowed to precess, whereas in the static-vector mode the pickup basis is used to reconstruct the instantaneous local vector components from the known bunch-passage pattern.

\subsection{Natural pickup basis on a cylindrical former}

On a cylindrical pickup surface surrounding the beam, the three dipole components separate into different angular symmetry classes. For a point dipole on the beam axis with moment $\vec m=(m_x,m_y,m_z)$, evaluated on a cylindrical pickup surface of radius $r_{\mathrm{coil}}$, the radial field components are
\begin{align}
	B_r^{(x)}(\phi,z)
	&=
	\frac{\mu_0 m_x}{4\pi}
	\frac{2r_{\mathrm{coil}}^2-z^2}
	{\left(r_{\mathrm{coil}}^2+z^2\right)^{5/2}}
	\cos\phi ,
	\label{eq:Br_x_static}
	\\
	B_r^{(y)}(\phi,z)
	&=
	\frac{\mu_0 m_y}{4\pi}
	\frac{2r_{\mathrm{coil}}^2-z^2}
	{\left(r_{\mathrm{coil}}^2+z^2\right)^{5/2}}
	\sin\phi ,
	\label{eq:Br_y_static}
	\\
	B_r^{(z)}(\phi,z)
	&=
	\frac{\mu_0 m_z}{4\pi}
	\frac{3r_{\mathrm{coil}}z}
	{\left(r_{\mathrm{coil}}^2+z^2\right)^{5/2}} .
	\label{eq:Br_z_static}
\end{align}
The transverse components are first-harmonic angular modes on the pickup cylinder, while the longitudinal component is independent of $\phi$. The natural correspondence is therefore
\begin{equation}
	m_x \leftrightarrow \cos\phi,
	\qquad
	m_y \leftrightarrow \sin\phi,
	\qquad
	m_z \leftrightarrow \text{azimuthally symmetric axial response}.
	\label{eq:static_vector_basis}
\end{equation}

\begin{figure*}[tb]
	\centering
	\begin{subfigure}[t]{0.315\linewidth}
		\includegraphics[width=\linewidth]{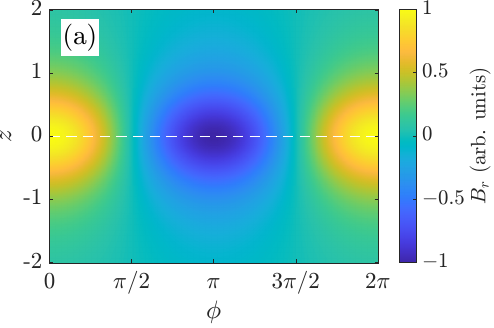}
  		 \caption{$\vec m = m_x\ex$, giving the $\cos\phi$ symmetry of Eq.~(\ref{eq:Br_x_static}).}
		\label{fig:unwrap-mx}
	\end{subfigure}
	\hspace{0.1cm}
	\begin{subfigure}[t]{0.315\linewidth}
		\includegraphics[width=\linewidth]{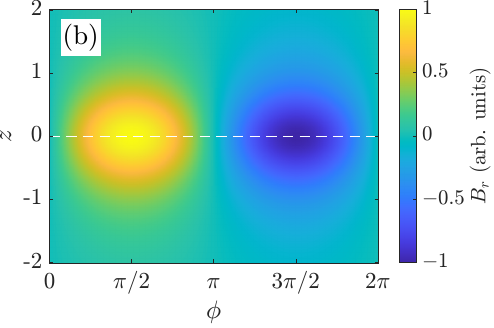}
    	\caption{$\vec m = m_y\ey$, giving the $\sin\phi$ symmetry of Eq.~(\ref{eq:Br_y_static}).}

		\label{fig:unwrap-my}
	\end{subfigure}
	\hspace{0.1cm}
	\begin{subfigure}[t]{0.315\linewidth}
		\includegraphics[width=\linewidth]{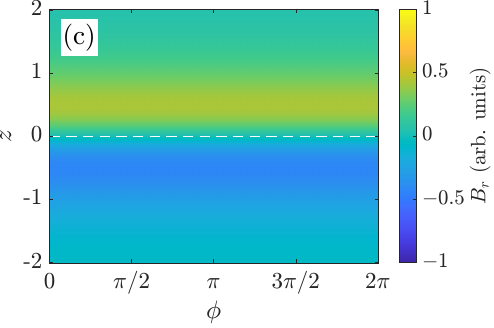}
		\caption{$m_z$: $B_r(\phi,z)$ independent of $\phi$.}
		\label{fig:unwrap-mz}
	\end{subfigure}
 	 \caption{Radial field component $B_r(\phi,z)$ on the cylindrical pickup surface, shown in unwrapped-cylinder coordinates for the three static dipole orientations. These three symmetry classes form the natural field basis for the cosine-$\theta$, sine-$\theta$, and axial pickup channels.}

	\label{fig:unwrapped_cylinder}
\end{figure*}

The three pickup channels were introduced in Sec.~\ref{sec:design-overview}; here we verify their mathematical orthogonality. The ideal cosine-$\theta$ and sine-$\theta$ windings are orthogonal by the angular integrals
\begin{equation}
	\int_0^{2\pi}\cos\phi\,\sin\phi\,d\phi = 0,
	\qquad
	\int_0^{2\pi}\cos\phi\,d\phi =
	\int_0^{2\pi}\sin\phi\,d\phi = 0.
	\label{eq:static_vector_orthogonality}
\end{equation}
Thus, in the ideal continuous-winding limit, the transverse channels are mutually orthogonal and are also blind to azimuthally symmetric fields. The axial channel belongs to a different symmetry class and is correspondingly orthogonal to the transverse first-harmonic channels in the ideal geometry. The corresponding symmetry patterns on the unwrapped cylindrical surface are shown in Fig.~\ref{fig:unwrapped_cylinder}.

\subsection{Three-channel single-passage response}

The ideal single-passage responses can be written in the form
\begin{align}
	S_x^{\mathrm{bunch}}(z_b) &= K_x\,m_x\,g_T(z_b;\sigma_L),\label{eq:static_vector_response_x}\\
	S_y^{\mathrm{bunch}}(z_b) &= K_y\,m_y\,g_T(z_b;\sigma_L),\label{eq:static_vector_response_y}\\
	S_z^{\mathrm{bunch}}(z_b) &= K_z\,m_z\,g_z(z_b;\sigma_L),\label{eq:static_vector_response_z}
\end{align}
where $z_b$ is the bunch-centre coordinate relative to the pickup, $K_x$, $K_y$, and $K_z$ are geometry and flux-transformer constants, and $g_T$ and $g_z$ are the finite-bunch-length waveform templates for the transverse and axial channels. For a mechanically symmetric implementation one expects $K_x\simeq K_y$; deviations from equality provide a direct calibration observable.

The transverse template $g_T$ is common to the cosine-$\theta$ and sine-$\theta$ channels because they differ only by a $90^\circ$ azimuthal rotation of the winding pattern. The axial template $g_z$ differs qualitatively because the first-order axial gradiometer produces a bipolar waveform as the bunch passes the two oppositely wound axial loops.

\subsection{Finite bunch-length convolution}

The bunch is modeled by a Gaussian longitudinal profile of rms length $\sigma_L$,
\begin{equation}
	\lambda(z')=\frac{1}{\sqrt{2\pi}\sigma_L}
	\exp\left(-\frac{z'^2}{2\sigma_L^2}\right).
	\label{eq:static_vector_gaussian_bunch}
\end{equation}
For each pickup channel, the point-dipole response is convolved with this longitudinal distribution. The transverse template can be written schematically as
\begin{equation}
	g_T(z_b;\sigma_L)=
	\int_{-\infty}^{+\infty}
	\lambda(z')\,h_T(z_b-z')\,dz',
	\label{eq:static_vector_transverse_template}
\end{equation}
where $h_T$ is the point-bunch transverse response of the cosine-$\theta$ or sine-$\theta$ winding. The axial template is similarly
\begin{equation}
	g_z(z_b;\sigma_L)=
	\int_{-\infty}^{+\infty}
	\lambda(z')\,h_z(z_b-z')\,dz',
	\label{eq:static_vector_axial_template}
\end{equation}
where $h_z$ is the point-bunch response of the first-order axial gradiometer. The explicit axial-loop expression is developed in Sec.~\ref{sec:pz_axial_channel}; in the present section we treat it abstractly as the $P_z$ component of the vector-pickup basis.

At injection, $\sigma_L$ is long compared with a pickup radius of order $4\,\mathrm{cm}$, and finite-bunch-length effects strongly modify the peak amplitude and waveform shape. At flattop the bunch is shorter, and the axial gradiometer becomes more favorable. This is why the axial-channel benchmark cannot be inferred from the point-dipole peak field alone.

\begin{figure*}[tb]
	\centering
	\begin{subfigure}[t]{0.48\linewidth}
		\includegraphics[width=\linewidth]{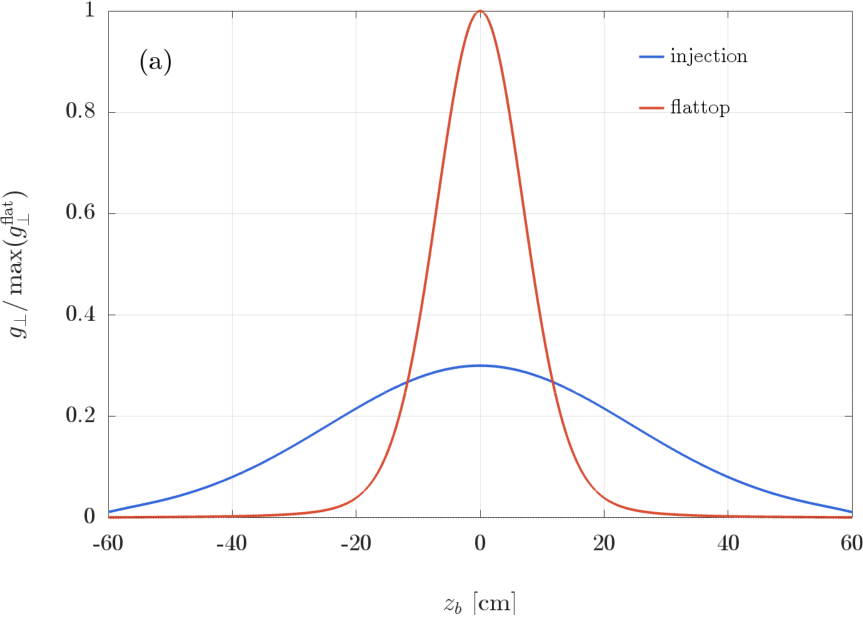}
		\caption{Transverse pickup template $g_\perp(z_b)$.}
		\label{fig:svw-perp}
	\end{subfigure}\hfill
	\begin{subfigure}[t]{0.48\linewidth}
		\includegraphics[width=\linewidth]{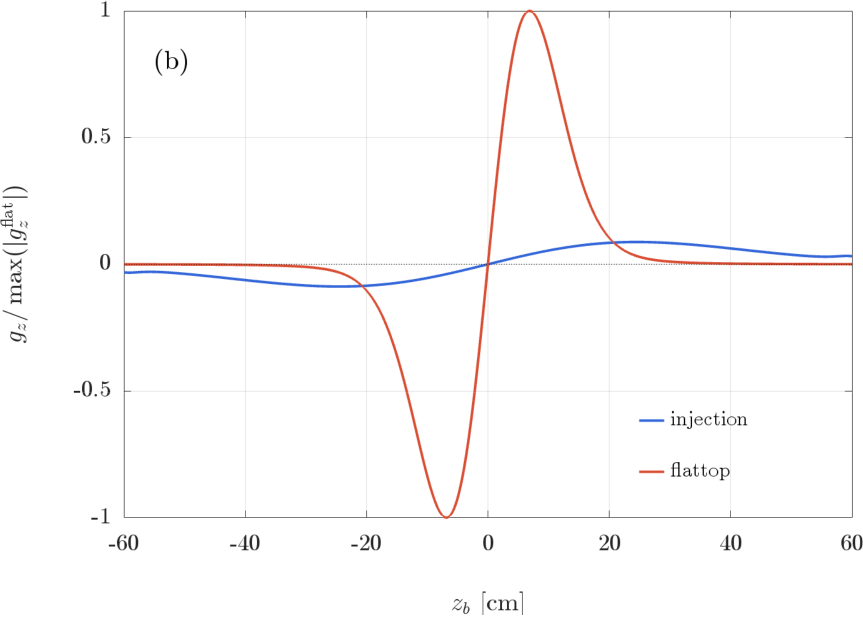}
		\caption{Axial-gradiometer template $g_z(z_b)$.}
		\label{fig:svw-axial}
	\end{subfigure}
		\caption{Single-passage waveform classes after finite bunch-length convolution: (a) unipolar transverse template $g_T$ from a cosine-$\theta$ or sine-$\theta$ saddle loop, and (b) bipolar axial template $g_z$ from the axial gradiometer. The two curves in each panel correspond to the injection and flattop rms bunch lengths $\sigma_L$ listed in Tab.~\ref{tab:params} and the pickup radius $\rcoil$ in Tab.~\ref{tab:three_channel_pickup_inputs}. Both panels are normalized to the flattop peak, so the injection curves directly show the amplitude reduction from bunch washout when $\sigma_L \gtrsim \rcoil$.}
	\label{fig:static_vector_waveforms}
\end{figure*}

\subsection{Beam-synchronous static signal formation}

The measured three-channel signal can be written as
\begin{equation}
	\vec{\Phi}^\mathrm{m}(t)=\sum_{n,b}\mathbf H(t-t_{n,b})\,c_b\,\vec m_b + \vec B^{\mathrm{bg}}(t),
	\label{eq:static_vector_pulse_train}
\end{equation}
where $t_{n,b}$ are the known bunch-passage times, $c_b$ is the known bunch-pattern or spin-sign factor, $\mathbf H$ is the three-channel response kernel, and $\vec B^{\mathrm{bg}}(t)$ denotes spin-independent and spin-correlated backgrounds. In the ideal basis the kernel is diagonal in the channel index, with the transverse channels sharing the same waveform template and the axial channel using the gradiometer template.

For a known bunch pattern, the same phase-corrected coherent summation used in the FID analysis can be generalized to three simultaneous matched filters. The estimated vector amplitude is obtained by applying the channel-specific template and the bunch-pattern phase factor to each channel separately,
\begin{equation}
	\hat m_i \propto
	\sum_{n,b}\int dt\,
	\Phi^\mathrm{m}_i(t)\,q_i(t-t_{n,b})\,c_b,
	\qquad i=x,y,z,
	\label{eq:static_vector_matched_filter}
\end{equation}
where $q_i$ is the matched-filter template for channel $i$. Calibration then converts $\hat m_i$ into $\hat P_i$ using Eq.~\eqref{eq:static_vector_moment}. This formulation also allows separate accumulation of bunch subsets, for example up/down spin classes or individual tagged bunches.

The three channels need not be operated in only one mode. They may be read out simultaneously if cross-talk is sufficiently small, or interleaved in time to reduce live coupling and facilitate systematic checks. Interleaved operation would still provide quasi-continuous vector reconstruction because the bunch-passage pattern and machine RF clock are known.

\subsection{Background rejection and cross-talk hierarchy}

The three channels have different dominant background and rejection mechanisms. For the axial gradiometer, the ideal charge-current rejection is geometric: the beam current generates an azimuthal magnetic field, while the axial loop measures axial flux. A single coaxial axial loop is therefore intrinsically blind to the ideal beam-current field, and the first-order gradiometer further rejects uniform environmental $B_z$ backgrounds.

For the transverse cosine-$\theta$ and sine-$\theta$ channels, the azimuthal symmetry of the beam-current field is rejected by the first-harmonic winding pattern in the ideal continuous limit. The rejection argument is sensitive to the azimuthal symmetry of the beam current itself. The EIC HSR beam, however, is deliberately flat by design: peak-luminosity operation requires a transverse emittance ratio of order $\varepsilon_x/\varepsilon_y \simeq 10$, produced by electron cooling at the IR2 cooling section at injection and maintained against intrabeam scattering by hadron cooling at top energy~\cite{eic_beam_dynamics}. Combined with the matched-optics $\beta$-function asymmetry at the pickup location, this yields rms beam sizes that are anisotropic by a factor of several (for example $\sigma_x \simeq \SI{3.5}{\milli\meter}$ versus $\sigma_y \simeq \SI{0.7}{\milli\meter}$ at injection in the matched optics at the HJET in IR4~\cite{Rathmann2026PRAB}). The wall image current of such a flat bunch therefore carries substantial higher-multipole ($m \geq 2$) azimuthal structure, while the $m = 1$ dipole component to which the saddle pickups respond is suppressed by parity for a centered beam. Residual coupling into the $m = 1$ channel comes from orbit-offset, dispersion, and small mismatches that break this parity, and its magnitude at the pickup radius is bounded by Ref.~\cite{Rathmann2026PRAB}, whose tabulated radial field profiles for symmetric and asymmetric beam distributions are visibly distinct near the beam axis but converge for radii of order the pickup radius $r_\mathrm{coil} = \SI{40}{\milli\meter}$. In addition, finite winding accuracy, mechanical asymmetry, image currents, feedthroughs, nearby conductive structures, and flux-transformer routing produce residual common-mode coupling. Together, these effects are the dominant engineering risk for the transverse static channels and set the required common-mode rejection and trim strategy.

The leading cross-talk classes are:
\begin{itemize}[leftmargin=2em]
	\item $x$--$y$ mixing from imperfect cosine/sine winding orthogonality, azimuthal placement errors, and unequal layer geometry;
	\item transverse--axial mixing from loop tilt, broken cylindrical symmetry, or common flux-transformer routing;
	\item spin-independent backgrounds from charge-current leakage, external fields, vibration in residual field gradients, and trapped flux;
	\item spin-correlated backgrounds from machine elements whose fields or beam conditions change with spin pattern.
\end{itemize}
These effects define the response-matrix problem below.

\subsection{Systematic effects and cross-talk}
\label{sec:static_vector_systematics_subsec}

In the ideal apparatus, the three pickup channels are mutually orthogonal by the angular-symmetry argument of Eqs.~\eqref{eq:static_vector_orthogonality}, and the relation between the reconstructed channel amplitudes and the local polarization vector is diagonal,
\begin{equation}
	\begin{pmatrix}
		\hat m_x\\
		\hat m_y\\
		\hat m_z
	\end{pmatrix}
	=
	\begin{pmatrix}
		R_{xx} & 0 & 0\\
		0 & R_{yy} & 0\\
		0 & 0 & R_{zz}
	\end{pmatrix}
	\begin{pmatrix}
		P_x\\
		P_y\\ 
		P_z
	\end{pmatrix}
	+
	\begin{pmatrix}
		B_x^{\mathrm{bg}}\\
		B_y^{\mathrm{bg}}\\
		B_z^{\mathrm{bg}}
	\end{pmatrix}.
	\label{eq:static_vector_response_matrix_ideal}
\end{equation}
In a realistic apparatus, geometric imperfections (winding-density errors, layer offsets, coil tilt, broken cylindrical symmetry), residual conductive structures (beam-pipe coatings, image currents, feedthroughs), and shared flux-transformer routing introduce off-diagonal cross-channel coupling, and the response matrix contains off-diagonal terms,
\begin{equation}
	\hat m_i = \sum_j R_{ij}P_j + B_i^{\mathrm{bg}},
	\qquad i,j\in\{x,y,z\},
	\label{eq:static_vector_response_matrix_general}
\end{equation}
where the off-diagonal elements quantify geometric cross-talk, alignment errors, and deviations from the ideal basis. The calibration program must therefore determine the diagonal gains $R_{ii}$ and bound or correct the off-diagonal terms $R_{ij}$.

A quantitative evaluation of the off-diagonal response matrix requires detailed mechanical, electromagnetic, and prototyping inputs that are beyond the scope of the present concept paper. We defer this evaluation, together with the related systematics analysis (mechanical alignment tolerances, common-mode rejection floor of the cos-$\theta$ and sin-$\theta$ windings, residual environmental backgrounds, and spin-correlated machine couplings), to a dedicated engineering paper. Equations~\eqref{eq:static_vector_response_matrix_ideal} and~\eqref{eq:static_vector_response_matrix_general} establish the framework within which the systematic-effect categories of Tab.~\ref{tab:static_vector_systematics} will be quantified.

\begin{table}[tb]
	\centering
	\caption{Principal classes of systematic effects for static vector SQUID polarimetry. Quantitative evaluation of each category is deferred to a dedicated engineering paper; the present table serves as a forward pointer for the categories that must be addressed.}
	\label{tab:static_vector_systematics}
	\begin{tabular}{p{3.5cm}p{5cm}p{6cm}}
		\hline\hline
		Category & Origin & Suppression or calibration strategy \\
		\hline
		Geometric cross-talk & Winding errors, layer offsets, loop tilt & Mechanical survey, bench calibration, response matrix \\
		Charge-current leakage & Broken symmetry, image currents, feedthroughs & Gradiometry, trim coils, reference pickup, shielding \\
		Environmental background & External fields, vibration, trapped flux & SC shield, quiet location, cooldown procedure \\
		Spin-correlated background & Machine or bunch-pattern correlations & Spin-pattern reversal, bunch-subset tests, HJET/pC comparison \\
		Gain and phase errors & SQUID/FLL/electronics response & Injected calibration signals, RF-clock phase reference \\		\hline\hline
	\end{tabular}
\end{table}

\section{Longitudinal component and axial gradiometer channel}
\label{sec:pz_axial_channel}

\subsection{Physical picture}
\label{pz:sec:Pz_picture}

The local polarization vector $\vec P = (P_x, P_y, P_z)$ introduced in Sec.~\ref{sec:static_vector_extension} produces a bunch magnetic moment $\vec m = N_p\mu_p\vec P$ whose three orthogonal components are read out by the three pickup channels of Fig.~\ref{fig:three_channel_pickup_basis}. The cosine-$\theta$ and sine-$\theta$ saddle channels read out the transverse components $m_x$ and $m_y$ via Eqs.~\eqref{eq:static_vector_response_x}--\eqref{eq:static_vector_response_y}; the axial gradiometer reads out the longitudinal component $m_z$ via Eq.~\eqref{eq:static_vector_response_z}, and is the subject of this section.

At an arc pickup location with $\vec n_0 \simeq \ey$, residual or deliberately excited in-plane and longitudinal components $m_x$ and $m_z$ precess about $\vec n_0$ at $\nu_s = 1/2$, so both the cosine-$\theta$ and the axial channels see oscillating signals at the spin-precession frequency $f_s = f_\mathrm{rev}/2 \simeq \SI{39.1}{\kilo\hertz}$. The cosine-$\theta$ saddle is by construction insensitive to azimuthally symmetric fields and does not couple to $m_z$; the coaxial-loop gradiometer geometry is required to resolve the longitudinal component. Two channel-specific properties developed in this section set the axial channel apart from the saddle channels: a geometric charge-current rejection mechanism that requires no winding-balance tolerancing (Sec.~\ref{pz:sec:Pz_charge}), and a flux-signal kernel $g_z$ whose finite-bunch-length suppression differs qualitatively from the saddle template $g_T$ (Sec.~\ref{pz:sec:Pz_flux}).

\subsection{Charge current rejection}
\label{pz:sec:Pz_charge}
	
The beam charge current $I_\mathrm{beam}(t)$ flowing along $\ez$ produces only an azimuthal magnetic field,
	\begin{equation}
		\vec{B}_\mathrm{charge}(\rho,t) =
		\frac{\mu_0\,I_\mathrm{beam}(t)}{2\pi\rho}\,\ephi,
		\label{pz:eq:Bphi}
	\end{equation}
where $\rho$ is the transverse distance from the beam axis. This holds for every Fourier harmonic of the bunch train: whether the current is a DC component, an RF harmonic at $n\,f_\mathrm{rev}$, or the transient field of a single bunch passage, the field is purely azimuthal by the cylindrical symmetry of the problem.

A circular pickup loop coaxial with the beam, with its area normal along $\ez$, is threaded only by the axial field component $B_z$. Since $\vec{B}_\mathrm{charge}\cdot\ez = 0$ at every point on the loop, the charge-current flux is
	\begin{equation}
		\Phi_\mathrm{charge} =
		\iint_\mathrm{loop} \vec{B}_\mathrm{charge}\cdot d\vec{A} = 0
		\quad\text{exactly.}
		\label{pz:eq:charge_zero}
	\end{equation}
This rejection is purely geometric: it requires no special winding balance, holds independently of the beam current magnitude, energy, or transverse beam profile, and applies at all frequencies. A plain circular loop is intrinsically charge-blind.
	
This is in marked contrast to the cosine-$\theta$ coil, where the charge current couples into the winding through any imbalance $\epsilon$ in the winding density, and a common-mode rejection ratio $\mathrm{CMRR} = 1/\epsilon \gtrsim 10^{12}$ is required.
	
\subsection{Flux signal from the longitudinal magnetic moment}
\label{pz:sec:Pz_flux}

In the following, we consider the time-dependent longitudinal magnetic moment $m_z(t)$ of a bunch arising from spin precession. A “bunch” denotes the ensemble of particles in a given RF bucket. The detector does not resolve individual particles, but measures the total magnetic moment, i.e., the sum over all particle spins. The observed signal is therefore proportional to the oscillating component  $m_z(t)$ and reflects the bunch-averaged polarization.

\subsubsection{Point-dipole result}

After the small tipping pulse, the total magnetic moment of a bunch remains dominated by the main spin component, while only a small oscillating projection $m_z(t)$ appears along the detector axis $\ez$. It is this oscillating longitudinal component, not the full bunch magnetic moment, that generates the signal considered here.

Treating this oscillating component as a point dipole located at the bunch centre $z_b$ on the beam axis, the instantaneous flux threading a coaxial circular loop of radius $\rcoil$ centred at $z=z_\mathrm{loop}$ is
\begin{equation}
	\Phi_\mathrm{loop}(z_b,t) =
	\frac{\mu_0}{2}\,m_z(t)\,
	\frac{\rcoil^2}{\bigl[\rcoil^2 + (z_b-z_\mathrm{loop})^2\bigr]^{3/2}} \,.
	\label{pz:eq:Phi_axial_point}
\end{equation}

For a fixed value of $m_z(t)$ this is a unipolar pulse in the passage coordinate: as the bunch centre approaches the loop the flux rises, reaches its maximum when the bunch centre is in the plane of the loop ($z_b=z_\mathrm{loop}$),
\begin{equation}
	\Phi^\mathrm{pt}_\mathrm{peak}(t) =
	\frac{\mu_0}{2}\,\frac{m_z(t)}{\rcoil} \,,
	\label{pz:eq:Phi_axial_peak_pt}
\end{equation}
and falls symmetrically as the bunch recedes. The sign of $\Phi_\mathrm{loop}$ therefore does not change within a single passage. Since $\nu_s = 1/2$, the spin completes exactly half a precession cycle per revolution, so $m_z(t)$ changes sign on every successive bunch passage through the loop. This turn-by-turn sign alternation is the signature of the $\nu_s = 1/2$ spin tune and is what the matched filter locks onto at the sideband frequency $\nu_s\,f_\mathrm{rev} = \SI{39.1}{kHz}$.

\subsubsection{Finite bunch length}

At injection the rms bunch length $\sigma_L = \SI{24}{cm}$ exceeds $\rcoil = \SI{4}{cm}$ by a factor of six, so the point-dipole approximation significantly overestimates the peak flux. For a Gaussian bunch of rms length $\sigma_L$ with its centre at $z_b$, the total magnetic moment is distributed along $z$, and the instantaneous loop flux is obtained by convolution,
\begin{equation}
	\Phi_z(z_b,t) =
	\frac{\mu_0\,m_z(t)}{2\sqrt{2\pi}\,\sigma_L}
	\int_{-\infty}^{+\infty}
	\frac{\rcoil^2\;\exp\!\left(-z'^{\,2}/2\sigma_L^2\right)}
	{\left[\rcoil^2 + (z_b - z')^2\right]^{3/2}}\;dz'.
	\label{pz:eq:Phi_axial_gauss}
\end{equation}

In the limit $\sigma_L \gg \rcoil$ the convolution gives $\Phi_z^\mathrm{peak}(t) \approx \mu_0 m_z(t)/\sigma_L$: the same total magnetic moment is spread over a longer bunch, and the peak flux therefore falls as $1/\sigma_L$, suppressing the point-dipole result by a factor of $\sim\!6$ at injection. At flattop ($\sigma_L = \SI{6}{cm}$, comparable to $\rcoil$) the suppression is only a factor of $\sim\!2$. Equation~(\ref{pz:eq:Phi_axial_gauss}) is evaluated numerically using the beam parameters of Tab.~\ref{tab:params} and the axial-gradiometer geometry of Tab.~\ref{tab:three_channel_pickup_inputs}; the resulting peak fluxes at the SQUID input are given in Eq.~\eqref{eq:flux_pickup_axial_numeric}.
	
\subsection{Gradiometer configuration}
	\label{pz:sec:Pz_gradiometer}

A single axial loop is immune to the beam charge current but would pick up any uniform external $B_z$ (residual Earth's field, stray fields from ring magnets, low-frequency environmental noise). A first-order axial gradiometer -- two identical coaxial loops wound in series opposition and separated by $\Delta z$ -- cancels any field component uniform over $\Delta z$, while retaining the bunch signal because the non-uniform dipole field differs between the two loop positions.

With loop~1 at $z = +\Delta z/2$ and loop~2 at $z = -\Delta z/2$, the gradiometer output is
	\begin{equation}
		\Phi_\mathrm{grad}(z_b) =
		\Phi_z\!\left(z_b;\,z_\mathrm{loop} = +\tfrac{\Delta z}{2}\right) -
		\Phi_z\!\left(z_b;\,z_\mathrm{loop} = -\tfrac{\Delta z}{2}\right),
		\label{pz:eq:Phi_grad}
	\end{equation}
where each term is given by Eq.~(\ref{pz:eq:Phi_axial_gauss}). As the bunch sweeps through the gradiometer, $\Phi_\mathrm{grad}$ traces a bipolar waveform: it peaks positively near loop~1, passes through zero at the midpoint $z_b = 0$, and peaks negatively near loop~2. A uniform ambient $B_z$ gives zero gradiometer output.
	
In the point-dipole limit the peak of $|\Phi_\mathrm{grad}|$ is maximised at $\Delta z/2 = \rcoil/\sqrt{2} \approx \SI{2.8}{cm}$. We adopt $\Delta z = \SI{30}{cm}$, constrained by the centre-to-centre separation of coils~A and~B in the cryostat. Despite being far from the point-dipole optimum, numerical evaluation of Eq.~(\ref{pz:eq:Phi_grad}) with the full Gaussian bunch profile shows that the gradiometer retains $\sim\!65\%$ of the single-loop peak signal: the long injection bunch ($\sigma_L = \SI{24}{cm}$) already spans both loop positions, so the penalty for large $\Delta z$ is modest. The resulting post-pickup peak fluxes are quoted in Eq.~\eqref{eq:flux_pickup_axial_numeric}.
	
	

The per-bunch peak flux delivered to the SQUID input by the axial
gradiometer evaluates to
\begin{equation}
	\Phi_\mathrm{pickup}^{\mathrm{axial,\,inj}}
	\approx
	\SI{174.9}{\micro\Phi_0},
	\qquad
	\Phi_\mathrm{pickup}^{\mathrm{axial,\,flat}}
	\approx
	\SI{221.3}{\micro\Phi_0},
	\label{eq:flux_pickup_axial_numeric}
\end{equation}
to be compared with the cosine-$\theta$ values
$\SI{1236}{\micro\Phi_0}$ (injection) and $\SI{309}{\micro\Phi_0}$
(flattop) of Eq.~\eqref{eq:flux_pickup_numeric_full}. The flux ratio
$\Phi_\mathrm{pickup}^\mathrm{axial}/\Phi_\mathrm{pickup}^\mathrm{cos\theta}$ is
$0.142$ at injection and $0.716$ at flattop, consistent with the
sensitivity-ratio discussion of Sec.~\ref{sec:bunch-resolved-vec-polarimetry}.
	
\subsection{Signal-to-noise ratio}
	\label{pz:sec:Pz_SNR}

\subsubsection{Signal accumulation: how SNR is built up}
	
Each time a bunch passes through the gradiometer it produces a flux signal of peak amplitude $\Phi_z^\mathrm{peak}$ [Eq.~\eqref{eq:flux_pickup_axial_numeric}]. This per-bunch signal is well below the instantaneous noise floor of the SQUID: for the canonical white flux noise spectral density $S_\Phi^{1/2} = \SI{0.4}{\micro\Phi_0\per\sqrt{Hz}}$, consistent with the performance of modern wideband LTS DC SQUID systems~\cite{fagaly} at the \SI{39}{kHz} signal frequency where the noise is white and the $1/f$ corner of LTS devices lies well below, and a pulse bandwidth $\sim 1/\sigma_t \sim \SI{1}{GHz}$, the single-pulse noise is $\sim 1.3\times10^4\,\mu\Phi_0$, exceeding the per-bunch signal by several orders of magnitude. The signal is recovered by coherent summation: each bunch produces a pulse at a known time, and the pulses from successive revolutions carry an amplitude modulated at the spin-tune sideband $\nu_s\,f_\mathrm{rev} = \SI{39.1}{kHz}$. A matched filter locked to this sideband sums the contributions coherently,
so the signal grows linearly with the number of bunch passages while the noise grows only as the square root. After $N_\mathrm{pass}$ coherently summed bunch passages,
\begin{equation}
	\mathrm{SNR} =
	\frac{\Phi_z^\mathrm{peak}\,\sqrt{N_\mathrm{pass}}}
	{S_\Phi^{1/2}},
	\label{pz:eq:SNR_general}
\end{equation}
since $N_\mathrm{pass} \propto T$, the SNR grows as $\sqrt{T}$. The two strategies below differ only in how many bunches are summed per revolution.
	
\subsubsection{Strategy~1: all-bunch coherent summation}
	\label{pz:sec:strat1}
	
All $N_b$ bunches in the fill are summed coherently on each revolution. The spin orientation of each bunch is known from the injection pattern and the spin tune $\nu_s = 1/2$: consecutive bunches have alternating spin signs, and the matched filter applies the correct phase weight to each. The number of bunch passages accumulated in time $T$ is $N_\mathrm{pass} = N_b\,f_\mathrm{rev}\,T$, giving
	\begin{equation}
		\mathrm{SNR}_z^{(1)}(T) =
		\frac{\Phi_z^\mathrm{peak}\,\sqrt{N_b\,f_\mathrm{rev}\,T}}
		{S_\Phi^{1/2}}.
		\label{pz:eq:SNR_z_strat1}
	\end{equation}
	
\subsubsection{Strategy~2: single-bunch tracking}
\label{pz:sec:strat2}

A single tagged bunch is followed turn by turn, ignoring all other bunches. Here $N_\mathrm{pass} = f_\mathrm{rev}\,T$ (one bunch passage per revolution), and the SNR is lower than strategy~1 by $\sqrt{N_b}$:
\begin{equation}
	\mathrm{SNR}_z^{(2)}(T) =
	\frac{\Phi_z^\mathrm{peak}\,\sqrt{f_\mathrm{rev}\,T}}
	{S_\Phi^{1/2}}
	= \frac{\mathrm{SNR}_z^{(1)}(T)}{\sqrt{N_b}}.
	\label{pz:eq:SNR_z_strat2}
\end{equation}
At injection $\sqrt{N_b} = \sqrt{290} \approx 17$. The same single-bunch matched-filter formulation extends to every bunch in the fill via parallel matched-filter channels on the digitized SQUID stream, and is not specific to the axial channel: the cos-$\theta$ and sin-$\theta$ channels admit the same per-bunch decomposition. The full bunch-resolved vector polarimetry capability (joint three-channel reconstruction of $\vec P(t)$ for each bunch, per-bunch precision as a function of averaging window, and the operational comparison with existing polarimetry) is developed as the joint deliverable of Sec.~\ref{sec:bunch-resolved-vec-polarimetry}.

\subsubsection{Coherence-time interaction and multi-cycle accumulation}
\label{sec:pz-decoherence-time}

The axial-gradiometer channel is subject to the same reversible coherence-time limit as the transverse FID and matched-filter measurements of Secs.~\ref{sec:fid_spin_tune_phase_lock}--\ref{sec:coherent-integration}. Tipped or residual perpendicular spin components dephase with the rms spin-tune spread $\sigma_{\nu_s}$ on the timescale $\tau_\mathrm{coh} = 1/(2\pi f_\mathrm{rev}\,\sigma_{\nu_s})$ defined in Eq.~\eqref{eq:t2_star_definition}. The canonical operating-point value, $\sigma_{\nu_s}^\mathrm{eff} = 10^{-3} \to \tau_\mathrm{coh}^\mathrm{eff} \simeq \SI{2.04}{\milli\second}$ (Tab.~\ref{tab:pi_pulse_field_requirements}), is used throughout the present analysis; the DLC and LC test scales of Tab.~\ref{nzero:tab:matlab_params} give correspondingly shorter coherence times.

The axial channel is no different in this respect from the transverse matched-filter channel: both extract signals whose envelopes decay on $\tau_\mathrm{coh}$. Any matched-filter integration time $T > \tau_\mathrm{coh}$ must therefore proceed as a sequence of $N \simeq T/\tau_\mathrm{coh}$ tip-$\pi$-echo-restore cycles, each within one coherence window, using the spin-echo extension of Sec.~\ref{sec:fid-echo-sequence}. The per-cycle integration times tabulated below are noise-limited single-cycle values; their realization at $T > \tau_\mathrm{coh}$ incurs the bounded per-cycle depolarization budget of Sec.~\ref{sec:fid-echo-sequence}.
	
\subsubsection{Numerical performance and measurement strategies}
\label{sec:pz-numerical-performance}

The sensitivity of the axial $P_z$ channel can be compared directly with
the transverse cosine-$\theta$ reference. For fixed $N_b$, $f_\mathrm{rev}$,
and integration time $T$, the SNR is proportional to the peak flux. The
ratio of the axial and transverse sensitivities is therefore simply
\begin{equation}
	\frac{\mathrm{SNR}_z}{\mathrm{SNR}_x}
	=
	\frac{\Phi_z^\mathrm{grad}}{\Phi_x}.
	\label{pz:eq:snr_ratio}
\end{equation}
At injection this ratio is $0.142$, while at flattop it is $0.716$. The
axial gradiometer is less sensitive than the cosine-$\theta$ coil at
both energies, but the gap is much smaller at flattop than at injection.
The reduction reflects the strong bunch-length dependence of the axial
signal: in the long-bunch limit it scales approximately as
$1/\sigma_L$, whereas the cosine-$\theta$ signal is much less sensitive
to $\sigma_L$. The shorter flattop bunch,
$\sigma_L=\SI{6}{cm}\simeq r_\mathrm{coil}$, is therefore strongly
favourable for the axial gradiometer.

The all-bunch matched-filter integration times for the axial channel are
therefore obtained directly by rescaling the cosine-$\theta$ values of
Sec.~\ref{sec:coherent-integration} by the inverse of these ratios. The
resulting per-cycle times exceed the reversible coherence time
$\tau_\mathrm{coh}$ at every relevant $\sigma_{\nu_s}$ scale, so the axial
measurement must be accumulated as a sequence of tip-$\pi$-echo-restore
cycles using the spin-echo extension of Sec.~\ref{sec:fid-echo-sequence},
matching the multi-cycle strategy already established for the transverse
matched-filter channel. Single-bunch tracking is one operational mode of a
per-bunch matched-filter framework that applies equally to the cos-$\theta$
and sin-$\theta$ channels (Sec.~\ref{pz:sec:strat2}); the bunch-resolved
vector polarimetry deliverable and its comparison with existing
storage-ring polarimetry are developed in
Sec.~\ref{sec:bunch-resolved-vec-polarimetry}.

\subsubsection{SQUID readout bandwidth}
	
The spin signal arrives as a train of bunch-passage signals at the revolution frequency $f_\mathrm{rev} = \SI{78}{kHz}$, with the amplitude modulated at the spin-tune sideband $\nu_s f_\mathrm{rev} = \SI{39}{kHz}$. A DC SQUID operated in a wideband flux-locked loop (FLL) readily handles this: modern LTS SQUID FLL systems achieve bandwidths of $\sim\!1\,\mathrm{MHz}$ \cite{fagaly}, well above the $\SI{39}{kHz}$ signal frequency. The SQUID sensor itself is a DC device, but the FLL electronics extend the linear dynamic range and tracking bandwidth into the RF range required here.
	
\subsection{Real-world effects not yet addressed}
	\label{pz:sec:Pz_realworld}
	
The arguments above establish the ideal geometric sensitivity of the axial gradiometer, but several non-ideal effects remain to be quantified before this channel can be promoted from a sensitivity estimate to an engineering design. The most important ones are listed here for later follow-up.
	
\begin{itemize}[itemsep=4pt]
		
	\item \textbf{Mechanical alignment of the pickup loop.} As argued in Section~\ref{pz:sec:Pz_charge}, the charge-current rejection is exact for any $\ez$-directed current, regardless of the transverse beam position, beam shape, or current density profile. The only misalignment that breaks this rejection is a tilt of the pickup loop relative to the beam axis, which introduces a component of the loop normal in the azimuthal or radial direction that couples to $\vec{B}_\mathrm{charge}$. The scaling of the residual flux with tilt angle should be worked out explicitly, preferably with a 3D field model that includes the actual beam-pipe geometry, in order to establish the mechanical alignment tolerance for the loop.
		
	\item \textbf{Image currents and nearby conductive structures.} The ideal argument assumes free space. In reality the beam tube, copper coating, bellows, transitions, and neighbouring superconducting leads reshape the field lines and can convert part of the beam-current field into a small but non-zero axial pickup. The copper layer must also be thick enough to guide the image currents over the relevant RF harmonics; a rough skin-depth estimate suggests that $\SIrange{50}{100}{\micro m}$ of Cu should already be sufficient in the tens-of-MHz range, but this should be checked with a realistic field model of the local environment.\footnote{The skin depth in copper is $\delta = \sqrt{2\rho/(\omega\mu_0)}$ with $\rho \approx 1.7\times10^{-8}\,\Omega\,\mathrm{m}$, giving $\delta \approx \SI{20}{\micro m}$ at \SI{10}{MHz} and $\delta \approx \SI{6.5}{\micro m}$ at \SI{100}{MHz}. A coating of $5$--$10$ skin depths, i.e.\ \SIrange{50}{100}{\micro m}, is therefore sufficient; a somewhat thicker coating of \SIrange{0.1}{0.2}{mm} would provide margin for fabrication tolerances and higher frequency content.}
		
	\item \textbf{Ambient and non-uniform $B_z$ backgrounds.} A first-order axial gradiometer cancels only the component of $B_z$ that is uniform across the two-loop baseline $\Delta z$. Longitudinal field gradients, vibration-driven motion in residual fields, and stray fields from nearby magnets or pulsed devices can leak into the signal channel. The expected environmental rejection therefore needs to be estimated for the actual cryostat and ring location.
		
	\item \textbf{Cross-talk from the $P_x$ and $P_y$ channels.} In the full three-channel concept the cosine-$\theta$, sine-$\theta$, and axial loops share the same cryostat and former. Mutual inductive coupling between channels, imperfect orthogonality of the pickup geometries, and common flux-transformer routing may produce cross-talk. This should be quantified experimentally on a prototype former.
		
	\item \textbf{Bunch-to-bunch variations.} In the present estimate all bunches are assumed to have identical intensity and bunch length. In practice, the bunch charge may vary across the fill because the bunches are injected individually from the AGS. This leads primarily to a bunch-dependent signal amplitude, but not to a corresponding variation of the signal frequency. By contrast, substantial bunch-to-bunch variations of the bunch length are not expected on the relevant timescale, since all bunches follow the same RF program of the machine. Thus, the dominant effect of bunch-to-bunch non-uniformity is expected to be amplitude weighting in the coherent sum, rather than loss of phase coherence.
		
	\item \textbf{Absolute calibration.} The present note focuses on relative sensitivity. The full readout chain (loop geometry, flux-transformer coupling, SQUID gain, matched-filter normalization, and the effective FID tipping angle $\alpha$) introduces multiplicative factors that propagate into the polarization scale and cannot be eliminated by internal characterization alone. Absolute polarimetry therefore requires cross-calibration of the SQUID against an external polarization standard; the atomic-hydrogen-jet (HJET) polarimeter, with its proven absolute-polarization capability, will provide this standard at the EIC~\cite{Rathmann2026PRAB}.
		
	\item \textbf{Validity of the assumed coherence model.} The coherence estimate above is intentionally simple and is adequate for this brief writeup. A more realistic treatment should use the actual spin-tune distribution expected for the machine optics and bunch parameters, including any time dependence across the store.
		
	\end{itemize}
	
\subsection{Implementation}
	\label{pz:sec:Pz_implementation}
	
The axial gradiometer requires two plain circular NbTi loops wound directly on the same Macor ceramic former as the cosine-$\theta$ and sine-$\theta$ coils. The loops are placed at the axial positions of coils~A and~B ($\Delta z = \SI{30}{cm}$ centre-to-centre), with loop radius $\rcoil = \SI{40}{mm}$ matching the saddle-pickup inner radius of the former. They are connected in series opposition and coupled through a superconducting flux transformer to one additional wideband FLL SQUID channel. No special winding geometry is required.

The present design assumption is that the beam tube itself must be cold and located inside the cryostat, together with the pickup loops. Proceeding radially outward from the beam axis, the layout is as follows: the beam travels through the hollow ceramic former; the superconducting NbTi pickup loops are wound on the outer surface of the former; a vacuum gap separates the former from the surrounding stainless steel beam tube, whose inner (vacuum-facing) surface carries a thin copper coating that guides the beam image currents in a well-defined way. This fully cryogenic layout is expected to reduce uncontrolled magnetic pickup, preserve the cylindrical symmetry underlying the charge-current rejection argument, and keep the electromagnetic environment of the gradiometer as simple as possible.
	
\subsection{Conclusions for the axial channel}
	\label{pz:sec:Pz_summary}
	
The axial loop gradiometer provides a viable and geometrically elegant route to $P_z$ measurement. Its key properties are:
	
\begin{itemize}[itemsep=3pt]
		
	\item \textbf{Exact charge current rejection by geometry.} $B_\phi$ from the beam current is tangential to the axial loop normal at every point, giving zero flux at all frequencies with no winding balance required.

	\item \textbf{Per-bunch flux of the same order as the cosine-$\theta$ coil at flattop.} At the SQUID input, the per-bunch peak flux is $\Phi_\mathrm{pickup}^\mathrm{axial} \approx \SI{174.9}{\micro\Phi_0}$ (injection) and $\SI{221.3}{\micro\Phi_0}$ (flattop). Relative to the cosine-$\theta$ coil [Eq.~\eqref{eq:flux_pickup_numeric_full}], this is $\sim\!14\%$ at injection and $\sim\!72\%$ at flattop: the axial channel is significantly weaker than the saddle channel at injection but approaches it at flattop, where the shorter bunch length ($\sigma_L=\SI{6}{cm}\simeq r_\mathrm{coil}$) is favourable for the axial geometry.
	
	\item \textbf{Per-cycle precision and multi-cycle accumulation (Strategy~1).} At the canonical operating point ($\alpha = \SI{30}{\milli\radian}$, $P = 0.70$, $S_\Phi^{1/2} = \SI{0.4}{\micro\Phi_0\per\sqrt{Hz}}$), the all-bunch matched-filter sensitivity is $K_z = \SI{156.8}{\per\sqrt{\second}}$ at injection and $\SI{198.4}{\per\sqrt{\second}}$ at flattop, giving noise-limited $1\%$ $P_z$ measurement times $T_{1\%} \simeq \SI{7.53}{\minute}$ at injection (for a residual $P_z \simeq 0.03$ anchored on the pC transverse-tilt observations of Ref.~\cite{Schmidke:2018}) and $T_{1\%} \simeq \SI{25.4}{\second}$ at flattop (for a residual $P_z \simeq 0.10$). Both exceed the canonical coherence time $\tau_\mathrm{coh}^\mathrm{eff} \simeq \SI{2.04}{ms}$ (Tab.~\ref{tab:pi_pulse_field_requirements}) and are realized as multi-cycle tip-$\pi$-echo-restore sequences per Sec.~\ref{sec:fid-echo-sequence}.

	\item \textbf{Bunch-resolved vector polarimetry (Strategy~2).} The single-bunch matched-filter framework extends to all three channels and provides continuous bunch-by-bunch reconstruction of $\vec P(t)$ throughout the fill; the joint three-channel capability is developed in Sec.~\ref{sec:bunch-resolved-vec-polarimetry}.

	\item \textbf{Minimal hardware.} One wideband FLL SQUID channel; two plain circular NbTi loops on the existing former; no cryostat modification.

\end{itemize}
	
Together with the cosine-$\theta$ coil ($P_x$) and the sine-$\theta$ coil ($P_y$), the axial gradiometer completes a three-component noninvasive spin vector polarimeter in a single cryostat.

\section{Bunch-resolved vector polarimetry}
\label{sec:bunch-resolved-vec-polarimetry}

\subsection{Three-channel measurement modes}
\label{sec:sec10:modes}

The three pickup channels of Sec.~\ref{sec:design-overview} admit both \emph{static} (no kicker firing) and \emph{dynamic} (with kicker firing) operating modes. In static mode, each component of the local polarization vector produces a signal at its natural frequency, fixed by the precession geometry: $P_y$, aligned with $\vec n_0 \simeq \vec e_y$, is constant in time and produces a bunch-passage pulse train at the revolution frequency $f_\mathrm{rev}$; the perpendicular components $P_x$ and $P_z$ precess about $\vec n_0$ at the spin tune $\nu_s = 1/2$ and produce oscillating signals at $f_s = f_\mathrm{rev}/2$. The signal frequencies are physically determined by the precession and cannot be shifted by operational choice.

In dynamic mode, the longitudinal kicker fires a small-angle pulse synchronized to the phase-locked spin precession established in Sec.~\ref{sec:fid_spin_tune_phase_lock}. The rotation $R_z(\alpha)$ mixes the $(P_x, P_y)$ components in the $(\vec e_x, \vec e_y)$ plane and leaves $P_z$ unchanged in the same instant. The free precession between successive kicks rotates the perpendicular components $(P_x, P_z)$ continuously in the $(\vec e_x, \vec e_z)$ plane at $f_s$, so the in-plane component aligned with $\ex$ at any chosen moment is determined by the firing phase $\varphi_k = 2\pi f_s\, t_k$ relative to the phase-locked precession. By selecting $\varphi_k$, the kicker tips either residual $P_x$ ($\varphi_k = 0$) or residual $P_z$ ($\varphi_k = \pi/2$) into the $\vec e_y$ direction, where it appears as a small additive contribution to the sin-$\theta$ channel signal at $f_\mathrm{rev}$. Independently, the same kick rotates stored $P_y$ into the $-\vec e_x$ direction with amplitude
$P_y \sin\alpha$; this contribution precesses and is read out by the cos-$\theta$ channel at $f_s$ regardless of firing phase, providing the Sec.~\ref{sec:fid_spin_tune_phase_lock} FID readout simultaneously with the in-plane projection extracted via sin-$\theta$.

The six accessible (channel, mode) combinations are summarized in Table~\ref{tab:vector_polarimetry_modes}. Each row corresponds to a distinct matched-filter readout of one polarization component on one channel at one signal frequency. The static rows operate by default throughout the fill; the dynamic rows fire periodically and serve dual roles as $\alpha$ calibration and cross-checks against the static readouts.

A representative operating point can be established using the transverse-spin tilt measured at RHIC by the pC polarimeters~\cite{Schmidke:2018}. The observed tilts of $\phi_\mathrm{pC} \simeq 3^\circ$ at $\SI{100}{GeV}$ and $\simeq 8^\circ\text{--}16^\circ$ at $\SI{255}{GeV}$ imply residual transverse polarization components of order $P_x \simeq 0.03$ at injection and $P_x \simeq 0.10$ at flattop. Although no comparable measurement of $P_z$ exists, we adopt the same numerical range as a conservative working estimate for the longitudinal residual; this should be regarded as a working assumption rather than a quantitative claim about $P_z$. The resulting operating-point integration times for the six (channel, mode) combinations are summarized in Tab.~\ref{tab:vector_polarimetry_modes}. All static-mode integration times lie at or below a few minutes; the longest, for $P_z$ at injection, is $\sim 3$\,min and corresponds to many tip-$\pi$-echo-restore cycles per Sec.~\ref{sec:fid-echo-sequence}. Dynamic-mode integration times incur an additional $\sin^2\alpha$ suppression and are tabulated for cross-calibration purposes. The static-mode saddle-channel signal is linear in the residual in-plane polarization component, so a fixed integration time yields a fractional precision $\delta P_x/P_x$ that scales as $1/P_x$; in HSR configurations where the equilibrium $\vec n_0$ is tuned close to the vertical and the residual in-plane component is small by design, the integration time required for a target fractional precision grows accordingly, and the dynamic-mode kicker readout (which transfers a controlled fraction of the dominant $P_y$ into the in-plane channel as a calibrated probe) becomes the natural complement for resolving the smallest residual components.

\begin{table*}[tb]
	\centering
	\caption{Three-channel vector polarimetry measurement modes. Static operation (no kicker firing) is the default and reads all three polarization components simultaneously and continuously, each on its dedicated channel at its natural signal frequency. Dynamic operation fires the longitudinal kicker (tip axis $\vec e_z$) at a chosen phase $\varphi_k$ of the phase-locked spin precession to tip the selected in-plane component into the $\vec e_y$ direction, where it is read on the sin-$\theta$ channel at $f_\mathrm{rev}$. The same firing simultaneously produces the FID readout on cos-$\theta$ at $f_s$ via the $-P_y \sin\alpha$ contribution. Alternating firings between $\varphi_k = 0$ and $\varphi_k = \pi/2$ provides dynamic readouts of the entire perpendicular plane $(P_x, P_z)$ from a single set of kicker pulses. Per-cycle integration times $T_{1\%}$ are quoted from the per-channel matched-filter sensitivities of Secs.~\ref{sec:static_vector_extension}--\ref{sec:pz_axial_channel}, evaluated at the operating point $P=0.7$, $\alpha=\SI{30}{\milli\radian}$, $S_\Phi^{1/2}=\SI{0.4}{\micro\Phi_0\per\sqrt{Hz}}$. Static-mode signal amplitudes for $P_x$ and $P_z$ adopt residual polarization values of $0.03$ at injection and $0.10$ at flattop, anchored on the RHIC pC tilt observations of Ref.~\cite{Schmidke:2018}; the longitudinal residual is assumed equal to the transverse residual in the absence of any prior measurement.}
	\label{tab:vector_polarimetry_modes}
	\renewcommand{\arraystretch}{1.15}
	\setlength{\tabcolsep}{4pt}
	\begin{tabular}{l l c c l c c}
		\hline\hline
		comp. & mode & channel & sig.\ freq & kicker phase $\varphi_k$ & $T_{1\%}$ (inj) & $T_{1\%}$ (flat) \\
		\hline
		$P_y$ & static  & sin-$\theta$ & $f_\mathrm{rev}$ & ---     & \SI{16.6}{\milli\second} & \SI{266}{\milli\second} \\
		$P_x$ & static  & cos-$\theta$ & $f_s$            & ---     & \SI{9.05}{\second}        & \SI{13}{\second}        \\
		$P_z$ & static  & axial        & $f_s$            & ---     & \SI{7.53}{\minute}        & \SI{25.4}{\second}      \\
		$P_y$ & dynamic & cos-$\theta$ & $f_s$            & any     & \SI{18.5}{\second}        & \SI{4.93}{\minute}      \\
		$P_x$ & dynamic & sin-$\theta$ & $f_\mathrm{rev}$ & $0$     & \SI{2.79}{\hour}          & \SI{4.03}{\hour}        \\
		$P_z$ & dynamic & sin-$\theta$ & $f_\mathrm{rev}$ & $\pi/2$ & \SI{2.79}{\hour}          & \SI{4.03}{\hour}        \\
		\hline\hline
	\end{tabular}
\end{table*}

The duty cycle between static monitoring and dynamic calibration is an operational choice. A representative pattern is continuous static monitoring throughout the fill, interrupted periodically (e.g.\ every $T_\mathrm{cal}$ minutes) by short dynamic-mode bursts that refresh the $f_s$ phase lock and update the in-situ $\alpha$ calibration. Each burst contains a small number of kicker firings at prescribed $\varphi_k$. The cumulative depolarization cost over the full store is bounded by the noninvasive tip-$\pi$-echo-restore budget established in Sec.~\ref{sec:fid-echo-sequence}, evaluated for the chosen $T_\mathrm{cal}$.

\subsection{Comparison with existing polarimetry}
\label{sec:sec10:comparison}

At the EIC, absolute proton-beam polarization will be measured by the same atomic-hydrogen-jet (HJET) and proton-carbon (pC) instruments developed and operated at RHIC~\cite{Rathmann2026PRAB}. Three-dimensional reconstruction of the stable spin direction in principle is possible with these instruments by combining the transverse measurements with controlled local spin rotations and spin transport, as demonstrated at RHIC~\cite{Schoefer:2024TECH}, but only as a time-consuming dedicated procedure, providing a fill-averaged result. The three-channel SQUID polarimeter proposed in this paper supplements that suite by delivering continuous, noninvasive, bunch-resolved measurement of the full polarization vector $\vec P(t)$ throughout the store. The SQUID output is qualitatively distinct from the established scattering-based instruments, particularly for the longitudinal component $P_z$, as discussed below.

The pC polarimeter operates with an unpolarized carbon target. Only single-spin observables are accessible, and the longitudinal analyzing power $A_z \propto \vec S \cdot \vec k$ is a parity-violating pseudoscalar that vanishes identically under parity-conserving strong interactions. The longitudinal polarization $P_z$ is therefore inaccessible to pC polarimetry by parity conservation, irrespective of statistical precision.

The HJET polarimeter uses a polarized atomic-hydrogen jet target and provides access to single-spin and double-spin observables. In its present RHIC operating configuration the target is vertically polarized, so HJET measures the vertical beam polarization $P_y$ via the analyzing power $A_y$ and the spin-correlation parameter $A_{yy}$. The $A_{yy}$-based determination requires knowledge of both the target polarization $Q_y$ and the analyzing power $A_{yy}$ itself; the latter is itself a calibrated quantity~\cite{PhysRevLett.123.162001}. Rotation of the target polarization to other directions gives access to $P_x$ (via $A_{xx}$ with a transversely polarized target) and to $P_z$ (via $A_{zz}$ with a longitudinally polarized target); these target-rotation techniques are established~\cite{PhysRevC.58.658} and are expected to form part of the upgraded HJET planned for the EIC, but they were not the routine RHIC operating mode.

The SQUID polarimeter is not subject to either limitation. The longitudinal magnetic moment $m_z = N_p\mu_p P_z$ is sensed directly by the axial-gradiometer geometry of Sec.~\ref{sec:pz_axial_channel}, with no scattering process and no target manipulation required. The output is continuous throughout the store and resolved bunch by bunch. To our knowledge, the SQUID concept provides the first practical route to continuous, bunch-resolved, noninvasive $P_z(t)$ monitoring of a stored proton beam.

A related development worth noting is the framework of Ref.~\cite{NIKOLAEV2020135983}, in which deliberately precessing in-plane deuteron polarizations are used as a means to search for $T$-violating asymmetries in $pd$ transmission, exploiting Fourier analysis at $f_s$ and $2f_s$ to expose oscillating signals against backgrounds. That approach relies on active RF spin rotation to generate the in-plane components and on beam-current attenuation to extract asymmetries. The SQUID polarimeter provides the natural detector counterpart: continuous direct measurement of the precessing in-plane and longitudinal magnetic moments, bunch-resolved, with no active manipulation needed. The two approaches are complementary, one generates oscillating signals to expose new physics in scattering, the other senses oscillating signals continuously, and the SQUID polarimeter substantially relaxes the operational requirements for the precessing-polarization-based
searches developed in Ref.~\cite{NIKOLAEV2020135983}.

\subsection{Bunch-resolved polarization history}
\label{sec:sec10:bunch_history}

The combination of the three-channel pickup geometry and the matched-filter analysis of Sec.~\ref{sec:matched-filter-coherent-summation} delivers a distinctive operational output: the time evolution of the full polarization vector $\vec P(t)$ for every bunch in the stored beam, continuously throughout the fill, without disturbing the beam. We illustrate this output with a synthetic but representative simulation of an eight-hour flattop fill ($N_b = 1160$ bunches, $T_\mathrm{avg} = \SI{180}{s}$ averaging window per bunch and per time bin), in which $P_y$ undergoes a slow $T_1$-type decay, residual $P_x$ and $P_z$ follow few-percent drifts over the fill, and two localized anomalous events affect distinct subsets of bunches.

Fig.~\ref{fig:bunch_polarization_history} shows the three resulting heatmaps. Panel~(a) tracks the dominant stored vertical component $P_y$ and exhibits the expected slow fill-averaged decay. Panels~(b) and (c) track the residual transverse and longitudinal components and reveal features that no fill-averaged polarimeter can resolve: bunch-to-bunch variation in the steady-state offsets, slow drifts of the offsets over hours, and the two localized anomalous events as compact patches against the smoother background. Panel~(c) in particular constitutes the first practical demonstration of bunch-resolved longitudinal polarization reconstruction; the same information is not currently extractable from any combination of pC and HJET measurements (see Sec.~\ref{sec:sec10:comparison}).

\begin{figure}[tb]
	\centering
	\includegraphics[width=0.85\linewidth]{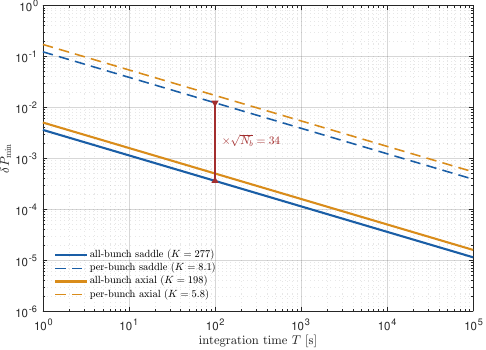}
	\caption{Per-bunch versus all-bunch matched-filter precision at flattop ($N_b = 1160$). Solid curves: all-bunch matched filter, which uses every bunch passage. Dashed curves: per-bunch matched filter, which uses only the passages of a single bunch and is therefore $\sqrt{N_b} \simeq 34$ worse for the same integration time. Blue: saddle channels (per-bunch $K = \SI{8.13}{\per\sqrt{\second}}$). Red: axial gradiometer (per-bunch $K = \SI{5.83}{\per\sqrt{\second}}$). The $\sqrt{N_b}$ vertical offset between the dashed and solid curves is the only factor by which bunch-resolved reconstruction is statistically penalized relative to fill-averaged measurement.}
	\label{fig:per_bunch_precision}
\end{figure}

\begin{figure}[tb]
	\centering
	\includegraphics[width=0.8\linewidth]{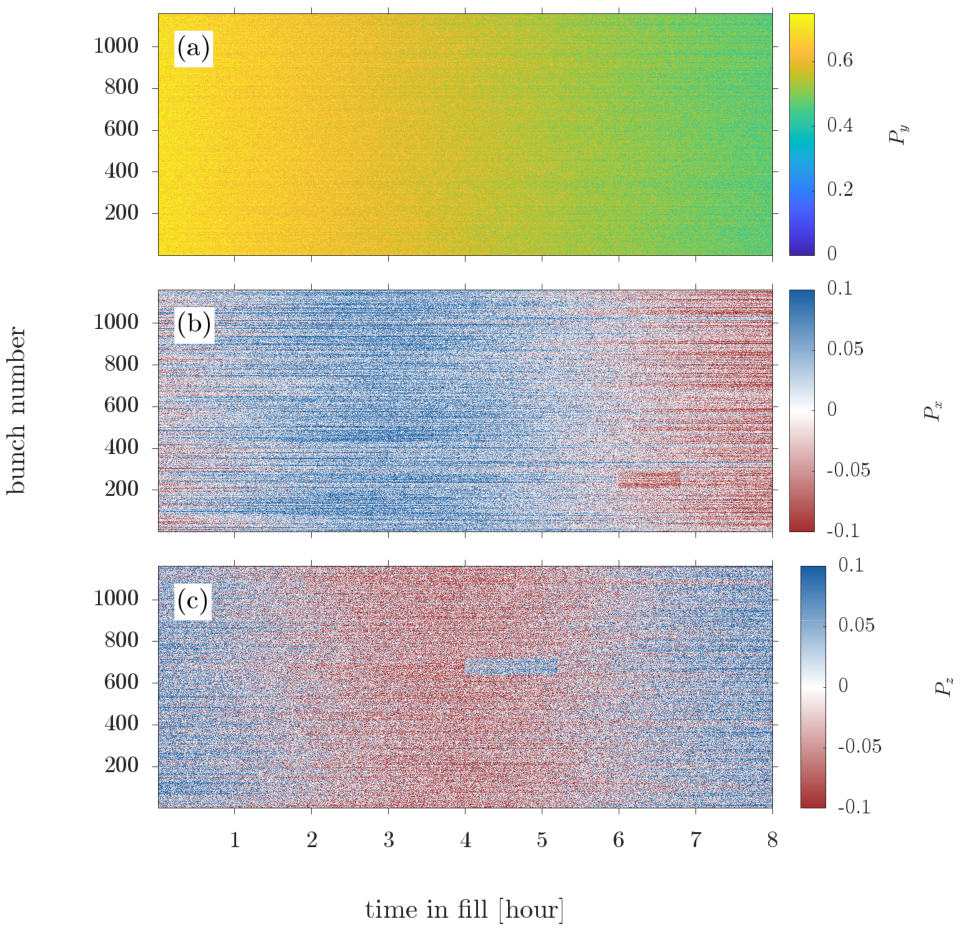}
	\caption{Bunch-resolved polarization history over an 8-hour EIC HSR fill at flattop. (a) Dominant component $P_y$. (b) Residual in-plane $P_x$. (c) Residual longitudinal $P_z$, the first-ever bunch-resolved measurement.}
	\label{fig:bunch_polarization_history}
\end{figure}

The achievable per-bunch precision is set by the SNR available to one specific bunch as a function of the averaging window $T$ devoted to that bunch. Both the saddle and axial channels reach a $1\%$ per-bunch precision in a few hundred seconds at flattop and in correspondingly shorter times at injection. Fig.~\ref{fig:per_bunch_precision} makes this explicit. The all-bunch matched filter (solid curves) uses every bunch passage and reaches $\delta P_\mathrm{min} \sim 10^{-4}$ within seconds; the per-bunch matched filter (dashed curves) uses only the passages of a single bunch and is therefore $\sqrt{N_b}$ worse for the same integration time. The $\sqrt{N_b} \simeq 34$ at flattop sets the vertical separation of the dashed and solid curves and is the only factor by which per-bunch reconstruction is statistically penalized relative to fill-averaged measurement. Selecting $T_\mathrm{avg}$ per bunch and per time bin is then a direct trade-off between bunch time-resolution and per-cell polarization precision: the $\SI{180}{s}$ value used in Fig.~\ref{fig:bunch_polarization_history} gives roughly $1\%$ per-cell precision on the saddle channels and $1.5\%$ on the axial channel at flattop, sufficient to resolve the illustrated drifts and anomalous events while still providing $160$ independent time bins per fill.

\section{Extensions and outlook}
\label{sec:extensions-outlook}

Although the quantitative treatment in this paper uses the polarized proton beam as its worked example, the concept is not specific to protons and extends naturally to other particle species and to other storage rings. Two directions are of particular interest: storage-ring searches for electric dipole moments, for which the JEDI collaboration has recently reported the first direct limit on the deuteron EDM at COSY~\cite{Andres2026DeuteronEDM}, and self-polarized electron beams.

\subsection{Storage-ring electric dipole moment searches}
\label{subsec:edm-outlook}

A frozen-spin storage ring stores light ions at the magic momentum $p_\mathrm{magic} = mc/\sqrt{G} \approx \SI{0.70}{\giga\electronvolt}/c$ for protons, at which the in-plane $g{-}2$ precession is frozen and the spin tracks the velocity~\cite{CPEDM:2019nwp}. A nonzero electric dipole moment then tilts the spin slowly out of the ring plane, so that a small vertical polarization $P_y(t)$ grows almost linearly in time: this growing out-of-plane projection is the EDM signal. This invariant-spin-axis tilt is precisely the observable that the JEDI collaboration recently determined for deuterons at COSY, combining a radio-frequency Wien filter, a Siberian snake, and an electron-cooler solenoid to set a first direct limit on the deuteron electric dipole moment, $|d_d| < 2.5\times10^{-17}\,e\,\mathrm{cm}$ at the $95\%$ confidence level~\cite{Andres2026DeuteronEDM}. Counter-propagating clockwise and counter-clockwise beams share the same magnetic guide field but sample radial fields of opposite sign, so their difference cancels the dominant fake-EDM systematic associated with radial magnetic fields.

The SQUID readout is a natural match for this signal. The observable is a slowly growing out-of-plane moment, precisely the quantity read by the static transverse and axial channels of Secs.~\ref{sec:static_vector_extension} and~\ref{sec:pz_axial_channel}. Because the pickup is noninvasive and continuous, both counter-rotating beams can be read in real time, replacing the slow destructive scattering polarimetry otherwise used to follow the spin. The clockwise-minus-counter-clockwise difference isolates the EDM term, while the stable revolution frequency of a dedicated EDM ring enables the same coherent, phase-locked integration developed here. A quantitative treatment of the frozen-spin operating point, the dual-beam systematics, and the achievable EDM sensitivity is the subject of a dedicated follow-up paper.

\subsection{Electron storage rings and self-polarized beams}
\label{subsec:electron-outlook}

The matched-filter reconstruction and the FID-echo cycle both rely on a stable revolution frequency, so that the turn-by-turn precession phase can be predicted and the per-bunch contributions summed coherently. Among electron machines, a storage ring with a stable $f_\mathrm{rev}$ is therefore the natural fit. For the EIC electron storage ring~\cite{EIC_YellowReport2022}, the large electron magnetic moment ($\mu_e/\mu_p \approx 658$), together with a tighter pickup coil, yields orders of magnitude more flux per passage; since the integration time scales as $\Phi_\mathrm{pickup}^{-2}$, the corresponding measurement times collapse to the millisecond range.

The electron spin dynamics differ in two respects. The spin tune is $\nu_s = G_e\gamma \approx 41$ at top energy, set directly by the beam energy with no Siberian snakes, so the half-integer working point used in the HSR does not apply. In addition, Sokolov-Ternov self-polarization~\cite{SokolovTernov1964} is fast and asymmetric between spin states: bunches anti-aligned with the guide field depolarize faster than the aligned ones, with a characteristic time of order minutes at \SI{18}{\giga\electronvolt} and a steep $E^{-5}$ energy dependence. Machines without a stable revolution period are not direct users of the coherent method: in a rapid-cycling synchrotron the ramp leaves no stable $f_\mathrm{rev}$ and the spin tune sweeps through resonances, leaving no coherent window, while a recirculating linac such as CEBAF has no revolution frequency at all and admits at best single-pass detection, aided by the large electron moment.

\section{Summary}
\label{sec:summary}

We have developed a quantitative design and feasibility analysis of a SQUID-based polarimeter for the EIC Hadron Storage Ring that exploits the six-snake $\nu_s = 1/2$ spin lattice to read out the macroscopic magnetic dipole moment of polarized bunches at the spin precession frequency $f_s = f_\mathrm{rev}/2$. The pickup couples three local channels at a single station: a saddle cosine-$\theta$ loop sensitive to the horizontal transverse dipole component, a saddle sine-$\theta$ loop sensitive to the vertical component, and a coaxial axial gradiometer sensitive to the longitudinal component. The three channels are read out simultaneously by independent dc-SQUID front ends and reconstruct the full polarization vector $(P_x, P_y, P_z)$. The measurement proceeds as a sequence of phase-locked tip-$\pi$-echo-restore cycles synchronized to the SQUID reference oscillator, with the closing restore pulse returning the tipped in-plane component to the
longitudinal stable spin axis so that the cumulative polarization cost across a full measurement is negligible.

At the canonical operating point of $\alpha = \SI{30}{\milli\radian}$, $P = 0.7$, $\sigma_{\nu_s}^\mathrm{eff} = 10^{-3}$, and $S_\Phi^{1/2} = \SI{0.4}{\micro\Phi_0\per\sqrt{Hz}}$, the matched-filter sensitivities reach $K_x = K_y \simeq \SI{1108}{\per\sqrt{s}}$ at injection and $\simeq\SI{277}{\per\sqrt{s}}$ at flattop on the saddle channels, and $K_z \simeq \SI{156.8}{\per\sqrt{s}}$ (injection) and $\simeq\SI{198.4}{\per\sqrt{s}}$ (flattop) on the axial gradiometer. With residual transverse and longitudinal polarizations anchored on the RHIC pC tilt observations of Ref.~\cite{Schmidke:2018}, all static-mode integration times for $1\%$ precision lie between $\sim\SI{17}{\milli\second}$ (stored $P_y$ at injection) and $\sim\SI{7.5}{\minute}$ (residual $P_z$ at injection); at flattop, all three components reach $1\%$ precision in under $\SI{30}{\second}$. Free-precession coherence is preserved over the canonical $\tau_\mathrm{coh}^\mathrm{eff} \simeq \SI{2.04}{\milli\second}$ by phase-locked
operation; longer integrations are realized through the synchronized $\pi$ rephasing pulse delivered as $N \simeq 80$ per-turn pulses with per-pass integrated field $\int B_z\,dl|_{\pi,\text{per-pass}} \simeq \SI{1.1}{\tesla\meter}$ at injection and $\SI{12.9}{\tesla\meter}$ at flattop (Tab.~\ref{tab:pi_pulse_field_requirements}).

Beyond the absolute calibration use case, the continuous and noninvasive character of the SQUID readout supports a class of measurements that no existing storage-ring polarimeter delivers. Per-bunch matched filtering on the saddle and axial channels reconstructs $\vec P(t)$ for every bunch throughout the fill, at a per-bunch precision set by $K/\sqrt{N_b}$. At flattop this is $\sim\SI{8.1}{\per\sqrt{s}}$ on the saddle channels and $\sim\SI{5.8}{\per\sqrt{s}}$ on the axial channel, sufficient to resolve few-percent bunch-by-bunch variations of all three components in averaging windows of order one minute. The longitudinal projection $P_z$, inaccessible to single-spin pC polarimetry as a matter of parity conservation and not routinely measured by HJET, becomes a routine monitoring observable on every bunch and time bin (Sec.~\ref{sec:bunch-resolved-vec-polarimetry}). The same readout provides a noninvasive diagnostic of the spin tune, its linewidth-derived spread, and the coherence envelope.

The matched-filter sensitivities quoted above derive from a conservative baseline operating point. Three modest hardware optimizations, each individually within current SQUID engineering practice, would compound to deliver substantially shorter integration times. From the scaling $K_\mathrm{template} \propto (\eta\sqrt{N_\mathrm{SQUIDs}})/S_\Phi^{1/2}$ [Eq.~\eqref{eq:K_template}], lowering the white flux-noise floor from $\SI{0.4}{\micro\Phi_0\per\sqrt{Hz}}$ to $\SI{0.3}{\micro\Phi_0\per\sqrt{Hz}}$ (via multi-stage SQUID arrays or operation at sub-K temperatures), improving the flux-transformer coupling efficiency from $\eta=0.7$ to $\eta=0.9$ (via optimized inductive matching of the pickup loop to the SQUID input coil), and doubling the number of independent SQUID channels per pickup from $N_\mathrm{SQUIDs}=4$ to $8$, each contributes independently. The compound enhancement is $K_\mathrm{template}\mapsto 2.4\,K_\mathrm{template}$, equivalent to a factor $\sim 6$ shorter integration time for the same target precision. Under this optimized scenario the matched-filter $1\%$ integration times of Tab.~\ref{tab:snr} would shorten to $\sim\SI{3}{\second}$ at injection and $\sim\SI{50}{\second}$ at flattop, bounding the practical performance ceiling for the architecture proposed in this paper.

The same architecture applies without modification to deuteron and $^3$He beams via species-specific spin-magnetic factors, with the storage-ring electric-dipole-moment and electron-beam extensions outlined in Sec.~\ref{sec:extensions-outlook}. Engineering details of the saddle-coil fabrication, the axial gradiometer mechanical layout, the common-mode-rejection budget, the SQUID front-end electronics, and cryogenic and shielding requirements are treated in a companion engineering paper. The present work establishes the principle, the sensitivity budget, and the operational concept, demonstrating that noninvasive, bunch-resolved, three-component polarimetry of the EIC proton beam is feasible with currently available SQUID technology.

\section*{Acknowledgments}
\noindent The author thanks Helmut Soltner (Forschungszentrum J\"ulich, Germany) for useful discussions.

\medskip
\noindent This manuscript has been authored by an employee of Brookhaven Science Associates, LLC under Contract No.\ DE-SC0012704 with the U.S.\ Department of Energy. The publisher, by accepting the manuscript for publication, acknowledges that the United States Government retains a non-exclusive, paid-up, irrevocable, world-wide license to publish or reproduce the published form of this manuscript, or allow others to do so, for United States Government purposes.

\vspace{1em}
\bibliographystyle{unsrturl}
\bibliography{squid_eic_polarimetry_v2}

\appendix

\section{Spin manipulation with an RF Wien filter}
\label{app:spin-manip-rf-WF}

The longitudinal kicker introduced in Sec.~\ref{sec:fid-echo-sequence} provides the tipping, $\pi$, and restore pulses required for the SQUID measurement cycle. The same spin-manipulation functions can in principle be realized with an RF Wien filter operated at the spin-resonance frequency, with the operational advantage that the Wien-filter configuration preserves the closed orbit by construction. A recent COSY experiment by Slim \textit{et al.}~\cite{PhysRevResearch.7.023257} demonstrated that fast rf switches integrated into the Wien-filter driving circuit allow the device to be gated bunch by bunch. In their two-bunch demonstration with polarized deuterons, the signal bunch received the resonant RF and underwent multiple continuous spin flips with a fitted per-flip spin-flip efficiency $\varepsilon_\mathrm{SF} = 0.9954 \pm 0.0037$, while the pilot bunch was shielded from the RF during its passage through the device with a gate efficiency $\varepsilon_\mathrm{gate} = 0.9921 \pm 0.0136$ consistent with unity, leaving its idle precession unperturbed. The two efficiencies quantify distinct aspects of the demonstration: $\varepsilon_\mathrm{SF}$ measures how cleanly the resonant RF reverses the polarization of the driven bunch, while $\varepsilon_\mathrm{gate}$ measures how completely the masked bunch is isolated from the RF field during the gated interval. That experiment was performed at the natural deuteron spin tune $\nu_s = \gamma G_d$ (with $G_d \simeq -0.143$) in a ring with no Siberian snakes; the two-branch resonance degeneracy that arises at $\nu_s = 1/2$ and which is the subject of Sec.~\ref{nzero:sec:field-structure} below is therefore specific to the EIC HSR six-snake lattice considered here. A bunch-gated RF Wien filter pairs naturally with the bunch-resolved SQUID readout developed in Sec.~\ref{sec:bunch-resolved-vec-polarimetry}: the combination of per-bunch $(P_x, P_y, P_z)$ measurement and per-bunch polarization reversal provides the natural
systematic-control mode for the precision spin-physics applications, including the storage-ring electric-dipole-moment searches and related beyond-the-Standard-Model studies treated in a dedicated follow-up paper.

Having established the spin lattice of the LC configuration, we now consider its response to an external RF perturbation.

To investigate controlled spin manipulation in the LC configuration introduced in Sec.~\ref{sec:lc_dlc_config}, we consider the effect of a transverse RF Wien filter providing a horizontal magnetic field. The spin transport is described using the one-turn map of Eq.~(\ref{nzero:eq:M_oneturn}) together with the sector-wise rotations of Eqs.~(\ref{nzero:eq:R_snake}) and (\ref{nzero:eq:R_arc}), with the RF Wien filter treated as a localized time-dependent perturbation.

The RF Wien filter is operated at the spin-resonance frequency,
\begin{equation}
	f_{\mathrm{RF}} = (k + \nu_s)\, f_{\mathrm{rev}}, \qquad k \in \mathbb{Z},
\end{equation}
such that the perturbation is phase-locked to the spin precession. For the present lattice with full Siberian snakes, $\nu_s = \tfrac{1}{2}$, and the resonance condition reduces to
\begin{equation}
	f_{\mathrm{RF}} = \left(k + \tfrac{1}{2}\right) f_{\mathrm{rev}}.
\end{equation}

Under these conditions, the spin motion can be described in a rotating frame in which the resonant component of the RF field appears as an effective static transverse field. A continuous sequence of small spin rotations then adds coherently, leading to a gradual reorientation of the polarization.

The driven motion can be expressed in terms of a turn-dependent spin map,
\begin{equation}
	\mathbf{M}_{\mathrm{tot}}(n) = \mathbf{M}_{\mathrm{ring}}\, \mathbf{R}_{\mathrm{WF}}(\phi_n),
\end{equation}
with RF phase
\begin{equation}
	\phi_n = 2\pi \frac{f_{\mathrm{RF}}}{f_{\mathrm{rev}}} n + \phi_0.
\end{equation}
Tracking the spin turn-by-turn allows one to monitor the projection onto the invariant spin axis,
\begin{equation}
	P_n = \vec{S}_n \cdot \vec{n}_0.
\end{equation}

On resonance, the RF field generates a coherent buildup of spin rotation, leading to a continuous change of $P_n$ and, for sufficient interaction time, a reversal of the polarization. Off resonance, the phase coherence is lost and the oscillation amplitude is reduced, preventing full spin flip.

To illustrate the resonant mechanism, we perform a numerical tracking calculation in which a small RF-induced rotation is applied once per turn. Starting from a vertically polarized beam, $\vec{P} = (0,1,0)$, the spin is propagated using the same one-turn map as in the unperturbed case, supplemented by the RF kick with time-dependent phase.

This simplified model represents the idealized resonant component of the spin motion and assumes a well-defined effective rotation axis in the rotating frame.

\begin{figure}[!]
	\centering
	\includegraphics[width=0.85\linewidth]{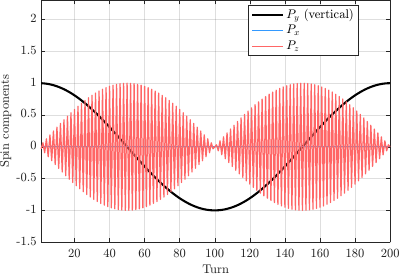}
	\caption{Turn-by-turn evolution of the spin components under the action of a resonant RF Wien filter. The vertical polarization $P_y$ (black) is gradually reversed, while the longitudinal component $P_z$ (red) oscillates and the radial component $P_x$ (blue) stays essentially constant. The RF field is synchronized with the spin precession according to $f_{\mathrm{RF}} = (k + \nu_s) f_{\mathrm{rev}}$, enabling coherent buildup of the spin rotation.}
	\label{nzero:fig:rf_wien_flip}
\end{figure}

The result is shown in Fig.~\ref{nzero:fig:rf_wien_flip}. The vertical polarization component $P_y$ decreases continuously and changes sign, demonstrating a full reversal of the spin direction. The longitudinal component $P_z$ oscillates while the radial component $P_x$ remains essentially constant, consistent with a rotation about the radial axis in the rotating frame. The observed evolution confirms that, under resonant conditions, an RF Wien filter can provide coherent spin manipulation, with polarization reversal arising from the accumulation of small rotations over many turns.

\subsection{Field structure}
\label{nzero:sec:field-structure}

The origin of the two resonance conditions can be understood by expressing a linearly oscillating RF magnetic field in terms of its rotating components. For a transverse field,
\begin{equation}
	B_x(t) = B_0 \cos(\omega t)
	= \tfrac{B_0}{2}\left(e^{+i\omega t} + e^{-i\omega t}\right),
\end{equation}
which shows explicitly that the field consists of two circularly rotating components with frequencies $+\omega$ and $-\omega$, corresponding to opposite rotation senses.

The spin precesses with angular frequency
\begin{equation}
	\omega_s = 2\pi \nu_s f_{\mathrm{rev}}.
\end{equation}
Resonance occurs when one of the two rotating components becomes stationary in the rotating spin frame. This leads to the conditions
\begin{equation}
	\omega - \omega_s = k\,\omega_{\mathrm{rev}}
	\qquad \text{or} \qquad
	-\omega - \omega_s = k\,\omega_{\mathrm{rev}},
\end{equation}
where $\omega_{\mathrm{rev}} = 2\pi f_{\mathrm{rev}}$. Dividing by $\omega_{\mathrm{rev}}$ gives
\begin{equation}
	\frac{\omega}{\omega_{\mathrm{rev}}} = k + \nu_s
	\qquad \text{or} \qquad
	\frac{\omega}{\omega_{\mathrm{rev}}} = k + (1 - \nu_s).
\end{equation}
In terms of the RF tune $Q_{\mathrm{osc}} = f_{\mathrm{RF}}/f_{\mathrm{rev}}$, this corresponds to the two resonance conditions
\begin{equation}
	\nu_s = Q_{\mathrm{osc}}
	\qquad \text{and} \qquad
	\nu_s = 1 - Q_{\mathrm{osc}}.
\end{equation}

Away from $\nu_s = \tfrac{1}{2}$, the two branches are separated and only one of the two rotating components is resonant. For the present lattice with $\nu_s = \tfrac{1}{2}$, however, the two conditions become degenerate. In this case, both the $+\omega$ and $-\omega$ components are simultaneously resonant, and a purely linearly oscillating field does not define a unique rotation sense in the rotating frame.

A well-defined coherent spin rotation is obtained only if the RF field provides a single rotating component. This can be achieved by combining two orthogonal magnetic-field components with a relative phase shift of $90^\circ$,
\begin{equation}
	B_x(t) \propto \cos(\omega t), \qquad
	B_z(t) \propto \sin(\omega t),
\end{equation}
so that
\begin{equation}
	B_x(t) + i B_z(t) \propto e^{+i\omega t},
\end{equation}
which corresponds to selecting only one of the two rotation senses. In this case, only a single resonance branch is excited and the spin experiences a well-defined rotation in the rotating frame. A different implementation of this rotation-sense selection, using multiple AC dipoles arranged in two closed vertical orbit bumps interleaved with four DC spin rotators rather than a single rotating-field device, was developed for the RHIC spin flipper. First commissioning measurements with this design demonstrated significant suppression of the unwanted resonance branch though without full cancellation~\cite{Bai:2012spinflipper}; subsequent operation of the device in the snake ring at 255 GeV achieved a high single-flip efficiency~\cite{Huang:2018spinflipefficiency}, although routine operational use for systematic-error reduction during physics running has remained limited.

\subsection{Idealized spin reversal in the rotating frame}
\label{nzero:sec:ideal-reversal}

Once the rotation-sense degeneracy is broken and a single rotating
component is selected, the driven spin motion takes a simple closed
form. In the frame rotating at the spin-precession frequency
$\omega_s = 2\pi\nu_s f_\mathrm{rev}$, the resonant component of the
combined Wien-filter/longitudinal field is stationary and acts as a
static effective field along a fixed axis $\vec e$, while the
counter-rotating component averages to zero over many turns. For the
field phasing $B_x(t)\propto\cos\omega t$, $B_z(t)\propto\sin\omega t$
of Sec.~\ref{nzero:sec:field-structure}, this effective axis lies along
the radial direction, $\vec e = \ex$.

The spin then undergoes uniform rotation about $\ex$,
\begin{equation}
	\vec P(N) = \Rmat_{\ex}\!\bigl(\Theta(N)\bigr)\,\vec P(0),
	\qquad
	\Theta(N) = N\,\delta\theta,
	\label{nzero:eq:reversal_rotation}
\end{equation}
where $N$ is the number of turns the combined field is gated on and
$\delta\theta$ is the spin rotation accumulated per turn. The per-turn
rotation angle is set by the integrated transverse field through the
same T-BMT relation as the tipping kick from Eq.\,\eqref{eq:alpha},
\begin{equation}
	\delta\theta = \frac{(1+G)}{B\rho}\int B_\perp\,\dd l,
	\label{nzero:eq:reversal_perturn}
\end{equation}
with $B\rho = p/e$ the magnetic rigidity and $\int B_\perp\,\dd l$ the
integrated resonant transverse field per revolution.

Writing the initial polarization as
$\vec P(0) = (P_x, P_y, P_z)$, the rotation about $\ex$ leaves the
radial projection unchanged and rotates the $(P_y, P_z)$ pair,
\begin{equation}
	\Rmat_{\ex}(\Theta)\,\vec P
	=
	\bigl(
	P_x,\;
	P_y\cos\Theta + P_z\sin\Theta,\;
	-P_y\sin\Theta + P_z\cos\Theta
	\bigr).
	\label{nzero:eq:reversal_components}
\end{equation}
A clean reversal of the stored vertical polarization is obtained at
$\Theta = \pi$,
\begin{equation}
	\Rmat_{\ex}(\pi)\,(P_x, P_y, P_z)
	= (P_x,\,-P_y,\,-P_z),
	\label{nzero:eq:reversal_pi}
\end{equation}
so that $P_y \to -P_y$ with its magnitude preserved, the longitudinal
component is likewise reversed, and the radial component on the rotation
axis is conserved. During the buildup, $P_z$ oscillates as
$\Theta$ advances from $0$ to $\pi$ while $P_y$ reverses monotonically,
as shown in Fig.~\ref{nzero:fig:rf_wien_flip}; the radial component
$P_x$ remains constant throughout. A single fixed-axis rotation
therefore reverses the two components transverse to $\ex$ but not
the on-axis component, which is the operationally relevant behavior
here: the goal is to reverse the stored vertical polarization $P_y$, and
hence, through the interaction-region spin rotators that map the stable
vertical direction onto the longitudinal direction at the interaction
point, to reverse the sign of the longitudinal polarization delivered to
the ePIC detector for a selected bunch. Reversing $P_y$ on a single
bunch in this way provides a powerful handle on the systematics of
spin-dependent observables, since bunches of opposite longitudinal
polarization sign traverse the same detector under otherwise identical
conditions.

The reversal is dosed by the accumulated angle $\Theta = N\,\delta\theta$.
The combined field is gated on for the number of turns required to reach
$\Theta = \pi$,
\begin{equation}
	N_\pi = \frac{\pi}{\delta\theta}
	= \frac{\pi\,B\rho}{(1+G)\int B_\perp\,\dd l},
	\label{nzero:eq:reversal_Nturns}
\end{equation}
which fixes the relation between the integrated field per turn and the number of turns of operation: a larger integrated field per turn reaches the reversal in fewer turns, and the product $N_\pi\,(1+G)\int B_\perp\,\dd l / B\rho = \pi$ is held fixed. Operating for $\Theta > \pi$ rotates the polarization back past the reversal, while $\Theta < \pi$ leaves a partial flip with a residual in-plane projection; the reversal fidelity is therefore set by how accurately $N\,\delta\theta$ is tuned to $\pi$, the same quantity characterized by the per-flip efficiency $\varepsilon_\mathrm{SF}$ reported in the COSY bunch-gated demonstration~\cite{PhysRevResearch.7.023257}.

\subsection{Implementation options}

In contrast to an RF dipole flipper, which necessarily generates a transverse Lorentz force and drives coherent betatron oscillations, the RF Wien filter preserves the beam orbit by construction,
\begin{equation}
	\vec{F}_L = q\bigl(\vec{E} + \vec{v} \times \vec{B}\bigr) = 0.
\end{equation}
For an RF dipole, the driven orbit motion couples the spin dynamics to the transverse beam motion, leading to amplitude-dependent resonance strengths and phase decoherence across the beam, which limits the achievable spin-flip efficiency, as observed in RHIC spin-flipper studies~\cite{Bai:2012spinflipper}. By eliminating orbit excitation, the RF Wien filter decouples spin manipulation from the transverse dynamics and provides a more uniform and controlled spin rotation across the beam ensemble.

Three practical implementation options can be considered. The first is to combine a conventional RF Wien filter providing the transverse field $B_x(t)$ with a second device generating a longitudinal RF magnetic field $B_z(t)$, phase-locked with a relative phase shift of $90^\circ$. The second is to develop an integrated RF structure that generates both $B_x(t)$ and $B_z(t)$ within a single device. The third is to operate the RF Wien filter in a pulsed, phase-synchronous mode:
\begin{enumerate}
	\item \textbf{RF Wien filter plus longitudinal-field device.} A conventional RF Wien filter provides the transverse magnetic field $B_x(t)$ together with the compensating electric field $E_y(t) = -v_z\, B_x(t)$ that enforces $\vec F_L = q\,(\vec E + \vec v \times \vec B) = 0$ on the nominal closed orbit, so that the device acts purely on the spin without exciting orbit motion. A second device, phase-locked to the first with a relative phase of $90^\circ$, provides the longitudinal magnetic field $B_z(t)$. Because the longitudinal field is parallel to the beam velocity, $\vec v \times \vec B_z = 0$ and no compensating electric field is required for the second device: the orbit is preserved automatically. The combination \begin{equation*} 
		B_x(t) \propto \cos(\omega t), \qquad B_z(t) \propto \sin(\omega t) 
	\end{equation*} 
	realizes a single rotating field component $e^{+i\omega t}$ in the horizontal plane, which selects one of the two resonance branches and produces a unique direction of cumulative spin rotation, as required to break the $\nu_s = 1/2$ degeneracy of Sec.~\ref{nzero:sec:field-structure}. Each device is a well-characterized stand-alone component; the cost of this configuration is the mechanical layout and phase-stability requirement of operating two synchronized RF assemblies. A natural hardware-reuse possibility is to operate the longitudinal kicker of Sec.~\ref{sec:fid-echo-sequence} as the second device, driven in continuous-wave mode at the spin-resonance frequency rather than in the pulsed mode used for the tip-$\pi$-restore cycle.
	
	\item Integrated multi-electrode structure:  
	A dedicated RF structure can be designed to generate both $B_x(t)$ and $B_z(t)$ within a single device, for example using an extended multi-electrode geometry with independently driven field components.
	
	\item Pulsed RF Wien filter.
\end{enumerate}

\subsection{Pulsed-mode RF Wien filter}

In the pulsed mode, the RF Wien filter is operated using short, phase-synchronized excitations rather than a continuous wave. In continuous operation, as discussed in Sec.\,\ref{nzero:sec:field-structure}, a linearly oscillating field $B_x(t) \propto \cos(\omega t)$ contains two counter-rotating components $e^{+i\omega t}$ and $e^{-i\omega t}$. At $\nu_s = \tfrac{1}{2}$, both components satisfy the resonance condition $f_{\mathrm{RF}} = (k+\nu_s)f_{\mathrm{rev}}$, and therefore no unique rotation sense is defined.

In contrast, the pulsed scheme samples the RF field at discrete times $t_n = n/f_{\mathrm{rev}}$ and enforces a controlled phase relation with respect to the spin precession. The phase of the excitation,
\begin{equation}
	\phi_n = 2\pi Q_{\mathrm{osc}} n + \phi_0, \qquad Q_{\mathrm{osc}} = k + \nu_s,
\end{equation}
is matched to the spin phase advance such that the kicks add coherently from turn to turn. Under this condition, only the component co-rotating with the spin motion contributes to the cumulative rotation, while the counter-rotating component does not build up.

The spin manipulation is thus realized as a sequence of discrete kicks applied once per turn, whose coherent sum produces a net rotation over many turns. In this way, the pulsed scheme effectively selects a single resonance branch in the time domain, without requiring a physically rotating magnetic field or additional field components such as $B_z(t)$.

The numerical example shown in Fig.~\ref{nzero:fig:rf_wien_flip} corresponds to this idealized situation in which a single effective rotation sense is selected. The figure therefore illustrates the coherent spin evolution obtained when either the $+\omega$ or the $-\omega$ branch is effectively isolated. A purely linearly oscillating RF field at $\nu_s = \tfrac{1}{2}$ does not automatically realize this condition, and additional measures such as the pulsed operation described here are required.

\end{document}